\documentclass[onecolumn]{article}

\usepackage{a4,set_habil,epsfig,rotating,latexsym,color}
\usepackage[sort,noadjust]{cite}
\usepackage{mathrsfs}
\usepackage{amssymb}
\usepackage{amsfonts}
\usepackage{varioref}
\usepackage{array}

\newcommand{\e}{{\rm e}}
\newcommand{\Tr}{{\rm Tr}}
\begin{document}
\title{Applications of Large Random Matrices in Communications Engineering}
\author{Ralf R.\ M\"uller, Giusi Alfano, Benjamin M.\ Zaidel, Rodrigo de Miguel}
\maketitle

%
\setlength\unitlength{1mm}

\newcommand{\insertfig}[3]{
\begin{figure}[tbp]\begin{center}\begin{picture}(120,90)
\put(0,-5){\includegraphics[width=12cm,height=9cm,clip=]{#1.eps}}\end{picture}\end{center}
\caption{#2}\label{#3}\end{figure}}

\newcommand{\inserttwofig}[4]{
\begin{figure}\begin{center}\begin{picture}(150,170)
\put(10,80){\includegraphics[with=8.5cm]{#1.ps}}
\put(10, 0){\includegraphics[with=8.5cm]{#2.ps}}
\end{picture}\end{center}
\caption{#3}\label{#4}\end{figure}}


\newfont{\bb}{msbm10 scaled 1100}
\newfont{\bbs}{msbm10 scaled 917}
\newcommand{\CC}{\mbox{\bb C}}
\newcommand{\CCs}{\mbox{\bbs C}}
\newcommand{\RR}{\mbox{\bb R}}
\newcommand{\RRs}{\mbox{\bbs R}}
\newcommand{\ZZ}{\mbox{\bb Z}}
\newcommand{\FF}{\mbox{\bb F}}
\newcommand{\GG}{\mbox{\bb G}}
\newcommand{\NN}{\mbox{\bb N}}


\newcommand{\av}{{\matr a}}
\newcommand{\bv}{{\matr b}}
\newcommand{\cv}{{\matr c}}
\newcommand{\dv}{{\matr d}}
\newcommand{\ev}{{\matr e}}
\newcommand{\fv}{{\matr f}}
\newcommand{\gv}{{\matr g}}
\newcommand{\hv}{{\matr h}}
\newcommand{\iv}{{\matr i}}
\newcommand{\jv}{{\matr j}}
\newcommand{\kv}{{\matr k}}
\newcommand{\lv}{{\matr l}}
\newcommand{\mv}{{\matr m}}
\newcommand{\nv}{{\matr n}}
\newcommand{\ov}{{\matr o}}
\newcommand{\pv}{{\matr p}}
\newcommand{\qv}{{\matr q}}
\newcommand{\rv}{{\matr r}}
\newcommand{\sv}{{\matr s}}
\newcommand{\tv}{{\matr t}}
\newcommand{\uv}{{\matr u}}
\newcommand{\wv}{{\matr w}}
\newcommand{\vv}{{\matr v}}
\newcommand{\xv}{{\matr x}}
\newcommand{\yv}{{\matr y}}
\newcommand{\zv}{{\matr z}}
\newcommand{\zerov}{{\matr 0}}
\newcommand{\onev}{{\matr 1}}


\newcommand{\Am}{{\matr A}}
\newcommand{\Bm}{{\matr B}}
\newcommand{\Cm}{{\matr C}}
\newcommand{\Dm}{{\matr D}}
\newcommand{\Em}{{\matr E}}
\newcommand{\Fm}{{\matr F}}
\newcommand{\Gm}{{\matr G}}
\newcommand{\Hm}{{\matr H}}
\newcommand{\Id}{{\matr I}}
\newcommand{\Jm}{{\matr J}}
\newcommand{\Km}{{\matr K}}
\newcommand{\Lm}{{\matr L}}
\newcommand{\Mm}{{\matr M}}
\newcommand{\Nm}{{\matr N}}
\newcommand{\Om}{{\matr O}}
\newcommand{\Pm}{{\matr P}}
\newcommand{\Qm}{{\matr Q}}
\newcommand{\Rm}{{\matr R}}
\newcommand{\Sm}{{\matr S}}
\newcommand{\Tm}{{\matr T}}
\newcommand{\Um}{{\matr U}}
\newcommand{\Wm}{{\matr W}}
\newcommand{\Vm}{{\matr V}}
\newcommand{\Xm}{{\matr X}}
\newcommand{\Ym}{{\matr Y}}
\newcommand{\Zm}{{\matr Z}}


\newcommand{\Ac}{{\cal A}}
\newcommand{\Bc}{{\cal B}}
\newcommand{\Cc}{{\cal C}}
\newcommand{\Dc}{{\cal D}}
\newcommand{\Ec}{{\cal E}}
\newcommand{\Fc}{{\cal F}}
\newcommand{\Gc}{{\cal G}}
\newcommand{\Hc}{{\cal H}}
\newcommand{\Ic}{{\cal I}}
\newcommand{\Jc}{{\cal J}}
\newcommand{\Kc}{{\cal K}}
\newcommand{\Lc}{{\cal L}}
\newcommand{\Mc}{{\cal M}}
\newcommand{\Nc}{{\cal N}}
\newcommand{\Oc}{{\cal O}}
\newcommand{\Pc}{{\cal P}}
\newcommand{\Qc}{{\cal Q}}
\newcommand{\Rc}{{\cal R}}
\newcommand{\Sc}{{\cal S}}
\newcommand{\Tc}{{\cal T}}
\newcommand{\Uc}{{\cal U}}
\newcommand{\Wc}{{\cal W}}
\newcommand{\Vc}{{\cal V}}
\newcommand{\Xc}{{\cal X}}
\newcommand{\Yc}{{\cal Y}}
\newcommand{\Zc}{{\cal Z}}


\newcommand{\alphav}{{\matr\alpha}}
\newcommand{\betav}{{\matr\beta}}
\newcommand{\gammav}{{\matr\gamma}}
\newcommand{\deltav}{{\matr\delta}}
\newcommand{\etav}{{\matr\eta}}
\newcommand{\lambdav}{{\matr\lambda}}
\newcommand{\epsilonv}{{\matr\epsilon}}
\newcommand{\nuv}{{\matr\nu}}
\newcommand{\muv}{{\matr\mu}}
\newcommand{\zetav}{{\matr\zeta}}
\newcommand{\phiv}{{\matr\phi}}
\newcommand{\psiv}{{\matr\psi}}
\newcommand{\thetav}{{\matr\theta}}
\newcommand{\tauv}{{\matr\tau}}
\newcommand{\omegav}{{\matr\omega}}
\newcommand{\xiv}{{\matr\xi}}
\newcommand{\sigmav}{{\matr\sigma}}
\newcommand{\piv}{{\matr\pi}}

\newcommand{\Gammam}{{\matr\Gamma}}
\newcommand{\Lambdam}{{\matr\Lambda}}
\newcommand{\Deltam}{{\matr\Delta}}
\newcommand{\Sigmam}{{\matr\Sigma}}
\newcommand{\Phim}{{\matr\Phi}}
\newcommand{\Pim}{{\matr\Pi}}
\newcommand{\Psim}{{\matrPsi}}
\newcommand{\Thetam}{{\matrTheta}}
\newcommand{\Omegam}{{\matr\Omega}}
\newcommand{\Xim}{{\matr\Xi}}
\newcommand{\Upsilonm}{{\matr\Upsilon}}

\def \nT        {n_\mathrm{T}}
\def \nR        {n_\mathrm{R}}
\newcommand{\ThetaR}{{\matr\Theta}_{\rm R}}
\newcommand{\ThetaT}{{\matr\Theta}_{\rm T}}
\def \NC       {{\sf NC }}

\newcommand{\norm}[1]{\left\Vert#1\right\Vert}
\newcommand{\abs}[1]{\left\vert#1\right\vert}
\newcommand{\set}[1]{\left\{#1\right\}}
\newcommand{\rD}{\mathrm{D}}
\newcommand{\rd}{\mathrm{d}}
\newcommand{\im}{{\scriptsize \textsf{im}}}
\newcommand{\re}{{\scriptsize \textsf{re}}}
\newcommand{\tot}{{\scriptsize \textsf{tot}}}
\newcommand{\B}{\mathscr{B}}
\newcommand{\bE}{\bar{\mathscr{E}}}

\newcommand{\sFyz}{\gimel_s(y,z)}
\newcommand{\tsFyz}{\tilde\gimel_s(y,z)}
\newcommand{\sFypz}{\gimel_s(\tilde y,z)}
\newcommand{\tsFyzup}{\tilde\gimel_{\upsilon}(y,z)}

\newcommand{\rsb}{{\scriptsize \textsf{rsb1}}}

\newcommand{\iint}{\int\!\!\! \int}

\newtheorem{thm}{theorem}[section]
\newtheorem{assumption}[thm]{Assumption}
\newtheorem{prop}[thm]{Proposition}


\newcommand{\Esc}{\mathscr{E}}
\newcommand{\Ssc}{{\mathscr{S}}}

\newcommand{\Rrm}{\mathrm{R}}


\newcommand{\fzv}[2]{\noindent #1 \hfill \parbox{13.2cm}{#2}}
\def\mathlette#1#2{{\mathchoice{\mbox{#1$\displaystyle #2$}}%
                               {\mbox{#1$\textstyle #2$}}%
                               {\mbox{#1$\scriptstyle #2$}}%
                               {\mbox{#1$\scriptscriptstyle #2$}}}}
\newcommand{\matr}[1]{\mathlette{\boldmath}{#1}}
\newcommand{\SINR}{{sig\-nal--to--dis\-tor\-tion ratio}}
\newcommand{\SNR}{{sig\-nal--to--noise ratio}}
\newcommand{\snr}{{\rm SNR}}
\newcommand{\ebno}{{\frac{\eb}{\no}}}
\newcommand{\ebnozf}{\left(\ebno\right)_{\rm ZF}}
\newcommand{\MAC}{{multiple--ac\-cess chan\-nel}}
\newcommand{\vX}{\matr{\X}}
\newcommand{\vY}{\matr{\Y}}
\newcommand{\seq}{b}
\newcommand{\co}{g}
\newcommand{\mseq}{\matr{B}}
\newcommand{\tcp}{{\Sigmas}}
\newcommand{\x}{x}
\newcommand{\y}{y}
\newcommand{\df}{decision feedback}

\def\argmin{\mathop{\rm argmin}}
\def\argmax{\mathop{\rm argmax}}
\newcommand{\diag}{{\rm diag}}
\def\expect{\mathop{\mbox{\large $\E$}}}
\newcommand{\Hilbert}{{\cal H}}
\newcommand{\Landau}[1]{{\cal O}\left(#1\right)}
\newcommand{\landau}[1]{o\left(#1\right)}
\newcommand{\Var}[1]{{\rm Var}\left\{#1\right\}}
\newcommand{\comp}[1]{{#1^{\rm c}}}
\newcommand{\hermite}[1]{{#1^{\rm H}}}
\newcommand{\transp}[1]{{#1^{\rm T}}}
\newcommand{\conj}[1]{{#1^{\ast}}}
\newcommand{\iu}{{\rm j}}
\newcommand{\vnull}{{\rm \bf 0}}
\newcommand{\I}{{\rm \bf I}}
\newcommand{\1}{{\rm \bf 1}}
\newcommand{\0}{{\rm \bf 0}}
\newcommand{\chol}[1]{{\rm chol}\left(#1\right)}
\newcommand{\h}{{\rm ent}}
\newcommand{\prob}[2]{{\rm p}_{#1}\!\left( #2 \right) }
\newcommand{\Prob}[2]{{\rm P}_{#1}\!\left( #2 \right) }
\newcommand{\proba}[2]{\breve{\rm p}_{#1}\!\left( #2 \right) }
\newcommand{\Proba}[2]{\breve{\rm P}_{#1}\!\left( #2 \right) }
\newcommand{\Probi}[2]{{\rm P}_{#1}^{-1}\!\left( #2 \right) }
\newcommand{\StT}[2]{{\rm G}_{#1}\!\left( #2 \right) }
\newcommand{\StTi}[2]{{\rm G}_{#1}^{-1}\!\left( #2 \right) }
\newcommand{\ST}[2]{{\rm S}_{#1}\!\left( #2 \right) }
\newcommand{\RT}[2]{{\rm R}_{#1}\!\left( #2 \right) }
\newcommand{\YT}[2]{{\Upsilont}_{#1}\!\left( #2 \right) }
\newcommand{\YTi}[2]{{\Upsilont}_{#1}^{-1}\!\left( #2 \right) }
\newcommand{\Q}{{\rm Q}}
\newcommand{\load}{{\beta}}
\newcommand{\quant}{{\rm quant}}
\newcommand{\sign}{{\rm sign}}
\newcommand{\dirac}[1]{\deltaf \! \left( #1 \right)} 
\newcommand{\kron}[1]{\deltaf \! \left[ #1 \right]} 
\newcommand{\MomGen}[2]{{\Phif}_{#1}\left( #2 \right) }
\newcommand{\st}{\sigma_0}
\newcommand{\X}{{\uppercase\expandafter{\x}}}
\newcommand{\Y}{{\uppercase\expandafter{\y}}}
\newcommand{\Noise}{Z}
\newcommand{\tsf}{\Theta}
\newcommand{\nmtsf}{{\theta_{\min}}}
\newcommand{\mlatt}{\mu_{\ln \! \Att}}
\newcommand{\slatt}{\sigma_{\ln \! \Att}}
\newcommand{\varnoise}{\sigma_\noise^2}
\newcommand{\aux}{\Upsilon}
\newcommand{\define}{\stackrel{\triangle}{=}}
\newcommand{\D}{\displaystyle}
\newcommand{\eq}[1]{(\ref{#1})}
\newcommand{\eqs}[2]{(\ref{#1}) and (\ref{#2})}
\newcommand{\eqd}[3]{(\ref{#1}), (\ref{#2}), and (\ref{#3})}
\newcommand{\eqv}[4]{(\ref{#1}), (\ref{#2}), (\ref{#3}), and (\ref{#4})}
\newcommand{\zf}{{\rm ZF}}
\newcommand{\cd}{{\rm CD}}
\newcommand{\cf}{\beta}
\newcommand{\mmse}{{\rm MMSE}}
\newcommand{\chisquare}{$\chisq^2$}
\newtheorem{theorem}{Theorem}[section]
\newtheorem{lemma}{Lemma}[section]
\newtheorem{corollary}{Corollary}[section]
\newtheorem{definition}{Definition}[section]
\newcommand{\azeta}{\left(1-\sqrt{\zeta}\right)^2}
\newcommand{\bzeta}{\left(1+\sqrt{\zeta}\right)^2}
\newcommand{\bin}[2]{{\left(\!\begin{array}{c} #1 \\ #2 \end{array}\!\right)}}
\newcommand{\xyplot}[3]{\begin{tabular}{c}\begin{turn}{90}\begin{tabular}{c}
	#3 \\[-2mm] \begin{turn}{-90} #1 \end{turn}\end{tabular}\end{turn}
	\\[-2mm] \hphantom{\begin{turn}{90} #3 \end{turn}{90}} 
	\hspace*{\parskip} #2 \end{tabular}}

\begin{abstract}This work gives an overview of analytic tools for the design, analysis, and modelling of communication systems which can be described by linear vector channels such as $\matr y=\matr{Hx}+\matr z$ where the number of components in each vector is large.
Tools from probability theory, operator algebra, and  statistical physics are reviewed.
The survey of analytical tools is complemented by examples of applications in communications engineering.

Asymptotic eigenvalue distributions of many classes of random matrices are given.
The treatment includes the problem of moments and the introduction of the Stieltjes transform.

Free probability theory, which evolved from non-commutative operator algebras, is explained from a probabilistic point of view in order to better fit the engineering community.
For that purpose freeness is defined without reference to non-commutative algebras.  
The treatment includes additive and multiplicative free convolution, the R-transform, the S-transform, and the free central limit theorem.

The replica method developed in statistical physics for the purpose of analyzing spin glasses is reviewed from the viewpoint of its applications in communications engineering.
Correspondences between free energy and mutual information as well as energy functions and detector metrics are established.

These analytic tools are applied to the design and the analysis of linear multiuser detectors, the modelling of scattering in communication channels with dual antennas arrays, and the analysis of optimal detection for communication via code-division multiple-access and/or dual antenna array channels.

\end{abstract}  

\section{Introduction}
\label{Int}

In a multi-dimensional communication system many data streams are transmitted from various sources to various sinks via a common medium called a channel.
This situation is almost as old as mankind and appears whenever more than two people start discussing with each other.
Technical systems handling this task are telephone networks, both fixed and wireless, the internet, local area networks, computers' data buses, etc.

The complexity of such communication systems increases with the number of people or data streams to be handled simultaneously.
This rise in complexity is not limited to the hardware to be deployed, but also affects the design, the modeling, and the analysis of the system.
From an engineering point of view, it is particularly important to be able to predict the behavior of a technical system before it is actually built.
With an ever increasing number of people using various kinds of communications technology, this challenge appears to become, sooner or later, a hopeless task. At first sight, that is.

In a combustion engine, many fuel and oxygen molecules interact with each other.
However, although we cannot control the individual behavior of each molecule (and do not attempt to do so), we can trust that the mixture of gas and air will explode upon ignition, heat up, expand, and drive the engine.
Physicists have successfully built a theory that links the evolution of macroscopic quantities such as temperature and pressure to the microscopic behavior of molecules, which is only described statistically. The theory that builds the bridge between quantum mechanics (statistically describing the behavior of individual molecules) and thermodynamics (deterministically describing the properties of large systems) is known as statistical physics.
Simply the fact that there are \textit{enough} objects interacting randomly makes the collection of these objects obey certain rules.
This collective behavior depends, amongst other things, on the kind of interactions present, and it can be understood as a generalization of the law of large numbers.

Communication systems for multiple data streams may also be modeled by the statistical interactions between signals belonging to the different streams.
Provided that the number of data streams transmitted simultaneously through the system is large enough, similar effects as in thermodynamics should occur.
Indeed recent research has found theoretical and numerical evidence for the occurrence of phase transitions in large code-division multiple-access (CDMA) systems, cf.\ Section~\ref{RepMet}, which correspond to the hysteresis behavior of magnetic materials.

The use of microscopic statistical models to predict macroscopic quantities in physics is not limited to thermodynamics.
It was already used by Wigner in the 1950s in order to predict the spacings of nuclear energy levels.
Nowadays, it has become a well-established method in quantum physics to describe the energy-levels of heavy nuclei by the eigenvalue distributions of large dimensional random matrices \cite{mehta:91}.
Random matrices and their applications in communications engineering are discussed in Section~\ref{RanMatThe}.

The interest in {\em free probability theory} was originally driven by mathematicians, particularly those who were interested in operator algebras and the Riemann hypothesis.
When numerical evidence showed a striking connection between the zeros of Riemann's zeta function and the spacings of adjacent eigenvalues of large random matrices, physicists became interested in the subject hoping for new insights into their problems.
However, the first concrete meaning for the {\em R-transform}, one of the most fundamental concepts in free probability theory, was found in the theory of large CDMA systems.
A detailed discussion on free probability and its applications in communications engineering is given in Section~\ref{FreProThe}.

Random matrix and free probability theory are concepts well suited to analyze the interaction of many Gaussian random processes.
However, modern communication systems operate in finite (often binary) fields rather than real or complex numbers. This is reminiscent of the intrinsic angular momentum of particles (spin), which, by the laws of quantum mechanics, can only have a finite number of projections along any given direction. To analyze the thermodynamic properties of magnetic materials comprising many such spins, physicists have developed a powerful analytic tool known as the {\em replica method}.
The replica method is able to predict the macroscopic behavior of spin glasses just as well as bit error rates of maximum a-posteriori detectors for CDMA signals.
A detailed discussion of the replica method and its applications in communications engineering is given in Section~\ref{RepMet}.

Mathematical formulas allow for simplications when variables approach either zero or infinity.
This work aims to show that engineering can also get easier as system size becomes either large or small.
Decades ago communication engineers already made use of this effect summarizing all kinds of additive distortion in the universal additive Gaussian noise.
The authors believe that Gaussian noise was just the beginning, and that the semicircle distribution and the R-transform will be become standard tools in multi-user communications, just as common as the Gaussian distribution and the Fourier transform are in single user digital communications.

\section{Communication Systems}
\label{ComSys}

The principles summarized in this paper apply to a broad class of communication channels.
We do not aim to cover all
of them, and restrict ourselves to the discrete vector-valued additive noise channel,
which is general enough to develop rich examples, and simple enough to keep equations illustrating.

In vector notation the vector-valued noise channel is given by
\begin{equation}
\label{AWGN}
\matr y[\nu]=\matr H[\nu] \matr x[\nu] + \matr n[\nu]
\end{equation}
with 
\begin{itemize}
\item the $K\times 1$ vector of transmitted symbols $\matr x[\nu]$, 
\item the $N\times 1$ vector of received symbols $\matr y[\nu]$,
\item the $N\times K$ channel matrix $\matr H[\nu]$,
\item the $N\times 1$ vector of additive noise $\matr n[\nu]$,
 \item and discrete time $\nu$.
\end{itemize}
In order to simplify notation, the time index will be dropped whenever it is not needed to explicitly express the dependence on discrete time.

It is well known in literature \cite{verdu:98} that, for Gaussian noise that is i.i.d., i.e. 
\begin{equation}
\label{iidnoise}
\expect \matr n[\nu] \matr n^{\dagger}[\nu^\prime]=\sigma_0^2 \I \delta_{\nu\nu^\prime}, 
\end{equation}
the signal
\begin{eqnarray}
\label{AWGN2}
\matr r[\nu] &=& \matr H^{\dagger}[\nu] \matr y[\nu]\\
&=& \matr H^{\dagger}[\nu] \matr H[\nu] \matr x[\nu] + \matr H^{\dagger}[\nu] \matr n[\nu]
\end{eqnarray}
provides sufficient statistics for the estimation of the signal $\matr x[\nu]$.
This means that all information about $\matr x[\nu]$ that could be extracted from the received signal $\matr y[\nu]$ can also be extracted from the signal $\matr r[\nu]$.
For i.i.d.\ Gaussian noise, the two channels \eqs{AWGN}{AWGN2} are actually equivalent in terms of all performance measures such as bit error rate, signal-to-noise ratio, channel capacity, etc.

Channels \eqs{AWGN}{AWGN2} appear in several areas of wireless and wireline communications:
\begin{itemize}
\item
In CDMA, the components of the vector $\matr x$ are regarded as the signals of $K$ individual users while the matrix $\matr H$ contains their spreading sequences as columns. See e.g.\ \cite{verdu:98}.
\item
In antenna array communications, the components of the vectors $\matr x$ and $\matr y$ represent the signals sent and received by the $K$ transmit and $N$ receive antenna elements, respectively. See e.g.\ \cite{tse:05}.
\item
In cable transmission, the components of the vector $\matr x$ contain the signals sent on the bundled twisted pairs within a cable. 
The coefficients in the matrix $\matr H^{\dagger}\matr H$ describe the electromagnetic crosstalk between the respective twisted pairs. See e.g.\ \cite{lee:07}.
\item
For block transmission over a dispersive channel, the components of the vectors $\matr x$ and $\matr y$ contain the symbols sent and received consecutively in time. 
Discrete time $\nu$ counts blocks, and the matrix $\matr H$ is a circulant matrix of the channel's discrete-time impulse response. See e.g.\ \cite{haykin:96}.
\item
In orthogonal frequency-division multiplexing (OFDM), the components of the vectors $\matr x$ and $\matr r$ represent the $K$ sub-carriers at transmitter and receiver site, respectively, and the matrix $\matr H^{\dagger}\matr H$ accounts for inter-carrier interference. See e.g.\ \cite{rugini:11}.
\item
In (linear)\footnote{Nonlinear codes have very limited practical importance.} forward-error correction coding, the components of the vectors $\matr x$ and $\matr y$ are the information symbols and code symbols, respectively. The matrix $\matr H$ is called the generator matrix of the code. Discrete time counts codewords. In this application, summation and multiplication run in finite, often binary, fields. See e.g.\ \cite{mackay:03}.
\item
In compressed sensing, the components of the vectors $\matr x$ and $\matr y$ are the ($K$-dimensional) true data and the ($N$-dimensional) sensed observations, respectively. The $N\times K$ matrix $\matr H$ is the sampling basis, characterized by the fact of being generally $N\ll K$. See e.g.\ \cite{donoho:06}. 

\end{itemize}

Irrespective of the application one has in mind, the performance of digital communication via channel \eq{AWGN} can be analyzed for a variety of receiver algorithms and assumptions on the properties of the channel matrix $\matr H$.
Numerous results are reported in literature \cite{verdu:98,proakis:00} and we do not attempt
to be fully comprehensive in this review.
In the following, we summarize only those of the known results which are needed in subsequent sections.

\subsection{Uncoded Error Probability}
\label{UncErrPro}

The purpose of a detector is to estimate a transmitted signal $\hat{\matr x}$ given the observation of the received signal $\matr y$ (or equivalently the sufficient statistics $\matr r$).
The larger the overlap between the estimated signals and the signals actually sent, the better the detector is.

When the estimated signal and the true signal do not fully overlap, an error occurs.
Two kinds of errors are defined:
\begin{description}
\item[]{\bf Symbol errors} occur if the estimated vector differs from the transmitted vector, i.e.\ $\hat{\matr x}\ne \matr x$.
\item[]{\bf Bit errors} occur if a component of the estimated vector differs from the respective component of the transmitted vector, i.e.\ $\hat{x}_k\ne x_k$ for some $k$.
\end{description}
In literature \cite{verdu:98}, the detectors minimizing symbol and bit error probability are called {\em jointly} and {\em individually} optimum detectors, respectively. 
The only difference between the cost functions of the two optimization criteria is that the individually optimum detector counts multiple differences within the components of the same symbol, while the jointly optimum detector does not do so.

Both detectors are special cases of the {\em marginal posterior mode} detector.
The marginal posterior mode detector tries to individually minimize bit error probability under the generally false assumption that the variance of the noise is $\sigma^2$ instead of $\st^2$. 
The assumed noise variance $\sigma^2$ is a control parameter of the marginal posterior mode detector.
If the assumed noise variance is the true noise variance, i.e.\ $\sigma=\st$, it becomes the individually optimum detector.
If the assumed noise variance tends to zero, i.e.\ $\sigma\to0$, it becomes the jointly optimum detector \cite{verdu:98}.

For white Gaussian noise, the received signal $\matr y$ conditioned on the transmitted signal $\matr x$ and the channel realization $\matr H$ follows a probability distribution with density
\begin{equation}
\prob{\matr y|\matr x,\matr H}{\matr y,\matr x,\matr H}={\frac{\e^{-\st^{-2}\left(\matr y-\matr H\matr x\right)^{\dagger}\left(\matr y-\matr H\matr x\right)}}{\left(\pic\st\right)^K}}.
\end{equation}
By Bayes' law, we find
\begin{equation}
\prob{\matr x|\matr y,\matr H}{\matr x,\matr y,\matr H}=\frac{\prob{\matr y|\matr x,\matr H}{\matr y,\matr x,\matr H}\prob{\matr x|\matr H}{\matr x,\matr H}}{\prob{\matr y|\matr H}{\matr y,\matr H}}.
\end{equation}
Thus, the marginal posterior mode detector with perfect channel state information at receiver side and parameter $\sigma$ is given by
\begin{equation}
\hat{x}_k(\sigma) = \argmax\limits_{\xi} \sum\limits_{\matr x:x_k=\xi} \e^{-\sigma^{-2}\left(\matr y-\matr H\matr x\right)^{\dagger}\left(\matr y-\matr H\matr x\right)}\prob{\matr x|\matr H}{\matr x,\matr H}.
\label{MPMdet}
\end{equation}
For general matrices $\matr H$, the optimization problem \eq{MPMdet} cannot be solved with polynomial complexity in $K$ \cite{verdu:98}.
Bit error probability is given by
\begin{equation}
P_{\rm b}(\sigma) = \frac1K\sum\limits_{k=1}^K\Pr\big(\hat{x}_k(\sigma)\ne x_k\big).
\label{Uncerr}
\end{equation}
\subsection{Linear Detection}
\label{LinDet}

Since the complexity of the optimum detectors is exponential in $K$, they are often infeasible in practice.
In order to circumvent this obstacle, suboptimum approaches were proposed, see e.g.\ \cite{moshavi:96,verdu:98,cottatellucci:04,honig:09} for a survey of literature. 
The most common family of suboptimum detectors are the linear detectors.
They implement the detection rule
\begin{equation}
\matr{\hat{x}} = \quant_{\cal X}\left(\matr{Ly}\right)
\end{equation}
where $\matr L$ is a $K\times N$ matrix whose construction characterizes the linear detector and the function $\quant_{\cal X}(\cdot)$ quantizes a vector component-wise towards the nearest element of the symbol alphabet $\cal X$.
For binary antipodal transmission, i.e.\ ${\cal X} = \{+1,-1\}$, the function $\quant_{\cal X}(\cdot)$ is the sign function. 

The best linear detector in terms of bit error probability is very difficult to construct \cite{verdu:98}.
Minimizing the mean-squared error
\begin{equation}
\label{defMSE}
\expect\limits_{\matr x,\matr n|\matr H}\left(\matr x-\matr{Ly}\right)^{\dagger}\left(\matr x-\matr{Ly}\right)
\end{equation}
instead, results in a linear detector performing similarly well.
This detector results from applying the optimum decision rule \eq{MPMdet} under the generally false assumption that the symbol alphabet is Gaussian distributed with true mean and true variance.
It can be expressed explicitly in terms of channel matrix $\matr H$, true noise variance $\st^2$, and signal power $P$ as \cite{verdu:98}
\begin{equation}
\matr{\hat{x}} = \quant_{\cal X} \left(\sqrt P\left(\st^2\mbox{\bf I}+P\matr H^{\dagger}\matr H\right)^{-1}\matr H^{\dagger}\matr y\right).
\label{detMMSE}
\end{equation}

The detector \eq{detMMSE} also maximizes the ratio of useful signal powerto interference and noise power (SINR).
For the $k^{\rm th}$ component of the vector $\matr x$, this ratio is given by \cite{verdu:98}
\begin{equation}
{\rm SINR}_k=P\matr h_k^{\dagger}\left(\st^2\mbox{\bf I}+P\matr {HH}^{\dagger}-P\matr h_k\matr h_k^{\dagger}\right)^{-1}\matr h_k, 
\label{eq2.17}
\end{equation}
where $P$ is the useful signal power, and $\matr h_k$ denotes the $k^{\rm th}$ column of $\matr H$.
With the singular value decomposition 
\begin{equation}
\matr H=\matr V^{\dagger}\sqrt{\matr\Lambda}\matr U
\end{equation} 
and the matrix inversion lemma, the SINR can also be written as
\begin{eqnarray}
{\rm SINR}_k&=&P\matr u_k^{\dagger}\left(\st^2\matr\Lambda^{-1}+P\,\mbox{\bf I}-P\matr u_k\matr u_k^{\dagger}\right)^{-1}\matr u_k\\
&=&\frac1{1-P\matr u_k^{\dagger}\left(\st^2\matr\Lambda^{-1}+P\,\mbox{\bf I}\right)^{-1}\matr u_k}-1.
\label{SINRlindet}
\end{eqnarray}
Note that the SINR depends on all eigenvalues of the matrix $\matr H^{\dagger}\matr H$, but only on the $k^{\rm th}$ column of the eigenvector matrix $\matr U$.
Note, however, that due to definition \eq{eigdecomp}, the eigenvectors are not the columns, but the rows of $\matr U$.

\subsection{Vector Precoding}
\label{VecPre}


Another example for a fundamental system model that falls within the framework of (\ref{AWGN}) is that of the \emph{MIMO Gaussian broadcast channel} (GBC). The model corresponds, in particular, to the downlink channel in cellular communication systems, when multiple transmit antennas are employed at the cell-sites. The MIMO GBC has drawn a lot of attention in recent years in view of the worldwide overwhelmingly increasing demand for high data-rate mobile communication services. For simplicity, we focus here on the more practical setting in which multiple transmit antennas are employed at the cell-site, while each mobile receiver is only equipped with a single receive antenna. It is noted, however, that multiple receive antennas at the mobile terminals can be straightforwardly incorporated into the model as well.

Referring to (\ref{AWGN}) 
in the context of the cellular downlink 
and considering the $\nu$th time instance, we have that 
\begin{itemize}
\item $\xv[\nu]$ represents the $K\times 1$ vector of symbols transmitted by the cell-site (where $K$ designates the number of transmit antennas),
\item $\yv[\nu]$ represents the vector of signals received by the $N$ (single antenna) mobile users,
\item $\Hm$ is the channel transfer matrix, where $[\Hm]_{i,j}$ is the channel coefficient between the $j$th transmit antenna and the $i$th user,
\item $\nv[\nu]$ is the vector of noise samples at the receivers of the $N$ users, assumed to be independent identically distributed zero-mean proper\footnote{A complex random variable is said to be {\em proper} if real and imaginary part are independent and identically distributed \cite{neeser:93}.} complex Gaussian random variables satisfying (\ref{iidnoise}).
\end{itemize}

The strategy which achieves the information theoretic capacity for the MIMO GBC is known as \emph{dirty paper coding} (DPC) \cite{Costa-83,Weingarten-Steinberg-Shamai-2006}. DPC is a transmission scheme that \emph{effectively} eliminates the impact of additive interference if known \emph{noncausally} at the transmitter, but unknown at the receiver. In the GBC setting, while employing a successive encoding procedure, codewords corresponding to previously encoded users' messages can be treated as noncausally known interference. Thus, applying DPC, the part of the transmitted signal corresponding to these codewords effectively induces no interference on subsequently encoded messages. The downside of DPC is its prohibitive computational complexity, which motivates the search for more simplified suboptimal alternatives (see, e.g., \cite{Zamir-Shamai-Erez-2002,Erez-ten-Brink-2005,Bennatan-Brustein-Caire-Shamai-2006} for some practically oriented attempts of implementing DPC).

One such alternative transmission scheme for the MIMO GBC is commonly referred to as \emph{vector precoding} (see, e.g., \cite{fischer:02,hochwald:05,mueller:08,zaidel:11} and references therein). The scheme is inspired by the idea of Tomlinson-Harashima Precoding (THP) \cite{tomlinson:71,harashima:72}, and can be generally described as follows, while adhering to the representation in \cite{mueller:08,zaidel:11}.

The precoding process, namely the construction of the signal vector $\xv$ transmitted by the cell-site, can be summarized in three consecutive stages.
The first stage performs encoding of the users' messages. The encoding stage comprises a bank of $N$ independent \emph{single-user} encoders. That is, each encoder produces a codeword per each of the corresponding user's messages, independently of the other users' codewords (in contrast to DPC).
We denote the code-symbols produced by the encoders at some arbitrary time-instance by $\set{s_n}_{n=1}^N$, and assume for simplicity that all symbols are taken from the same \emph{discrete} alphabet $\Ssc$. We also use $\sv_{[N\times 1]}$ to compactly denote the vector of the encoders' outputs, i.e., $\sv=[s_1, \dots,s_N]^T \in \Ssc^N$.
The code symbols are commonly treated as random variables, independent across users, and subject (by assumption) to some identical underlying discrete probability distribution $P_S(\tilde{s})$, $\tilde{s}\in\Ssc$. For convenience, the following probability density function (pdf) formulation shall be used henceforth
\begin{equation}\label{eq: Definition of dP_s}
\rd F_S(s) \triangleq f_S(s) \, \rd s \triangleq \sum_{\tilde{s} \in \Ssc} P_S(\tilde{s}) \delta(s-\tilde{s})\,  \rd s \quad .
\end{equation}

The second stage takes $\sv$ as input, performs a nonlinear operation, and outputs the $N$-dimensional vector $\wv$, a function of $\sv$ and $\Hm$ as described in detail below.  The third and final stage comprises a linear front-end. It multiplies the vector $\wv$ by a matrix $\Tm_{[K \times N]}$, which is, in general, a function of the channel transfer matrix $\Hm$. The result is then normalized so that the transmitted vector $\xv_{[K\times 1]}$ satisfies an instantaneous \emph{total} power (energy per symbol) constraint $P_\tot$. More specifically, $\xv$ is given by
\begin{equation}
\label{eq: Relation of transmitted vector to the vector of coded symbols}
\xv=\sqrt{P_\tot} \, \frac{\Tm\wv}{\norm{\Tm\wv}} \triangleq \sqrt{\frac{{P_\tot}}{{\mathscr{E}^\tot(\Tm,\wv)}}} \, \Tm\wv \quad,
\end{equation}
where $\mathscr{E}^\tot(\Tm,\wv)$ denotes the ``energy penalty" induced by the precoding matrix $\Tm$, and the particular choice of $\wv$, as well as the average symbol energy of the underlying alphabet $\Ssc$ (the explicit dependence on the arguments is henceforth omitted for simplicity). 
Denoting by $P$ the individual power constraint \emph{per user} (taken as equal for all), so that $P_\tot=N P$, we define  the \emph{transmit} SNR as 
\begin{equation}
\label{eq: Definition of snr}
\snr \triangleq \frac{P_\tot}{N \sigma^2} = \frac{P}{\sigma^2} \quad.
\end{equation}

The objective of the second (nonlinear) stage of the precoding scheme is to minimize the ``energy penalty" $\Esc^\tot$. This is performed by extending (``relaxing'') the original input alphabet $\Ssc$ to a larger alphabet set $\B=\bigcup_{\tilde{s}\in\Ssc} \B_{\tilde{s}}$, where the subsets $\set{\B_{\tilde{s}}}$ are \emph{disjoint}. The idea here is that every coded symbol $s\in \Ssc$ can be represented \emph{without ambiguity} using any element of $\B_s$  \cite{mueller:08}.
The vector $\wv=[w_1,\dots,w_N]^T$ thus satisfies
\begin{equation}
\label{eq: Definition of x as argmin}
\wv ={\argmin}_{{\tilde{\wv}} \in \B_{s_1}\times \cdots \times \B_{s_N}} \norm{\Tm{\tilde{\wv}}}^2 \quad .
\end{equation}
Commonly, $\B$ is taken as the \emph{complex integer lattice} in which case the minimization in (\ref{eq: Definition of x as argmin}) is NP-hard (often implemented using the sphere-decoding algorithm \cite{Agrell-Eriksson-Vardy-Zeger-2002}). However, a notable feature of vector precoding in its current formulation is that it allows for the use of \emph{convex} extended alphabets, which lend themselves to \emph{efficient} practical energy penalty minimization algorithms.

Let 
\begin{equation}
\label{eq: Definition of energy penalty per symbol}
\bE \triangleq \frac{\mathscr{E}^\tot}{N} \quad ,
\end{equation}
denote the energy penalty \emph{per symbol}. Then, in order to differentiate between the energy penalty induced by the precoding scheme, and the effect of the underlying symbol energy of the input alphabet $\Ssc$, one can alternatively represent the quantities of interest in terms of what we refer to here as the \emph{precoding efficiency}, defined through
\begin{equation}
\label{eq: Factorization of the energy penalty with sig-u}
\zeta \triangleq \frac{\bE}{\sigma_s^2} \quad ,
\end{equation}
where $\sigma_s^2 = \E\{\abs{s}^2\}$ (with the expectation taken with respect to (\ref{eq: Definition of dP_s})). Note in this respect that the normalization in (\ref{eq: Relation of transmitted vector to the vector of coded symbols}) makes the system insensitive to any scaling of the underlying alphabet $\Ssc$.

One should also note here that  a weaker \emph{average} transmit power constraint can also be straightforwardly applied in this setting, by replacing $\mathscr{E}^\tot$ with $\E\set{\mathscr{E}^\tot}$ in (\ref{eq: Relation of transmitted vector to the vector of coded symbols}). However, the two types of energy constraints yield the same asymptotic results in the large system limit, on which we focus in the sequel (see in particular Section \ref{Exa52}).  We thus restrict the discussion for convenience to the instantaneous power constraint implied by (\ref{eq: Relation of transmitted vector to the vector of coded symbols}).



\subsection{Compressed Sensing}
\label{ComSen}

Compressed Sensing is a branch of signal processing, devoted to the reconstruction of signals and fields from an \emph{apparently} under dimensioned number of collected samples. Indeed, while most of the signals arising from natural phenomena do present non-zero spectral components in a relatively large range of frequencies, there can be some which exhibit the feature of being the superposition of relatively few meaningful spectral components. These signals could be referred to, without loss of generality, as \emph{spectrally sparse} or, more in general, as \emph{sparse} signals. Of course, the concept of sparseness can be applied to any domain of the signal for which its samples are meaningfully collected, not just the frequency domain.  If  information on the  signal sparseness were  available already  at the sampling stage, one could think of  obtaining  a compressed representation of the signal by first computing the expansion coefficients in a properly chosen  basis, and then keeping only the largest coefficients while setting the rest to zero. This is, however, time and resource consuming, in that only few coefficients eventually survive the post-sampling processing, while the sampling process could, according to Nyquist theory, require the collection of a huge amount of data.

The main purpose of compressed sensing is obtaining a suitable compressed version of a signal by performing only a small number of linear and non-adaptive measurements. Compressed sensing theory, in its very basic definition,  predicts that recovering a sparse  signal from these undersampled measurements
can be done with computationally efficient methods, provided some conditions are fulfilled. Since we are dealing with linear measurements, with reference to the previously introduced model (\ref{AWGN}), the constraints leading to efficient reconstruction of sparse signals should be met by the measurements matrix $\matr H$.

Current applications of compressed sensing include  imaging \cite{duarte,romberg}, radar
\cite{strohmer} and wireless communication \cite{rauhutjstps}\footnote{Indeed, many  (lossy) compression techniques (such as JPEG or MP3) are based on the fact that  real-world signals can be
approximated by an expansion
over a suitably chosen basis, with only  few non-zero coefficients. Seminal papers in this area are \cite{donoho:06,rangan:09,candes:06}}. 
The purpose of this Section is to provide a general description of compressed sensing, that is the idea of \emph{collecting as few samples as possible (below the constraints of traditional signal theory) to recover a sparse signal behavior}. 

Referring to the linear channel (\ref{AWGN}), we can identify that the problem of recovering the data vector $\matr x$, of length $K$, from the measurement vector $\matr y$, of length $N\ll K$, in a Gaussian vector channel 
 would lead to strong ill-conditioning.
However, it can be solved even if $N \ll K$  provided that $\matr x$ is
$s$-\emph{sparse} (it has no more than $s \ll N$ non-zero entries), and the system matrix $\matr H$ satisfies the \emph{null-space property} \cite[Def.(2.1)]{rahuth:11} or, equivalently, the \emph{Restricted-Isometry-Property} (hereinafter referred to as RIP), i.e. \begin{definition}
For all $s$-\emph{sparse} $\matr x \in \CC^{K}$, the restricted isometry constant $\delta_s$ of the system matrix $\matr H$ is defined as the smallest $\delta_s$ for which
\[
(1-\delta_s)||\matr x||^2\leq ||\matr H \matr x ||^2\leq (1+\delta_s)||\matr x||^2\,.
\]
If $\delta_s$
is small for reasonably large $s$, then $\matr H$ is said to fulfill the RIP\footnote{As specified in \cite{rahuth:11}, \emph{small} and \emph{reasonably large} depend both on the application under exam.}.
\end{definition}
  By the above definition, we can say  that RIP requires   all column submatrices of the measurement matrix of a certain size to be well-conditioned, in relation with the sparsity index $s$ of the original data vector. While it turns out to be quite difficult
to check this condition for deterministic matrices, this can be quite satisfactorily performed for random matrices with independent entries, like Gaussian or Bernoulli matrices \cite{candes:06,donoho:06}, but also, more recently, for structured random matrices, like (partially) Fourier or circulant matrices, \cite[and references therein]{rahuth:11}. This allows for the embodiment  of the physical constraints (mostly of geometrical nature) in the system structure, rather than considering sampling coordinates to be decoupled from each other as in independent and identically distributed system matrices.

It is worth stressing that RIP is required to solve a relaxed version of the compressed sensing problem. The original (noiseless) statement would be
\[
\min_{\matr x \in \CC^K}||\matr x||_0\qquad \rm{subject\, to}\qquad \matr H \matr x=\matr y\,,
\]
with $||\cdot||_0$ the $\ell_0$-norm (the number of non-zero entries in the sparse vector). But by ensuring the system matrix $\matr H$ fulfills RIP one may instead solve
\[
\min_{\matr x \in \CC^K}||\matr x||_1\qquad \rm{subject\, to}\qquad \matr H \matr x=\matr y\,,
\]
which is a convex relaxation of the original statement, and whose solution is unfortunately NP hard.
The relaxed version solution of the compressed sensing problem thus depends only on the eigenvalues of well-conditioned submatrices of the system matrix $\matr H$, due to the definition of the RIP. On the other hand, recent attempts to solve the noisy problem in its original setting 
\[
\min_{\matr x \in \CC^K}||\matr x||_0\qquad \rm{subject\, to}\qquad \matr H \matr x+\matr n =\matr y\,
\]
\emph{on average} (i.e. giving an asymptotic estimate of the error rate in recovering the support of the data vector $\matr x$), show instead that the solution depends on both the eigenvalues and the eigenvectors of $\matr H$ \cite{TulinoCS}. 
\subsection{Low Density Parity Check Codes}
\label{LowDenParCheCod}

Low density parity check (LDPC) codes are a subclass of linear block codes. They were invented by Gallager \cite{gallager:62} in 1962 and forgotten for several decades before they were re-discovered in the 1990s. They are now the most widely used method of forward-error correction coding. In the sequel, we will focus on binary LDPC codes since they are most widely used in practice, although non-binary LDPC codes have, in theory, slightly better peformance.
 
The purpose of a binary linear block code is to protect a vector of binary data symbols $\matr x\in\{0,1\}^K$ against bit flips.
To achieve robustness against bit flips, redundancy is added by means of multiplying the vector of data symbols by a tall\footnote{With the notable exception of \cite{mackay:03}, most texts on coding theory use row vectors instead of column vectors resulting in not tall, but wide generator matrices.} matrix $\matr H\in\{0,1\}^{N\times K}$ over the finite field of binary numbers. The matrix $\matr H$ is called {\em generator matrix}. The components of the resulting (binary) vector are called {\em code} symbols. The ratio of the number of data symbols to the number of code symbols
\begin{equation}
R=\frac KN<1,
\end{equation}
i.e.\ the aspect ratio of the generator matrix, is called the {\em rate} of the code. 

The effect of bit-flips can be modeled by adding a binary noise vector to the code symbols.
Noise components that are ones flip the respective symbols, noise components that are zeros do not.   
The received vector $\matr y\in\{0,1\}^N$ is given by \eq{AWGN} where all additions and multiplications are performed modulo 2.
The task of reconstructing the data vector given the received vector is called {\em decoding} and various methods and algorithms dedicated to this task are reported in literature \cite{lin:04}.

For binary noise, \eq{AWGN2} are not sufficient statistics. However, sufficient statistics in $K$ dimensions can be obtained by
\begin{equation}
\matr r = \matr H^\perp \matr y = \matr H^\perp \matr n
\end{equation}
with $\matr H^\perp$ denoting the orthogonal complement of $\matr H$, i.e.\ $\matr H^\perp \matr H=\0_{K\times K}$. The matrix $\matr H^\perp$ is called {\em parity check matrix} and the vector of sufficient statistics $\matr r$ is called {\em syndrome}.

A generator matrix is said to be in {\em systematic form} if it obeys the structure
\begin{equation}
\matr H = \left[\begin{array}c \I_{K\times K} \\ \hline \matr P\end{array} \right]
\end{equation}
for some matrix $\matr P$.
For systematic generator matrices, the parity check matrix is easily found to be
\begin{equation}
\matr H^\perp = \left[ \matr P | \matr I_{N-K\times N-K}\right]
\end{equation}
since in modulo-two arithmetic $\matr P+\matr P=\0\, \forall \matr P$.

The following phase transition is the celebrated {\em channel coding theorem} in information theory:
For $K,N\to\infty$, with the rate $R=K/N$ fixed, there exists a sequence of generator matrices, such that the data vector can be reconstructed with arbitrarily low probability of error, if and only if the rate is below a certain threshold rate $C$. 
The threshold rate is called {\em channel capacity} and depends only on the probability of ones in the noise vector:
\begin{equation}
\sigma_0^2=\Pr(n_k=1) .
\end{equation}
For the channel in \eq{AWGN} which is called {\em binary symmetric channel}, the channel capacity is given by
\begin{equation}
C=1+\sigma_0^2\log_2(\sigma_0^2) +(1-\sigma_0^2)\log_2(1-\sigma_0^2).
\end{equation}
More general forms of the channel coding theorem exist for non-binary channels and non-linear codes \cite{mackay:03}.

While the channel coding theorem was known since 1948 \cite{shannon:48a}, explicit codes that work well for code rates close to the capacity $C$ of the binary symmetric channel were unknown until the 1990s.
A breakthrough occurred with the use of iterative decoding and the introduction of turbo codes proposed in \cite{berrou:96}. Soon, LDPC codes were re-discovered and {\em belief propagation} became the method of choice for decoding \cite{mackay:03}. 
The term LDPC codes does not refer to a specific code design, but to all linear block codes whose parity check matrix is sparse.
The sparsity of the parity check matrix reduces complexity but, even more importantly, ensures convergence of belief propagation to sensible data estimates even when the rate of the code is close to channel capacity \cite{richardson:08}. 

\subsection{Capacity of Gaussian Vector Channel}
\label{ChaCap}

Channel capacity is the maximum data rate which can be transmitted reliably over a channel.
For the Gaussian vector channel \eq{AWGN} and many other channels with continuous input and output alphabet, channel capacity is infinite unless a constraint on the input alphabet is imposed.
There are many ways to define such constraints.
The most widely used ones are upper bounds on the moments of the amplitude of the input signal, particularly on the second moment
\begin{equation}
\expect\limits_\nu \matr x^{\dagger}\matr x \le KP
\end{equation}
which is actually an average power constraint.
Note that $P$ is the average power per component of $\matr x$.

Channel capacity crucially depends on the statistics of the noise process $\matr n[\nu]$ and the availability of channel state information to either receiver or transmitter or both, that is whether the transmitter and/or receiver is aware of the realization of the channel matrix $\matr H$.
Perfect channel state information means full knowledge about $\matr H$.
Assuming perfect channel state information at the receiver, but no channel state information at the transmitter and a zero-mean ($\expect \matr n = \mbox{\bf 0}$) white Gaussian noise process \eq{iidnoise}, the
 channel capacity of \eq{AWGN} is given by \cite{cover:91}
\begin{equation}
C=\expect\limits_{\matr H}\log\det\left(\mbox{\bf I} + \frac P{\st^2} \matr H^{\dagger}\matr H\right).
\label{chcap}
\end{equation}
With the eigenvalue decomposition
\begin{equation}
\label{eigdecomp}
 \matr H^{\dagger}\matr H = \matr U^{\dagger} \matr\Lambda\matr U 
\end{equation}
where $\matr U$ is a unitary matrix containing the eigenvectors of $ \matr H^{\dagger}\matr H$ and $\matr\Lambda = \diag(\lambda_1,\dots,\lambda_K)$ is a diagonal matrix containing the eigenvalues of $ \matr H^{\dagger}\matr H$, the channel capacity can also be written as
\begin{eqnarray}
C &=& \expect\limits_{\matr\Lambda}\log\det\left(\mbox{\bf I} +\frac P{\st^2}\matr\Lambda\right)\\
&=& \sum\limits_{k=1}^K \expect\limits_{\lambda_k}\log\left(1+\frac P{\st^2}\lambda_k\right)\\
&=& K \expect\limits_\lambda \log\left(1+\frac P{\st^2}\lambda\right)
\end{eqnarray}
where $\lambda$ denotes a randomly chosen eigenvalue of $ \matr H^{\dagger}\matr H$.
Therefore, channel capacity is determined by the eigenvalue distribution of the channel, the power constraint, and the noise variance.

\section{Random Matrix Theory}
\label{RanMatThe}

The channel matrix $\matr H$ introduced in \eq{AWGN} is composed of $NK$ random elements.
Though it can be simply considered as an $NK$ dimensional random object, it has also some more interesting interpretations.

Consider a scalar zero-mean random process $H_\eta[\mu]$ over discrete time $\mu$.
Stack the time samples into the column dimensions of the matrix $\matr H$ and the ensembles $\eta$ into the row dimensions of $\matr H$ such that
\begin{equation}
\matr H = \left[\begin{array}{cccc}
H_1[0] & H_2[0] & H_3[0] & \cdots\\
H_1[1] & H_2[1] & H_3[1] & \cdots\\
H_1[2] & H_2[2] & H_3[2] & \cdots\\
\vdots & \vdots & \vdots & \ddots
\end{array}\right].
\end{equation}
If we let the dimensions $N,K\to\infty$, the matrix $\matr H$ describes a whole random process.
Nevertheless, we can still think of it as a single realization of a many-dimensional random variable.
This double interpretation results in the {\em self-averaging property} of many finite-dimensional functions of infinite-dimensional random matrices.

Consider the function ${\rm rowsum}: \matr X\in\CC^{N\times K}\mapsto K^{-\frac12}\matr{X1}\in\CC^{N}$ with $\matr 1$ denoting the all one vector.
It simply sums up the rows of its argument and normalizes the result.
As $K\to\infty$, the value of this function 
\begin{equation}
\label{rowsum}
\matr h = {\rm rowsum}(\matr H)
\end{equation}
is an $N$-dimensional Gaussian random vector due to the central limit theorem.
As $N\to\infty$, the empirical distribution function of its components $h_i$
\begin{equation}
\Prob{\matr h}{x}= \frac1N\,\big|\left\{ h_i:h_i<x \right\}\big|
\end{equation}
converges to a Gaussian distribution.
Communications engineering builds upon this result, whenever something is modeled as a Gaussian random process.

The type of distribution $\matr h$ follows does not depend on the distribution of $\matr H$ as $N\to\infty$.
Instead, the distribution of $\matr h$ is determined by the mapping from $N\times K$-dimensional space into $N$-dimensional space.
For many linear mappings, such as ${\rm rowsum}(\cdot)$, the projection follows a Gaussian distribution.
For non-linear mappings, however, a rich plurality of other limit distributions occurs. 

\subsection{Convergence of Eigenvalues}
\label{ConEig}

The eigenvalues of random matrices were found to be particularly important to characterize some performance measures in communications engineering, cf.\ Section~\ref{ComSys}.
Calculating the eigenvalues of (a function of) a random matrix, is a projection similar to the rowsum function in \eq{rowsum}.
The main difference between the two is the, in general, non-linear nature of the projection. 

Calculating the eigenvalues $\lambda_k$ of a  square $N\times N$ matrix $\matr H$ is a non-linear operation, in general.
The moments of the eigenvalue distribution are conveniently calculated by a normalized trace since
\begin{equation}
\frac1N\sum\limits_{k=1}^N\lambda_k^m = {\rm tr}\left(\matr H^m\right)
\end{equation}
with
\begin{equation}
{\rm tr}(\matr H) \define \frac1N\,{\rm trace}(\matr H).
\end{equation}
In the following, we also use
\begin{equation}
{\rm Tr}(\matr H) \define \lim\limits_{N\to\infty}{\rm tr}(\matr H).
\end{equation}
to denote the normalized trace in the large matrix limit.
The eigenvalue distributions of several types of random matrices are examined in greater detail in the following.

\subsubsection{Full Circle Law}
\label{FulCirLaw}

Let the random matrix $\matr H$ be square $N\times N$ with independent identically distributed entries with zero mean and variance $1/N$.
Let $\Lc$ denote the set containing the eigenvalues of $\matr H$.
Then, the empirical distribution of the eigenvalues 
\begin{equation}
\Prob{\matr H}{z} = \frac1N\,\big|\left\{ \lambda\in\Lc:\Re\lambda<\Re z \wedge \Im\lambda<\Im z \right\}\big|
\end{equation}
converges to a non-random distribution function as $N\to\infty$ whose density is given by
\begin{equation}
\prob{\matr H}{z} = \cases{\frac1\pic & $|z|<1$ \cr 0 & elsewhere}.
\end{equation}
The eigenvalues are uniformly distributed within the complex unit circle.

\subsubsection{Semi-Circle Law}
\label{SemCirLaw}
Let the random matrix $\matr H$ fulfill the same conditions as needed for the full circle law. 
Let
\begin{equation}
\matr K=\frac{\matr H+\matr H^{\dagger}}{\sqrt2}.
\end{equation}
Then, the empirical distribution of the eigenvalues of $\matr K$ converge to a non-random distribution function as $N\to\infty$ whose density is given by
\begin{equation}
\prob{\matr K}x = \cases{\frac1{2\pic}\, \sqrt{4-x^2} & $|x|<2$ \cr 0 & elsewhere}.
\end{equation}

The semicircle distribution plays a central role in free probability theory, cf.\ Section~\ref{FreProThe}, where it serves as the equivalent to the Gaussian distribution in classical probability theory.
Its even moments are the {\em Catalan} numbers ${\rm C}_n$
\begin{eqnarray}
{\rm Tr}\left(\matr K^m\right) &=& \frac1{2\pic}\int\limits_{-2}^{+2}x^m\sqrt{4-x^2}{\rm d}x\\
&=&{\rm C}_{m/2} \qquad\forall\,m\,\mbox{even}\\ &\define& \frac1{m/2+1}\bin{m}{m/2} 
\end{eqnarray}
which play a crucial role in combinatorics, see \cite{hilton:91} for further reading.
Since the semicircular density is symmetric, the odd moments vanish.
\subsubsection{Quarter Circle Law}
\label{QuaCirLaw}

Let the random matrix $\matr H$ fulfill the same conditions as needed for the full circle law. 
Then, the empirical distribution of the singular values of $\matr H$, i.e.\ the eigenvalues of 
\begin{equation}
\matr Q=\sqrt{\matr{HH}^{\dagger}},
\end{equation}
converge to a non-random distribution function as $N\to\infty$ whose density is given by
\begin{equation}
\prob{\matr Q}x = \cases{\frac1\pic\, \sqrt{4-x^2} & $0\le x <2$ \cr 0 & elsewhere}.
\end{equation}
This distribution is called the quarter circle distribution.
Obviously, its even moments are identical to those of the semicircular distribution. 
However, its odd moments do not vanish.
They are given by
\begin{eqnarray}
{\rm Tr}\left(\matr Q^m\right) &=& \frac1{\pic}\int\limits_0^2x^m\sqrt{4-x^2}{\rm d}x\\
&=&\frac{2^{2m}}{\pic m\left(\frac m2+1\right)\bin{m-1}{\frac{m-1}2}} \quad\forall\,m\,\mbox{odd}.
\end{eqnarray}

With standard methods for the transformation of probability densities, see \cite{papoulis:91}, the asymptotic eigenvalue distribution of $\matr Q^2=\matr{HH}^{\dagger}$ can be derived.
It reads
\begin{equation}
\prob{\matr Q^2}x = \cases{\frac1{2\pic}\, \sqrt{\frac{4-x}x} & $0< x <4$ \cr 0 & elsewhere}.
\end{equation}
Its $m^{\rm th}$ moments (even and odd) are the Catalan numbers ${\rm C}_m$.
This distribution is a special case of the Mar\v cenko-Pastur distribution (see next subsection) which also plays a central role in free probability theory.

\subsubsection{Deformed Quarter Circle Law}
\label{DefQuaCir}

The quarter circle law is part of a more general result for rectangular matrices:
Let the entries of the $N\times K$ matrix $\matr H$ be independent identically distributed with zero mean and variance $1/N$.
Then, the empirical distribution of the singular values of $\matr H$, i.e.\ the eigenvalues of  
\begin{equation}
\matr R=\sqrt{\matr{HH}^{\dagger}}
\end{equation} 
converges to a non-random distribution function as $N,K\to\infty$ with $\load=K/N$ fixed and its density is given by
\begin{equation}
\prob{\matr R}x=\cases{\frac{\sqrt{4\load-\left(x^2-1-\load\right)^2}}{\pic x\vphantom{\int\limits_1}} & $|1-\sqrt\load|<x<1+\sqrt\load$\cr  [1-\load]^+\,\deltaf(x) & elsewhere}.
\end{equation}

Again, we also consider the eigenvalue distribution of $\matr R^2=\matr{HH}^{\dagger}$ and find 
\begin{eqnarray}
\label{pdfDQCL}
\prob{\matr R^2}x=\cases{\frac{\sqrt{4\load-(x-1-\load)^2}}{2\pic x\vphantom{\int\limits_1}}  & $(1-\sqrt\load)^2<x<(1+\sqrt\load)^2$\cr [1-\load]^+\,\deltaf(x) & elsewhere}.
\end{eqnarray}
This distribution is known as the Mar\v cenko-Pastur distribution and its moments are given by
\begin{equation}
{\rm Tr}\left(\matr R^{2m}\right) = \frac1m\sum\limits_{i=1}^m\bin mi\bin m{i-1}\load^i. 
\label{momMP}
\end{equation}
It has been used in \cite{verdu:99a} to calculate channel capacity \eq{chcap} for CDMA with independent identically distributed random spreading.
\subsubsection{Haar Distribution}
\label{HaaDis}

Let the random matrix $\matr H$ be square $N\times N$ with independent identically complex Gaussian distributed  proper
entries with zero mean and finite positive variance.
Then, the empirical distribution of the eigenvalues of the unitary random matrix
\begin{equation}
\matr T=\matr H\left(\matr H^{\dagger}\matr H\right)^{-\frac12}
\end{equation} 
converges to a non-random distribution function as $N\to\infty$ whose density is given by
\begin{equation}
\prob{\matr T}z=\frac1{2\pic}\,\deltaf(|z|-1).
\label{haardis}
\end{equation}
Obviously, all of its moments $\expect|z|^m=1$ equal unity.

It is obvious that all eigenvalues lie on the complex unit circle, since the matrix $\matr T$ is unitary.
The essential statement of \eq{haardis} is that the eigenvalues are uniformly distributed.

Note that unlike the circle laws in previous subsections, \eq{haardis} demands Gaussian distributed entries in the random matrix $\matr H$.
This Gaussian condition does not seem to be necessary, yet allowing for any complex distribution with zero mean and finite variance is not sufficient \cite{silverstein:84}.

\subsubsection{Inverse Semi-Circle Law}
\label{InvSemCir}
Let the random matrix $\matr T$ fulfill the same conditions and be defined in the same way as in Section~\ref{HaaDis}. 
Let
\begin{equation}
\matr Y=\matr T+\matr T^{\dagger}.
\end{equation}
Then, the empirical distribution of the eigenvalues of $\matr Y$ converges to a non-random distribution function as $N\to\infty$ whose density is given by
\begin{equation}
\prob{\matr Y}x = \cases{\frac1\pic\frac1{\sqrt{4-x^2}} & $|x|<2$ \cr 0 & elsewhere}.
\end{equation}

\subsection{Stieltjes Transform}
\label{StiTra}

There are only few kinds of random matrices for which the corresponding asymptotic eigenvalue distributions are known explicitly.
For a wider class of random matrices, however, explicit calculation of the moments turns out to be feasible, see Section~\ref{Exa4} for an example. 

The task of finding an unknown probability distribution given its moments is known as the {\em problem of moments}.
It was addressed by Stieltjes in 1894 \cite{stieltjes:94} using the integral transform
\begin{equation}
\StT{}s\define\int\frac{{\rm d}\Prob{}{x}}{x-s}
\label{defSt}
\end{equation} 
with $\Im s>0$. 
It is now commonly referred to as the {\em Stieltjes transform}.
A simple Taylor series expansion of the kernel of the Stieltjes transform 
\begin{equation}
-\lim\limits_{s\to0}\frac{{\rm d}^m}{{\rm d}s^m}\,\frac{\StT{}{s^{-1}}}{s} =m! \int x^m{\rm d}\Prob{}x
\end{equation}
shows how the moments can be found given the Stieltjes transform without the need for integration.

In order to recover the probability density function from its Stieltjes transform, note that
\begin{equation}
\lim\limits_{\epsilon\to 0+} \Im \frac1{x - {\rm j} \epsilon} = \pi \delta(x). 
\end{equation}
The probability density function can thus be obtained from the Stieltjes transform simply taking the limit
\begin{equation}
\prob{}x = \lim\limits_{y\to 0+} \frac{1}\pic\,\Im\, \StT{}{x+{\rm j} y}
\label{Stinv}
\end{equation}
which is often called the {\em Stieltjes inversion formula} \cite{hiai:00}. 

\subsubsection{Products of Random Matrices}
\label{ProRanMat}

Let the random matrix $\matr H$ fulfill the same conditions as needed for the deformed quarter circle law.
Moreover, let $\matr X=\matr X^{\dagger}$ be an $N\times N$ Hermitian matrix, independent of $\matr H$, with an empirical eigenvalue distribution converging almost surely in distribution to a distribution function $\Prob{\matr X}x$ as $N\to\infty$. 
Then, almost surely, the eigenvalue distribution of the matrix product
\begin{equation}
\matr P = \matr{HH}^{\dagger}\matr X
\end{equation} 
converges in distribution, as $K,N\to \infty$, but $\load=K/N$ fixed, to a nonrandom distribution function whose Stieltjes transform satisfies
\begin{equation}
\label{PRM}
\StT{\matr P}s = \int\frac{{\rm d}\Prob{\matr X}x}{x\big(1-\load-\load s \StT{\matr P}s\big)-s}
\end{equation}
for $\Im s>0$.

This result was proven in its present form by Silverstein \cite{silverstein:95a}.
Under less general conditions on the statistics of $\matr H$ and $\matr X$, it can be found in the earlier work of Yin \cite{yin:86}.

\subsubsection{Sums of Random Matrices}
\label{SumRanMat}
Let the random matrix $\matr H$ fulfill the same conditions as needed for the deformed quarter circle law.
Let $\matr X=\matr X^{\dagger}$ be an $N\times N$ Hermitian matrix with an eigenvalue distribution function converging weakly to $\Prob{\matr X}x$ almost surely. 
Let $\matr Y=\diag\left(y_1,\dots,y_K\right)$ be a $K\times K$ diagonal matrix and the empirical distribution function of $\left\{y_1,\dots,y_K\right\}\in \RR^K$ converge almost surely in distribution to a probability distribution function $\Prob{\matr Y}x$ as $K\to\infty$.
Moreover, let the matrices $\matr H, \matr X, \matr Y$ be jointly independent. 
Then, almost surely, the empirical eigenvalue distribution of the random matrix
\begin{equation}
\matr S=\matr X+\matr{HYH}^{\dagger}
\end{equation}
converges weakly, as $K,N\to\infty$, but $\load=K/N$ fixed, to a nonrandom distribution function whose Stieltjes transform satisfies
\begin{equation}
\label{eq60}
\StT{\matr S}s = \StT{\matr X}{s-\load\int\frac{y\,{\rm d}\Prob{\matr Y}y}{1+y\,\StT{\matr S}s}}
\end{equation}
for $\Im s>0$.

This result was proven in its present form by Silverstein and Bai \cite{silverstein:95b}.
Under less general conditions on the statistics of $\matr H$ and $\matr X$, it can be found in the earlier work of Mar{\v c}enko and Pastur \cite{marcenko:67}.
It was used by Tse and Hanly \cite{tse:99b} to derive asymptotic results for the SINR of linear multiuser receivers.
Subsequently, it was used by Shamai and Verd\'u \cite{shitz:99} to derive the capacity of the flat fading Gaussian CDMA channel with several types of receivers.
\subsubsection{Girko's Law}
\label{GirLaw}

Let the $N\times K$ random matrix $\matr H$ be composed of independent entries $(\matr H)_{ij}$ with zero-mean and variances $w_{ij}/N$ such that all $w_{ij}$ are uniformly bounded from above.
Assume that the empirical joint distribution of variances $w:[0,1]\times[0,\beta]\mapsto\RR$ defined by $w(x,y) = w_{ij}$ for $i,j$ satisfying
\begin{equation}
\frac iN\le x\le \frac{i+1}N \qquad\mbox{and}\qquad \frac jN\le y\le\frac{j+1}N
\end{equation}
 converges to a bounded joint limit distribution $w(x,y)$ as $K=\beta N\to\infty$.
Then, for each $a,b\in[0,1], a<b$, and $\Im(s)>0$
\begin{equation}
\frac1N\sum\limits_{i=\lceil aN\rceil}^{\lceil bN\rceil}\left(\matr {HH}^{\dagger}-s\I\right)_{ii}^{-1} \longrightarrow \int\limits_a^bu(x,s){\rm d}x
\end{equation}
where convergence is in probability and $u(x,s)$ satisfies the fixed point equation
\begin{equation}
\label{fG}
u(x,s)=\left[-s+\int\limits_0^\beta\frac{w(x,y)\,{\rm d}y}{1+\int\limits_0^1u(x^\prime,s)w(x^\prime,y)\,{\rm d}x^\prime}\right]^{-1}
\end{equation}
for every $x\in[0,1]$.
The solution to \eq{fG} exists and is unique in the class of functions $u(x,s)\ge0$, analytic for $\Im(s)>0$ and continuous on $x\in[0,1]$.

Moreover, almost surely, the empirical eigenvalue distribution of $\matr {HH}^{\dagger}$ converges weakly to a limiting distribution whose Stieltjes transform is given by
\begin{equation}
G_{\matr{HH}^{\dagger}}(s)=\int\limits_0^1u(x,s)\,{\rm d}x.
\end{equation}
This theorem is due to Girko \cite{girko:90}. It has been used in \cite{hanly:99a} to prove certain properties of chips and receive antennas in CDMA systems with antenna diversity.
\subsubsection{Kronecker Products}

Let the $N\times K$ random matrix $\matr H$ fulfill the same conditions as needed for the deformed quarter circle law.
Let the entries of the $R\times K$ random matrix $\matr Y$ be i.i.d.\ distributed with a circularly symmetric distribution on $\CC$.
Denote the columns of $\matr H$ and $\matr Y$ by $\matr h_k$ and $\matr y_k$, respectively, and define the matrix
\begin{equation}
\matr {\tilde H} = \left[ \matr y_1 \otimes \matr h_1,\matr y_2 \otimes \matr h_2,\dots, \matr y_K\otimes \matr h_K\right]
\end{equation}
with $\otimes$ denoting the Kronecker product.
Let the empirical distribution of $\{ \matr y_1^{\dagger}\matr y_1,\dots,\matr y_K^{\dagger}\matr y_K\}$ converge to a limit distribution $\Prob{\matr Y}y$.
Then, the empirical eigenvalue distribution of the random matrix
\begin{equation}
\matr R = \matr{\tilde H\tilde H}^{\dagger}
\end{equation}
converges almost surely, as $K,N\to\infty$, but $\beta=K/N$ and $R$ fixed, to a nonrandom distribution function whose Stieltjes transform satisfies
\begin{equation}
\StT{\matr R}{s} = \left(s-\frac\beta R \int \frac{y{\rm d}\Prob{\matr Y}y}{1+y\,\StT{\matr R}s} \right)^{-1}
\end{equation}
for $\Im s>0$.

This result was proven by Hanly and Tse \cite{hanly:99a}.
It shows a striking similarity with \eq{eq60}. In fact, for $\matr X=\0$ and $R=1$, both results are identical. It is remarkable, however, that increasing $R$ beyond 1 has the same effect as dividing $N$ by $R$. This effect has been called {\em resource pooling} in \cite{hanly:99a} where $R$ takes the role of the number of antenna elements and $N$ the number of chips per symbol.

The entries of the random matrix $\matr {\tilde H}$ are correlated. However, the asymptotic eigenvalue distribution is the same as if they were statistically independent. Correlation among matrix entries may, but need not influence the asymptotic eigenvalue distribution.
If one gives up the i.i.d.\ assumption on $\matr Y$, the asymptotic eigenvalue distribution gets affected \cite{cottatellucci:07,bai:07}.

\subsubsection{Stieltjes Transforms of Some Distributions}
\label{StiTraSom}

The Stieltjes transforms of the distributions in Table~\ref{tabStT} can be found either via the defining integral or (often easier) via solving \eq{PRM}.
\begin{table}[h]
\caption{\label{tabStT} Table of Stieltjes transforms $(\Im s>0)$.}
\begin{eqnarray*}
\StT{\alpha\mbox{\bf I}}s &=& \frac1{\alpha-s}\\
\StT{\matr K}s &=&\frac s2\sqrt{1-\frac4{s^2}}-\frac s2\\
\StT{\matr Q}s &=&\frac{2\sqrt{4-s^2}}{\pic}\ln \left(\frac{2+\sqrt{4-s^2}}{-s}\right)-\frac{s}{2}-\frac{2}{\pic}\\
\StT{\matr Q^2}s &=&\frac12\sqrt{1-\frac4s}-\frac12\\
\StT{\matr Q^{-2}}s &=&\frac{1-2s-\sqrt{1-4s}}{2s^2}\\
\StT{\matr R^2}s &=&\sqrt{\frac{(1-\load)^2}{4s^2}-\frac{1+\load}{2s}+\frac14}-\frac12-\frac{1-\load}{2s}\\
\StT{\matr R^{-2}}s &=&\frac{1-(1+\beta)s-\sqrt{(1-\beta)^2s^2-2(1+\beta)s+1}}{2s^2}\\
\StT{\matr Y}s &=& \frac{1}{\sqrt{s^2-4}}
\end{eqnarray*}
\end{table} 
\begin{table}[h]
\caption{\label{PropStT} Properties of the Stieltjes transform (for any $N\times \beta N$ matrix $\matr X$).}
\begin{eqnarray*}
\StT{\lambda^2}s &=& \frac{\StT\lambda{\sqrt s}-\StT\lambda{-\sqrt s}}{2\sqrt s}\\
\StT{\lambda^{-1}}s &=& -s^{-2}\StT\lambda{s^{-1}}-s^{-1}\\
\StT{\matr{XX}^{\dagger}}s &=& \beta\StT{\matr X^{\dagger}\matr X}s + \frac{\beta-1} s\\
\Im \StT{}s &\ge& 0 
\end{eqnarray*}
\end{table}
The Stieltjes transforms in Table~\ref{tabStT} hold for all $s\in\CC\setminus\RR$.
Due to the multiple branches of the complex square root, some formulas can be further simplified in the local neighborhoods of particular $s$.

\subsection{Convergence Properties of Eigenvectors}
\label{ConProEig}

While there are many results known in literature about the eigenvalues of large random matrices, little is known about the eigenvectors.
However, there is one particular result which proves helpful for communications engineering applications:

Let $\matr H$ be an $N\times K$ random matrix with independent identically distributed real-valued random entries with zero mean and all positive moments bounded from above.
Let the orthogonal matrix $\matr U$ be defined be the eigenvalue decomposition
\begin{equation}
\matr U^{\rm T}\matr{\Lambda U} = \matr H^{\rm T} \matr H.
\end{equation}
Note that the rows of $\matr U$ are the eigenvectors of $\matr H^{\rm T}\matr H$.
Let $\matr x\in\RR^N$ with $||\matr x||=1$ be an arbitrary vector with unit Euclidean norm and the random vector $\matr y=[y_1,\dots,y_N]^{\rm T}$ be defined as
\begin{equation}
\label{defY}
\matr y=\matr{Ux}.
\end{equation}
Then, as $N,K\to\infty$, but $\load=K/N$ fixed,
\begin{equation}
\label{LLN}
\sum\limits_{k=1}^{\lceil tN\rceil} y_k^2 \longrightarrow t  
\end{equation}
almost surely for every $t\in[0;1]$ with $\lceil x \rceil$ denoting the nearest integer to $x$ which is not smaller than $x$ \cite{silverstein:84}.

This result is like a law of large numbers for the components of any linear combination of the components of the eigenvectors of $\matr H^{\rm T}\matr H$.
It is not obvious to hold, since the elements of the eigenvector matrix $\matr U^{\rm T}$ are not statistically independent.
However, this theorem shows that, for the purpose of summing its squared elements, we can assume they are statistically independent in the large matrix limit.

The convergence in \eq{LLN} straightforwardly implies the following extension:
Let the real-valued sequence $\alpha_k$ be uniformly bounded from below by a constant larger than zero.
Then, under the same conditions as required for \eq{LLN} to hold, we have almost surely
\begin{equation}
\label{GLLN}
\sum\limits_{k=1}^{\lceil tN\rceil} \alpha_k y_k^2 \longrightarrow \lim\limits_{N\to\infty} \frac 1N \sum\limits_{k=1}^{\lceil tN\rceil} \alpha_k. 
\end{equation}
This result can be used to show the almost sure convergence of the SINRs of linear minimum mean-squared error (MMSE) detectors, see Section~\ref{App3}.

\subsection{Applications}
\label{App3}

In the following, some examples are given to demonstrate the usefulness of the previous results form random matrix theory to communications engineering.

\subsubsection{Convergence of the SINR of the Linear MMSE Detector}
\label{ConSINLin}

Consider the linear MMSE detector studied in Section~\ref{LinDet} with a real-valued channel matrix $\matr H$.
Its SINR is given by \eq{SINRlindet} as
\begin{equation}
{\rm SINR}_k = \frac1{1-P\matr u_k^{\rm T}\left(\st^2\matr\Lambda^{-1}+P\,\mbox{\bf I}\right)^{-1}\matr u_k} -1.
\label{sinrk}
\end{equation}
In general, it is different for each user $k$.
However, if the channel matrix $\matr H$ is composed of independent identically random entries with zero mean, variance $1/N$ and all other moments finite, we can use the deformed quarter circle law and \eq{GLLN} to show almost sure convergence of the SINR of all users to the same deterministic limit, as the matrix size grows large.

If we let $x_k=1 \Rightarrow x_{\ne k}=0$ in \eq{defY}, we find from \eq{GLLN} the almost sure convergence
\begin{eqnarray}
1-P\matr u_k^{\rm T}\left(\st^2\matr\Lambda^{-1}+P\,\mbox{\bf I}\right)^{-1}\matr u_k \hspace*{-3cm}\nonumber\\ &\longrightarrow& 1-P\,{\rm Tr}\left(\st^2\matr\Lambda^{-1}+P\,\mbox{\bf I}\right)^{-1}\\
&=& \frac{\st^2}P\,{\rm Tr}\left(\frac{\st^2}P\,\mbox{\bf I}+\matr\Lambda\right)^{-1}.
\label{asc}
\end{eqnarray}
From \eq{defSt} and the deformed quarter circle law, we get the almost sure identity
\begin{equation}
\label{asi}
{\rm Tr}\left(\frac{\st^2}P\,\mbox{\bf I}+\matr\Lambda\right)^{-1} = \StT{\matr H^{\rm T}\matr H}{-\frac{\st^2}P}.
\end{equation} 
With Table~\ref{PropStT}, the Stieltjes transform $\StT{\matr H^{\rm T}\matr H}s$ can be expressed in terms of $\StT{\matr H\matr H^{\rm T}}s$. The latter is given explicitly in Table~\ref{tabStT}.
Thus, combining \eqd{sinrk}{asc}{asi} gives after some trivial algebra the large system limit for the SINR of user $k$
\begin{equation}
\label{sinrfix}
{\rm SINR}_k \longrightarrow \frac{(1-\load)P}{2\st^2}-\frac12+\sqrt{\frac{(1-\load)^2P^2}{4\st^4}+\frac{(1+\load)P}{2\st^2}+\frac14}. 
\end{equation}
This derivation was limited to real-valued channel matrices due to technical reasons (a convergence property for eigenvectors of complex matrices similar to \eq{GLLN} has not been established so far).
However, it can be shown (via more involved methods) that it also does hold for complex channel matrices.

If the powers of the users differ, an explicit expression for the asymptotic SINRs is not possible.
However, the SINRs still converge to asymptotic limits as shown by Tse and Hanly \cite{tse:99b}.
They turn out to be not identical, but linear functions of the power of the user of interest.
Denoting the power of user $k$ by $P_k$ and assuming the empirical distribution of powers over the user population to converge to a limiting distribution $\Prob{P}{p}$ as $K=\beta N\to\infty$, we find \cite{tse:99b}
\begin{equation}
{\rm SINR}_k \longrightarrow P_k\eta
\end{equation}
with
\begin{equation}
\label{tsehanlyequation}
\eta = \frac1{\sigma_0^2+\beta\int\limits_0^\infty \frac{ p{\rm d}\Prob Pp}{1+p \eta}}.
\end{equation}
Tse and Hanly's asymptotic result has become very common in wireless communications to estimate the system performance. It was used e.g.\ in \cite{caire:01b} to optimze power control in iteratively decoded CDMA systems which a large number of users. 
\subsubsection{Implementation of the Linear MMSE Detector}
\label{ImpLinMMS}

The linear MMSE detector, see Section~\ref{LinDet}, reduces complexity of detection from exponential to polynomial dependency on the system dimension $K$.
Nevertheless, it still requires a matrix inversion in \eq{detMMSE} for performing its task.
This matrix inversion is computationally very costly, if it is performed by algorithms which cannot be implemented on parallel hardware architecture such as Gauss elimination.

Parallel algorithms for matrix inversion operate iteratively.
The classical Gauss-Seidel algorithm approximates the inverse matrix by a matrix-valued Taylor approximation 
\begin{equation}
\label{GSiter}
\matr X^{-1} = \alpha \sum\limits_{i=0}^\infty (\I - \alpha \matr X)^i
\end{equation} 
in the neighborhood of some multiple $\alpha^{-1}$ of the identity matrix which converges if all eigenvalues of the matrix $\matr X$ lie within the interval $(0;\frac2\alpha)$ \cite{axelsson:94}.
By appropriate choice of the parameter $\alpha$, convergence can always be ensured for positive definite matrices.

In principle, $\alpha$ can be chosen arbitrarily tiny ensuring convergence for any positive definite matrix.
In practice, however, one would like to choose $\alpha$ not too small to avoid numerical inaccuracies due to quantization errors.
For the latter purpose an upper bound on the eigenvalues of the matrix $\matr X$ is helpful.
For asymptotically large random matrices such upper bounds are provided by random matrix theory.
If the channel matrix is composed of independent identically distributed random entries, for instance, the eigenvalues of $\matr{HH}^{\dagger}$ are asymptotically upper bounded by
\begin{equation}
\lambda_k < \left(1+\sqrt\beta\right)^2
\end{equation}
via \eq{pdfDQCL}.
The matrix to be inverted for the linear MMSE detector is $\sigma_0^2\I+P\matr{HH}^{\dagger}$.
Since addition of identity matrices increases all eigenvalues by 1, convergence is ensured if
\begin{equation}
\sigma_0^2+P\left(1+\sqrt\beta\right)^2<\frac2\alpha.
\end{equation}
In practice, one would like to fulfill this condition with some margin to speed up convergence and to cope with deviations of the eigenvalue distributions of finite-size random matrices from their asymptotic behavior.
A more comprehensive treatment of this matter can be found in the work of Trichard et al.\ \cite{trichard:02}.
\subsubsection{Polynomial Expansion Detectors}
\label{PolExpDet}

The iterative method for matrix inversion presented in the previous subsection can be parallelized to run on up to $K$ processors.
Though, it may still require many iterations to converge.
The convergence can be accelerated significantly making use of the asymptotic convergence of the eigenvalue distribution of the matrix to be inverted.

Assume you want to invert a $K\times K$ matrix $\matr X$ whose eigenvalues ${\cal L}=\{\lambda_1,\dots,\lambda_K\}$ are known to you.
Note that due to the Cayley-Hamilton Theorem \cite{horn:85} any matrix is a zero of its characteristic polynomial
\begin{equation}
\prod\limits_{k=1}^K \left(\matr X - \lambda_k\I\right) = \matr 0.
\end{equation}
Expanding the product into a sum, we find
\begin{equation}
\sum\limits_{k=0}^K c_k({\cal L}) \matr X^k=\matr 0
\end{equation}
with some coefficients $c_k$ depending on the eigenvalues of $\matr X$.
Solving this equation for $\matr X^0=\I$ and multiplying both sides $\matr X^{-1}$ gives the desired inverse matrix as a $(K-1)^{\rm st}$ order polynomial in $\matr X$
\begin{equation}
\matr X^{-1} = -\sum\limits_{k=0}^{K-1} \frac{c_{k+1}({\cal L})}{c_0({\cal L})}\matr X^k \define \sum\limits_{k=0}^{K-1} \tilde w_k({\cal L}) \matr X^k.
\label{mosinv}
\end{equation}
Since the eigenvalue distribution depends only on the statistics of $\matr X$, the coefficients $\tilde w_k=-c_{k+1}/c_0$ can be pre-computed for large-dimensional random matrices.

While the Gauss-Seidel iteration \eq{GSiter} requires, in principle, the summation of an infinite number of terms to achieve arbitrary precision, the knowledge of the eigenvalues reduces the number of terms to be summed to the dimension of the matrix.
 
Evaluating a polynomial of degree $K-1$ can still be a task too complicated to perform in real-time for large matrix dimension $K$.
Though polynomials with lower degrees can, in general, not equal the inverse of the matrix, they may be accurate approximations.
For expansions with reduced degree $D<K-1$, the approximation error is reduced by replacing the coefficients $\{\tilde w_0,\dots,\tilde w_{K-1}\}$ with different coefficients $\{w_0,\dots,w_D\}$. Thus, we get
\begin{equation}
\matr X^{-1} \approx \sum\limits_{i=0}^D w_i({\cal L}) \matr X^i.
\end{equation}
Depending on the cost function for the approximation error, various designs for the coefficients $w_i$ are sensible.
 
In communications engineering the most common cost function is the mean-squared error \eq{defMSE}.
Thus, we find the coefficients $w_i$ by minimizing \eq{defMSE} while imposing a polynomial expansion structure onto the linear detector
\begin{equation}
\label{polexpdet}
\matr L = \sum\limits_{i=0}^D w_i \matr R^i \matr H^{\dagger}
\end{equation}
with $\matr R=\matr H^{\dagger}\matr H$.
With the channel model \eq{AWGN}, i.i.d.\ noise \eq{iidnoise}, and standard arguments of linear algebra, the optimum coefficients are straightforwardly shown to be determined by a system of Yule-Walker equations \cite{moshavi:96b}
\begin{equation}
\hspace*{-2cm}
\left[\begin{array}c m_1\\ m_2\\ \vdots\\ m_{D+1}\end{array}\right] =
\left[\begin{array}{ccc}
m_2 + \sigma_0^2m_1& \dots & m_{D+2}+ \sigma_0^2m_{D+1}\\ 
m_3 + \sigma_0^2m_2&  \dots & m_{D+3}+ \sigma_0^2m_{D+2}\\ 
\vdots&  \ddots &\vdots \\ 
m_{D+2}  + \sigma_0^2m_{D+1}& \dots & m_{2D+2}+ \sigma_0^2m_{2D+1}
\end{array}\right]
\left[\begin{array}c w_0\\ w_1\\ \vdots\\ w_{D}\end{array}\right]
\label{YWeq}
\end{equation}
where
\begin{equation}
\label{momPED}
m_k \define \frac1K\sum\limits_{i=1}^K \lambda_i^k.
\end{equation}
Moshavi et al.\ \cite{moshavi:96b} suggested to track the empirical eigenvalue moments $m_k$ adaptively and to solve the Yule-Walker equations \eq{YWeq} in real-time.
The weights can also be directly tracked adaptively interpreting the polynomial expansion detector as a multistage Wiener filter \cite{goldstein:97,goldstein:98}.

The moments \eq{momPED} converge to non-random deterministic limits for a large class of random matrices, they can be computed analytically as functions of the channel statistics and so can the weights.
This approach was proposed by M\"uller and Verd\'u \cite{mueller:01a} and explicit expressions for the optimum weights and the achievable SINRs were given for channel matrices with independent identically distributed entries based on \eq{momMP}.
For i.i.d.\ entries and users with identical powers $P_k=P$, the large-system SINR is given by the recursion \cite{honig:01}
\begin{equation}
{\rm SINR_D} = \frac{P}{\displaystyle \sigma_0^2+\frac{\beta P}{1+{\rm SINR}_{D-1}  }}
\end{equation}
with SINR$_0=0$.
This recursion shows that the SINR of the polynomial expansion detector converges to the exact solution as a continued fraction.
A generalization for users with different powers can be found in \cite{trichard:05}.
Loubaton and Hachem \cite{loubaton:03} highlighted the connection between the Mar\v{c}enko-Pastur distribution, continued fractions, and orthogonal polynomials for the analysis of polynomial expansion detectors.
Such connection is well-established in mathematical literature \cite{deift:98}, but  took its time to find its way into the design and analysis of code-division multiple-access.
Asymptotic design and analysis of polynomial expansion detectors for random matrices with non-i.i.d.\ entries are well-studied in the literature of communications engineering, see e.g.\  \cite{li:01b,chaufray:02,cottatellucci:04,mouhouche:07,cottatellucci:10a}. Some of these works make use of results of Section~\ref{FreProThe}.

For random matrices with non-i.i.d.\ entries, the polynomial expansion detector in \eq{polexpdet} can be improved by allowing the coefficients $w_i$ to be diagonal matrices instead of scalars \cite{cottatellucci:04}.
This does hardly affect the complexity of implementation, but makes the coefficients to also depend on other statistical properties of the random matrix than the eigenvalue moments alone. Nevertheless, convergence of the optimal weights to deterministic limits could be proven.


\section{Free Probability Theory}
\label{FreProThe}

While random matrix theory considers a large random matrix as a whole ensemble and proves convergence results, free probability looks at a random matrix from a different point of view:
A random matrix is primarily seen as a linear random operator.
Free probability theory provides a framework for dealing with certain classes of linear random operators.

The essential feature that distinguishes random operators including random matrices from scalar random variables is the commutative law which, in general, does not hold for matrices and other operators.
In order to see why this causes problems for probability theory, think of a random matrix as a single (non-commutative) random variable and consider the expectations
\begin{eqnarray}
\expect\left\{(xy)^m\right\} &=& \expect\left\{x^my^m\right\}\\
\expect\limits_{\rm free}\left\{\matr{(XY)^m}\right\} &\ne& \expect\limits_{\rm free}\left\{\matr X^m\matr Y^m\right\}
\label{ncom}
\end{eqnarray}
where $x,y$ are standard scalar random variables, $\matr X,\matr Y$ are random operators, and $\expect_{\rm free}$ denotes {\em free expectation}, i.e.\ an expectation operator relevant to the probability theory of non-commutative random operators (see Section~\ref{FreExp}). 

For independent random variables, all joint moments must factorize which implies
\begin{equation}
\expect\left\{(xy)^m\right\} = \expect\left\{x^my^m\right\} = \expect\left\{x^m\right\}\expect\left\{y^m\right\}.
\end{equation}
However, for statistically independent random matrices $\matr X$ and $\matr Y$, such a factorization cannot hold in general due to the non-commutative nature of matrix multiplication.
Thus, the fundamental concept of statistical independence does not make sense if a random matrix is considered as a single random object\footnote{Note that the more common definition of statistical independence by factorization of densities fails even more obviously, since it is not possible to even define a joint density in a sensible way due to the lack of appropriate relation operators ``$\le$'' and ``$>$''.}.
Random matrix theory circumvented this problem by considering a random matrix as being composed of standard scalar random variables.
Thus, it defines statistical independence of two random matrices if all entries of the one matrix are jointly independent from all entries of the other matrix.
Both approaches (a single random object, and a collection of scalar random variables) make sense and can enrich each other.
In this section, we take the viewpoint of free probability.
However, we restrict ourselves to asymptotically large random matrices as free random variables.
Free probability theory also applies to other classes of random operators.
\subsection{Free Expectation}
\label{FreExp}

An expectation operator should be linear and should assign $1$ to the identity matrix (the unit element of the matrix algebra).
It turns out that 
\begin{equation}
\expect\limits_{\rm free} \{\cdot\} \define {\rm Tr}(\cdot)
\end{equation}
is the right definition for some random matrices to fit into the framework of free probability theory.
We call those random matrices {\em converging} random matrices.
All random matrices discussed in Section~\ref{RanMatThe} are converging.
For those random matrices, ${\rm Tr}(\cdot)$ is indeed (almost surely) a deterministic quantity (as any expectation should be) due to the asymptotic convergence of their eigenvalues.

\subsection{Freeness}
\label{Fre}

Freeness is the conceptual counterpart in free probability to independence in classical probability theory.
Unfortunately, defining freeness is considerably more involved than defining independence.
The reason for this is due to the non-commutative nature of matrix multiplication and will soon become clear.

Consider the following example of four random matrices and assume that they satisfy
\begin{eqnarray}
\label{ex1}
{\rm Tr}(\matr{ABCD}) & = & {\rm Tr}(\matr{AB}){\rm Tr}(\matr{CD})\\
\label{ex2}
{\rm Tr}(\matr{ACBD}) &\ne& {\rm Tr}(\matr{AB}){\rm Tr}(\matr{CD})\\
\label{ex3}
{\rm Tr}(\matr{ACBD}) &\ne& {\rm Tr}(\matr{AC}){\rm Tr}(\matr{BD}).
\end{eqnarray}
For classical random variables and classical expectation operators, \eqs{ex1}{ex2} would contradict each other, since \eq{ex1} implies that the random variables $\matr A\matr B$ and $\matr C\matr D$ are uncorrelated while \eq{ex2} implies that they are not.
For non-commutative multiplication, however, \eqd{ex1}{ex2}{ex3} can be true at the same time.
In fact, with an appropriate definition of freeness, we find that \eqd{ex1}{ex2}{ex3} are implied by the assumption that the sets $\{\matr A,\matr B\}$ and $\{\matr C,\matr D\}$ form a family of free random variables while the sets $\{\matr A,\matr C\}$ and $\{\matr B,\matr D\}$ do not.
Note, however, that \eqd{ex1}{ex2}{ex3} do not imply freeness of the family $\big(\{\matr A,\matr B\},\{\matr C,\matr D\}\big)$, but additional conditions need to be satisfied.
In particular, the definition of freeness must respect the ordering of matrix factors in all possible products (joint moments).

\subsubsection{Non-Commutative Polynomials}
\label{NonPol}

Due to the non-commutative nature of matrix-multiplication, there are more different matrix polynomials of two or more variables for a fixed degree than for commutative variables such as the real or complex numbers.
On one hand, let $x,y$ be real numbers. The set of all $n^{\rm th}$ order polynomials in two variables $x$ and $y$ is given by 
\begin{eqnarray}
{\cal P}_n(x,y) \define \left\{\sum\limits_{i=1}^\infty \alpha_i x^{\ell_i}y^{m_i} :\nonumber\right.\hspace*{-30mm}\\
&&\left.\vphantom{\sum\limits_1} \ell_i,m_i\in\{0,1,\dots,n\} \wedge \alpha_i\in\RR \, \forall i \right\}.
\end{eqnarray}
For order two for instance, it is canonically given by a sum with only nine terms
\begin{equation}
\begin{array}c
\alpha_1 x^2y^2 + \alpha_2 x^2y + \alpha_3 xy^2 + \alpha_4 x^2 + \alpha_5 xy + \alpha_6 y^2 + \\
\mbox{} + \alpha_7 x +\alpha_8 y + \alpha_9.
\end{array}
\end{equation}
On the other hand, let $\matr A,\matr B$ be real matrices.
The set of all $n^{\rm th}$ order non-commutative polynomials in two variables $\matr A$ and $\matr B$ is given by
\begin{eqnarray}
{\cal P}_n(\matr A,\matr B) \define \left\{\sum\limits_{i=1}^\infty \alpha_i \prod\limits_{k=1}^n \matr A^{\ell_{i,k}}\matr B^{m_{i,k}}: \right.\nonumber\hspace*{-60mm}\\
&& \left. \sum\limits_{k=1}^n \ell_{i,k},\sum\limits_{k=1}^n m_{i,k} \in\{0,1,\dots,n\} \wedge \alpha_i\in\RR\,\forall i \right\}.\hphantom{mm}
\end{eqnarray}
For order two for instance, it is canonically given by a sum of 19 terms
\begin{equation}
\begin{array}c
\alpha_1 \matr A^2\matr B^2 + \alpha_2 \matr{AB}^2\matr A + \alpha_3 \matr{ABAB} + \alpha_4 \matr{BABA} + \mbox{}\\ 
\mbox{}  \alpha_5 \matr{BA}^2\matr B + \alpha_6 \matr B^2\matr A^2 + \alpha_7 \matr{A}^2\matr B + \alpha_8\matr{ABA}+\alpha_9\matr{AB}^2+\mbox{}\\
\mbox{}  \alpha_{10}\matr{BA}^2 + \alpha_{11} \matr{BAB} + \alpha_{12} \matr B^2\matr A + \alpha_{13} \matr A^2 + \alpha_{14} \matr{AB} + \mbox{}\\
\mbox{}  \alpha_{15} \matr{BA}+ \alpha_{16} \matr B^2 + \alpha_{17} \matr A + \alpha_{18} \matr B + \alpha_{19} \mbox{\bf I}.
\end{array}
\end{equation} 
A non-commutative polynomial in $p$ variables of order $n$ can be defined by
\begin{eqnarray}
{\cal P}_n(\matr A_1,\dots, \matr A_p) \define \left\{\sum\limits_{i=1}^\infty \alpha_i \prod\limits_{k=1}^n \prod\limits_{q=1}^p \matr A^{\ell_{i,k,q}}_q :\right. \hspace*{-60mm}\nonumber\\
&&\left.\sum\limits_{k=1}^n \ell_{i,k,q} \in\{0,1,\dots,n\} \wedge \alpha_i\in\RR\,\forall i,q \right\}.
\end{eqnarray}
Note that the number of terms can be considerably large even for small values of $n$ and $p$.
\subsubsection{Definition of Freeness}
\label{DefFre}

In literature \cite{voiculescu:92,speicher:98,voiculescu:98c,hiai:00}, freeness is defined in terms of algebras and sub-algebras.
Here we avoid referring to algebras, and define freeness in terms of non-commutative polynomials.

\begin{definition}
Let $s_k\in\{1,2,\dots,r\} $ be a sequence of integers such that 
\begin{equation}
\label{alter}
s_k-s_{k-1}\ne0.
\end{equation}
Then, the sets ${\cal Q}_1\define \{\matr A_1,\dots,\matr A_a\}$, ${\cal Q}_2 \define \{\matr B_1,\dots,\matr B_b\}$, $\dots$, ${\cal Q}_r$ form a free family $({\cal Q}_1,\dots,{\cal Q}_r)$ if, for every sequence $s_k$ obeying \eq{alter}, any sequence of polynomials $\matr Q_k$ such that $\matr Q_{k} \in {\cal P}_\infty({\cal Q}_{s_k})$, and any positive integer $n$,
\begin{equation}
{\rm Tr}\left(\matr Q_1\right)=\dots= {\rm Tr}\left(\matr Q_n \right) = 0 \Longrightarrow {\rm Tr}\left(\matr Q_{1}\matr Q_{2}\cdots\matr Q_{n} \right)=0 .
\end{equation}
\label{deffree}
\end{definition}
Note that due to \eq{alter} adjacent factors in the product $\matr Q_1\matr Q_2\cdots\matr Q_n$ must be polynomials of different sets of the family.
This reflects the non-commutative nature in the definition of freeness.

\subsubsection{Calculation of Expectations}
\label{CalExp}

Though the definition of freeness is very implicit, it can be used to recursively calculate cross-family joint moments of non-commutative random variables out of inter-family joint moments.

For instance, consider \eq{ex2} and assume that the factors are chosen from the free family $\big(\{\matr A,\matr B\},\{\matr C,\matr D\}\big)$.
Choose the non-commutative polynomials
\begin{eqnarray}
\label{defQ1}
\matr Q_1&=&\matr A - {\rm Tr}(\matr A)\I\\
\matr Q_2&=&\matr C - {\rm Tr}(\matr C)\I\\
\matr Q_3&=&\matr B - {\rm Tr}(\matr B)\I\\
\matr Q_4&=&\matr D - {\rm Tr}(\matr D)\I.
\label{defQ4}
\end{eqnarray}
Note that polynomials with adjacent indices are built of matrices belonging to different sets of the family.
Since
\begin{equation}
{\rm Tr}(\matr Q_k) = {\rm Tr}\big(\matr X-{\rm Tr}(\matr X)\I\big) = 0
\end{equation}
by the linearity of the expectation operator, Definition~\ref{deffree} implies
\begin{equation}
{\rm Tr}(\matr Q_1\matr Q_2\matr Q_3\matr Q_4) = 0.
\label{prodnull}
\end{equation}
Plugging \eq{defQ1} to \eq{defQ4} into \eq{prodnull} and expanding the products, we find with the linearity of the trace
\begin{equation}
\begin{array}{rcl}
{\rm Tr}(\matr{ACBD})\!\! &=\!\!& {\rm Tr}(\matr B){\rm Tr}(\matr{ACD}) + {\rm Tr}(\matr D){\rm Tr}(\matr{ACB}) + \nonumber\\
&& \mbox{} + {\rm Tr}(\matr A){\rm Tr}(\matr{CBD}) +  {\rm Tr}(\matr C){\rm Tr}(\matr{ABD})\\
&&\mbox{} - {\rm Tr}(\matr B){\rm Tr}(\matr D){\rm Tr}(\matr{AC}) \nonumber\\
&&\mbox{} - {\rm Tr}(\matr A){\rm Tr}(\matr B){\rm Tr}(\matr{CD}) \nonumber\\
&&\mbox{} - {\rm Tr}(\matr A){\rm Tr}(\matr D){\rm Tr}(\matr{CB})\\
&&\mbox{} - {\rm Tr}(\matr C){\rm Tr}(\matr B){\rm Tr}(\matr{AD})\\
&&\mbox{} -{\rm Tr}(\matr C){\rm Tr}(\matr D){\rm Tr}(\matr{AB})\\
&&\mbox{} - {\rm Tr}(\matr A){\rm Tr}(\matr C){\rm Tr}(\matr{BD})+\\
&&\mbox{} + 3{\rm Tr}(\matr A){\rm Tr}(\matr B){\rm Tr}(\matr C){\rm Tr}(\matr D).
\end{array}
\end{equation}
Now, we have broken down an expectation of four factors into sums of expectations of up to three factors.

The expectations of three factors can be broken down into sums of expectations of up to two factors.
This is demonstrated at the example of ${\rm Tr}(\matr{ACD})$.
Since the neighboring factors $\matr C$ and $\matr D$ belong to the same set of the free family, we must define the non-commutative polynomials in a different way as in \eq{defQ1} to \eq{defQ4}.
Now, an appropriate definition is
\begin{eqnarray}
\label{defQ1n}
\matr Q_1 &=& \matr A-{\rm Tr}(\matr A)\I\\
\matr Q_2 &=& \matr{CD} -{\rm Tr}(\matr{CD})\I.
\label{defQ2n}
\end{eqnarray}
Note that with definitions \eqs{defQ1n}{defQ2n}, adjacent polynomials in ${\rm Tr}(\matr Q_1\matr Q_2)$ belong to different sets of the free family. 
Proceeding this way for all remaining matrix products involving factors belonging to different sets of the family $\big(\{\matr A,\matr B\},\{\matr C,\matr D\}\big)$, we finally arrive at
\begin{eqnarray}
{\rm Tr}(\matr{ACBD}) &=& {\rm Tr}(\matr A){\rm Tr}(\matr B){\rm Tr}(\matr{CD})+\nonumber\\
&& + {\rm Tr}(\matr C){\rm Tr}(\matr D){\rm Tr}(\matr{AB})\nonumber\\
&& - {\rm Tr}(\matr{A}){\rm Tr}(\matr B){\rm Tr}(\matr C){\rm Tr}(\matr D).\hphantom{mmm}
\end{eqnarray}
The procedure for products of more than four factors is obvious, but can be very tedious.
The key point to succeed with this procedure is to define the non-commutative polynomials in an appropriate way which simply consists of collecting all factors belonging to identical sets of the free family and subtract its expectation.
\subsection{Free Random Matrices}
\label{FreRanMat}

Random matrices are a very popular and practically relevant example of non-commutative random variables.
However, not all sets of statistically independent random matrices are capable of forming free families.
So far, only a few examples of random matrices are known which form free families as their dimensions grow large.
Most of them were discovered by Voiculescu \cite{voiculescu:91,voiculescu:92}.
His results were strengthened and extended by Thorbj{\o}rnsen and Hiai and Petz \cite{thorbjoernsen:99,hiai:00} later on.

\subsubsection{Gaussian Random Matrices}
\label{GauRanMat}

Let the random matrices $\matr H_i, \forall i$, be square $N\times N$ with independent identically complex Gaussian distributed proper\footnote{A complex random variable is said to be {\em proper} if real and imaginary part are independent and identically distributed \cite{neeser:93}.} entries with zero mean and variance $1/N$. 
Moreover, let $\matr X_j,\forall j$, be an $N\times N$ matrices with upper bounded norm and a limit distribution as $N\to\infty$. 
Then the family
\begin{equation}
\Big(\left\{\matr X_1,\matr X_1^{\dagger},\matr X_2,\matr X_2^{\dagger},\dots\right\}, \left\{\matr H_1,\matr H_1^{\dagger}\right\}, \left\{\matr H_2,\matr H_2^{\dagger}\right\}, \dots\Big) 
\end{equation}
is asymptotically free as $N\to\infty$ almost surely \cite{hiai:00,thorbjoernsen:99}.

\subsubsection{Hermitian Random Matrices}
\label{HerRanMat}

Let the random matrices $\matr H_i, \forall i$, be $N\times K$ with independent identically complex Gaussian distributed proper entries with zero mean and variance $1/N$. 
Moreover, let the matrices $\matr X_j,\forall j$, be as in Section~\ref{GauRanMat} and let $\matr S_i=\matr H_i\matr H_i^{\dagger},\forall i$. 
Then the family
\begin{equation}
\Big(\left\{\matr X_1,\matr X_1^{\dagger},\matr X_2,\matr X_2^{\dagger},\dots\right\}, \left\{\matr S_1\right\}, \left\{\matr S_2\right\}, \dots\Big) 
\end{equation}
is almost surely asymptotically free as $N,K\to\infty$ with $\beta=K/N$ fixed \cite{hiai:00,thorbjoernsen:99}.

The asymptotic freeness of Hermitian random matrices has been used by Biglieri et al.\ \cite{biglieri:01} to calculate error rates of space-time codes.
\subsubsection{Unitary Random Matrices}

Let the random matrices $\matr T_i, \forall i$, be $N\times N$ Haar distributed random matrices as defined in Section~\ref{HaaDis}. 
Moreover, let the matrices $\matr X_j,\forall j$, be as in Section~\ref{GauRanMat}. 
Then, the family
\begin{equation}
\Big(\left\{\matr X_1,\matr X_1^{\dagger},\matr X_2,\matr X_2^{\dagger},\dots\right\}, \left\{\matr T_1,\matr T_1^{\dagger}\right\}, \left\{\matr T_2,\matr T_2^{\dagger}\right\}, \dots\Big)
\end{equation}
is almost surely asymptotically free as $N\to\infty$ \cite{hiai:00,thorbjoernsen:99}.

The asymptotic freeness of such unitary random matrices has been used by Debbah et.\ al.\ \cite{debbah:03} for the analysis of multi-carrier systems. 
\subsubsection{Similar Random Matrices}

Let the $N\times N$ matrices $\matr X_i$ be such that their inverses $\matr X_i^{-1}$ exist for all $i$ and 
\begin{equation}
\Tr\left(\matr X_i\matr X_j^{-1}\right) = 0 \qquad \forall i\ne j
\end{equation}
holds almost surely.
Moreover, let there be an $N\times N$ matrix $\matr H$ such that the family 
\begin{equation}
\label{defXi}
\left(\left\{\matr X_1,\matr X_1^{-1},\matr X_2,\matr X_2^{1},\dots \right\},\left\{ \matr H\right\}\right)
\end{equation}
is almost surely asymptotically free, as $N\to \infty$.
Then, the family
\begin{equation}
\label{freesimilar}
\left(\left\{\matr X_1\matr {HX}_1^{-1}\right\}, \left\{\matr X_2\matr {HX}_2^{-1}\right\}, \dots \right)
\end{equation}
is almost surely asymptotically free, too, as $N\to\infty$.

A weaker version of this result (without almost sure convergence) was proven by Evans and Tse \cite{evans:00} to analyze multiuser channel estimation in large CDMA systems. Note that this results implies that the identity matrix is free from itself which can also be easily proven directly from the definition of freeness. It also shows that the entries of free random matrices need not necessarily be independent:
If one uses permutation matrices for the matrices $\matr X_i$ in \eq{defXi}, one obtains free random matrices in \eq{freesimilar} that only differ by a re-ordering of their entries.

\subsection{R-Transform}
\label{RTra}

The calculation of distributions of functions of several free random variables via the definition of freeness is often a very tedious task.
However, for some operations such as summation, significant simplifications are possible.

Let $\matr A=\matr A^\dagger$ and $\matr B=\matr B^\dagger$ be two non-commutative random variables belonging to different sets of a free family.
Further, let
\begin{equation}
\label{defC}
\matr C \define \matr A+\matr B.
\end{equation}
Then, we call the probability measure (asymptotic eigenvalue distribution) $\prob{\matr C}{x}$ the {\em additive free convolution} of the probability measures $\prob{\matr A}x$ and $\prob{\matr B}x$.
Unlike classical convolution which provides the distribution of a sum of independent commutative random variables, additive free convolution is a highly non-linear operation and, thus, cannot be performed by simple integration.

In principle, the moments of $\prob{\matr C}x$ could be found from the moments of $\prob{\matr A}x$ and $\prob{\matr B}x$ via the definition of freeness.
Then, the distributions could be recovered from the moments solving the problem of moments via the Stieltjes transform.
However, this is a very tedious task.
Significant simplification is achieved via the {\em R-transform}.
The R-transform is defined in terms of the Stieltjes transform as
\begin{equation}
\RT{}w \define \StTi{}{-w}-\frac1w
\label{defRT}
\end{equation}
where $G^{-1}(\cdot)$ denotes the inverse function of the Stieltjes transform with respect to composition (this should not be confused with the Stieltjes inversion formula defined in \eq{Stinv}).

The R-transform linearizes additive free convolution of two non-commutative probability measures \cite{voiculescu:92}.
Thus, we have
\begin{equation}
\label{RTinv}
\RT{\matr C}w = \RT{\matr A}w + \RT{\matr B}w.
\end{equation}
Then, the distribution of $\matr C$ can be recovered from the R-transform via
\begin{equation}
\label{invRT}
\StT{\matr C}{\RT{\matr C}{-w}-w^{-1}}=w
\end{equation}
which follows directly from \eq{defRT} and the Stieltjes inversion formula \eq{Stinv}.

Tse \cite{tse:99c} discovered that the additivity of the R-transform is responsible for the decoupling of interference powers in the SINRs of asymptotically large random CDMA with linear multiuser receivers.
 

In Table~\ref{tabRT}, R-transforms of some of the random matrices introduced in Section~\ref{ConEig} are listed.
\begin{table}[tbh]
\caption{\label{tabRT} Table of R-transforms.}
\begin{eqnarray*}
\RT{\alpha\mbox{\bf I}}w &=& \alpha\\
\RT{\matr K}w &=&w\\
\RT{\matr Q^2}w &=& \frac 1{1-w}\\
\RT{\matr Q^{-2}}w &=& \frac 1{\sqrt{-w}}\\[2mm]
\RT{\matr R^2}w &=&\frac{\load }{1-w}\\[2mm]
\RT{\matr R^{-2}}w &=& \frac{\beta-1-\sqrt{(\beta-1)^2-4w}}{2w}\\[2mm]
\RT{\matr Y}w &=&\frac{-1+\sqrt{1+4w^2}}w
\end{eqnarray*}
\end{table} 
Table~\ref{PropRT} shows some general properties of the R-transform.
\begin{table}[tbh]
\caption{\label{PropRT} Properties of the R-transform.}
\begin{eqnarray*}
\RT{\alpha\matr X}w &=& \alpha\RT{\matr X}{\alpha w}\\
\frac1{\RT{\matr X}w} &=& \RT{\matr X^{-1}}{-\RT{\matr X}w\left(1+w\RT{\matr X}w\right)}\\ 
\lim\limits_{w\to 0}\RT{}w &=& \int x\,{\rm d}\Prob{}x
\end{eqnarray*}
\end{table}
The R-transform is decreasing on the real line and strictly decreasing unless the distribution is a single point mass \cite{zaidel:11}.


\subsection{Free Convolution of Binary Measures}
\label{AddFreCon}

Assume that the two free random variables in \eq{defC} are binary distributed 
\begin{equation}
\prob{\matr A}x=\prob{\matr B}x=\frac12\,\deltaf(x-1)+\frac12\,\deltaf(x+1).
\end{equation} 
Then, by \eq{defSt}, the Stieltjes transforms of their distributions are
\begin{equation}
\StT{\matr A}s=\StT{\matr B}s = \frac s{1-s^2}.
\end{equation}
With definition \eq{defRT}, we find that the R-transform is given by 
\begin{equation}
\RT{\matr A}w=\RT{\matr B}w = \frac{-1+\sqrt{1+4w^2}}{2w}.
\label{exrt}
\end{equation}
Here, we have discarded the other branch of the square root, since $\lim_{w\to0}\RT{\matr A}w = 0$ by Table~\ref{PropRT}.

The R-transform of the distribution of $\matr C$ is the sum of the R-transforms of the distributions of the matrices $A$ and $B$. 
Thus, we find
\begin{equation}
\RT{\matr C}w=\frac{-1+\sqrt{1+4w^2}}w.
\end{equation}
Though it differs only by a factor of 2 from the R-transforms in \eq{exrt}, it corresponds to an essentially different distribution function. 
Using Table~\ref{tabRT}, we find that the additive free convolution of two binary measures gives the {\em inverse} semicircle law
\begin{equation}
\prob{\matr C}x=\cases{\frac1\pic \frac1{\sqrt{4-x^2}} & $|x|<2$\cr
0 & elsewhere}.
\end{equation}

This procedure can be used, in principle, for the additive free convolution of any distributions.
Practical problems occur, when calculating the inverse functions in \eq{defRT} and \eq{invRT}, since these function can happen to not allow for an inverse function in closed form, e.g.\ since they are polynomials of fifth or higher order.
In some of these cases, it helps to apply the Stieltjes inversion formula \eq{Stinv} to an implicit equation for the Stieltjes transform $\StT{\matr C}s$.
This results in solving a system of two equations for the real and imaginary part of $\StT{\matr C}s$.
In the worst case, one might succeed to develop those functions which cause trouble into power series.
The latter approach, however, is rather complicated.
It can be simplified, making use of the Kreweras complement, see Section~{\ref{NonCroPar}.
  
The additive free convolution of two binary distributions has turned out to be a continuous distribution.
This is in sharp contrast to what we are used to from adding (classical) commutative random variables.

\subsection{Free Central Limit Theorem}
\label{FreCenLim}

In classical (commutative) probability theory, the sum of an infinite number of independent identically distributed zero-mean terms is Gaussian distributed.
This is well-known as the central limit theorem.
It seems obvious that there should be some counterpart in free probability as well.
It is intuitive that the notion of statistical independence should be replaced by freeness, but what is the counterpart of the Gaussian distribution?

Consider again the additive free convolution of free binary measures as in the previous subsection.
However, let there be now $n$ terms to be summed and the sum to be normalized to unit variance
\begin{equation}
\label{defCn}
\matr C_n\define\frac1{\sqrt n}\sum\limits_{k=1}^n\matr A_k.
\end{equation}
We know from \eq{exrt} and Table~\ref{PropRT} that 
\begin{equation}
\label{trn}
\RT{\matr C_n}w = \frac{\sqrt{n^2+4nw^2}-n}{2w}.
\end{equation}
In the Stieltjes domain, this reads
\begin{equation}
\StT{\matr C_n}s = \frac12 \frac{(n-2)s-\sqrt{n^2s^2-4n^2+4n}}{s^2-n}
\end{equation}
which, for $n>1$, corresponds to the density
\begin{equation}
\label{dfg}
\prob{\matr C_n}x = \cases{\frac1{2\pic} \frac{\sqrt{4n^2-4n-n^2x^2}}{n-x^2} & $|x|<2\sqrt{1-1/n}$ \cr
0 & elsewhere}.
\end{equation}
The densities for $n<8$ are shown in Fig.~\ref{CLT}.
\begin{figure}
\centerline{\epsfig{file=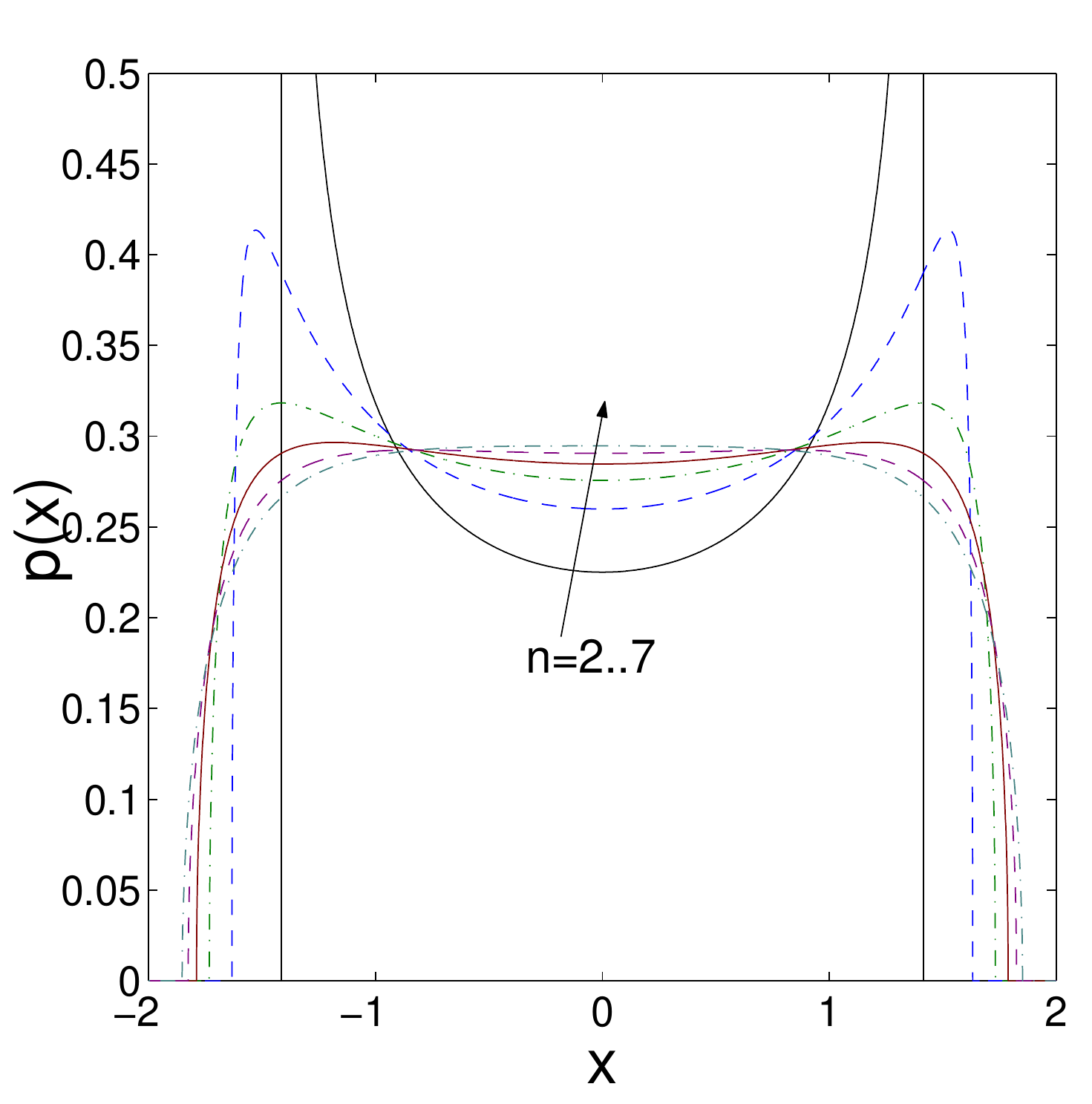,scale=.6}}
\caption{\label{CLT} Additive free convolution of several binary measures.}
\end{figure}
One can observe that they approach the semicircular distribution.
This can be easily verified from \eq{trn} since
\begin{equation}
\lim\limits_{n\to\infty}\RT{\matr C_n}w = w
\end{equation} 
and $w$ is the R-transform of the semicircle law by Table~\ref{tabRT}.
It can also be seen, taking directly the limit $n\to\infty$ in \eq{dfg}.

The semicircular distribution is not only the limit distribution for the additive free convolution of many binary measures, but any appropriately scaled zero-mean measures with finite moments, as proven in the following.

The R-transform is an analytic function within the neighborhood of $w=0$.
Thus, we can write it as
\begin{equation}
\label{eq118}
\RT{\matr A_k}w = \sum\limits_{\ell=0}^\infty \alpha_{\ell,k} w^\ell
\end{equation}
where $\alpha_{0,k}=0$ since we assume ${\rm Tr}(\matr A_k)=0$.
From \eq{exrt}, \eq{defCn}, and Table~\ref{PropRT}, we find
\begin{eqnarray}
\lim\limits_{n\to\infty}\RT{\matr C_n}w &=& \lim\limits_{n\to\infty}\frac1{\sqrt n} \sum\limits_{k=1}^n\RT{\matr A_k}{\frac w{\sqrt n}}\\
&=& \lim\limits_{n\to\infty}\frac1{\sqrt n} \sum\limits_{k=1}^n \sum\limits_{\ell =1}^\infty \alpha_{\ell,k} \left(\frac w{\sqrt n}\right)^\ell\\
&=& \lim\limits_{n\to\infty} \frac1n\sum\limits_{k=1}^n \alpha_{1,k} w.
\label{eq121}
\end{eqnarray}
Using Tables~\ref{tabRT} and \ref{PropRT}, this can be identified as a scaled semicircle distribution regardless of the higher order moments of the distributions of the terms to be summed.

The coefficients of the series expansion of the R-transform are the free cumulants.
In analogy to classical (commutative) probability theory, all cumulants of the limit distribution $\Prob{\matr C_\infty}x$ except for mean and variance vanish.
Therefore, the classical central limit theorem can easily be proven along the same lines as \eqs{eq118}{eq121}.
The limit distribution is called the {\em normal} distribution, in general, and {\em Gaussian} and {\em semicircle} distribution for commutative and non-commutative random variables, respectively.

According to the classical definition of the cumulant generating function, the mean of the distribution is the coefficient of the linear term.
For the R-transform, however, the mean is given as $\RT{}0$, i.e.\ the coefficient of the constant term $w^0$.
Some authors, e.g.\ \cite{hiai:00}, dislike that, and use the R-series defined as $w\RT{}w$ instead.
In communications engineering, Voiculescu's \cite{voiculescu:91} original notation in terms of R-transform is common.
Despite the discrepancy to the definition of the cumulant generation function in cummutative probability theory, we will call it {\em free cumulant generating function} in the sequel.

\subsection{Free Fourier Transform}
\label{FreFouTra}

The moment generating function is defined as the Fourier transform of the probability distribution (for both commutative and non-commutative random variables).
In classical probability theory, the cumulant generating function is simply given as the logarithm of the moment generating function.
In free probability theory, the similar interrelationship
\begin{eqnarray}
\lim\limits_{K\to\infty} \frac1K \log \expect\limits_{\matr J} \e^{ K {\rm trace} [\matr {JQ}]} &=& \int\limits_0^1{\rm trace} \left[\matr Q \RT{\matr J}{\matr Q w} \right]{\rm d}w \\
&=&\sum\limits_{k=1}^K \int\limits_0^{\lambda_k(\matr Q)} \RT{\matr J}w{\rm d}w
\end{eqnarray}
holds, if ${\rm rank}(\matr Q) = {\rm o}(\sqrt K)$ \cite{guionnet:05}.

This result is referred to as the Harish-Chandra \cite{harish:57} and as the Itzykson-Zuber \cite{itzykson:80} integral in mathematics and physics, respectively. Its representation in terms of the R-transform by \cite{guionnet:05} is rather recent.
It was re-derived in wireless communications literature in \cite{hassibi:02} and has helped to derive general large-system results for CDMA and MIMO systems in dependency of the systems eigenvalue distributions, see e.g.\ \cite{takeda:06,mueller:08}.
An example for its application will be given in Section~\ref{Exa52}. Recently, a new non-asymptotic version of the free Fourier transform has been reported in context of diversity-multiplexing trade-offs in MIMO systems \cite{leveque:10}.

\subsection{S-Transform}
\label{STra}

In analogy to additive free convolution, we define
\begin{equation}
\matr D\define\matr{AB}
\end{equation}
and call the probability measure $\prob{\matr D}x$ the {\em multiplicative free convolution} of the two probability measures $\prob{\matr A}x$ and $\prob{\matr B}x$, again under the restriction that $\matr A=\matr A^\dagger$ and $\matr B=\matr B^\dagger$ belong to different sets of a free family of non-commutative random variables.
Although the factors are non-commutative operators, multiplicative free convolution is commutative \cite{voiculescu:92}.

Under the additional assumption that the probability measures of both factors have non-zero mean, i.e.\ ${\rm Tr}(\matr A)\ne0\ne{\rm Tr}(\matr B)$, we can linearize multiplicative free convolutional via the definition of an appropriate transform such that
\begin{equation}
\ST{\matr D}z = \ST{\matr A}z\ST{\matr B}z
\end{equation}
where $\ST{}\cdot$ is called the {\em S-transform}.
In order to define the S-transform explicitly, we first introduce an auxiliary transform
\begin{eqnarray}
\label{defYT}
\YT{}s &\define& \int\frac{sx}{1-sx}\,{\rm d}\Prob{}x\\
&=&-s^{-1}\StT{}{s^{-1}}-1
\label{defYTS}
\end{eqnarray}
which can be obtained either directly by integrating an appropriate kernel with respect to the measure of interest \eq{defYT} or in terms of the Stieltjes transform \eq{defYTS}.
Since the definition in terms of the Stieltjes transform is rather straightforward, we do not provide a table for the auxiliary transform $\YT{}\cdot$.
Calculating the inverse with respect to composition of this auxiliary transform, the S-transform is given as
\begin{equation}
\label{defST}
\ST{}z \define \frac{1+z}z \, \YTi{}z. 
\end{equation}
In order to return to the probability distribution, you return to the Stieltjes domain via \eq{defST} and \eq{defYTS} and then apply the Stieltjes inversion formula \eq{Stinv}.

S-transforms of some of the random matrices introduced in Section~\ref{ConEig} are listed in Table~\ref{tabST}. General properties of the S-transform and its connection to the R-transform are shown in Table~\ref{PropST}.
\begin{table}[thb]
\caption{\label{tabST} Table of S-transforms.}
{\begin{eqnarray*}
\ST{\alpha\mbox{\bf I}}z &=& \frac1\alpha\\
\ST{\matr K}z &=& \frac1{\sqrt z}\\
\ST{\matr Q^2}z &=& \frac1{1+z}\\
\ST{\matr R^2}z &=&\frac1{\load+z}\\
\ST{\matr R^{-2}}z &=&\beta-1-z\\
\ST{\matr Y}z &=& \sqrt{\frac14+\frac1{2z}}
\end{eqnarray*}}
\end{table} 
\begin{table}[tbh]
\caption{\label{PropST} Properties of the S-transform.}
\begin{eqnarray*}
\ST{\alpha\matr X}z &=& \alpha\ST{\matr X}{z}\\
\ST{}z &=& \frac1{\RT{}{z\ST{}z}}\\
\lim\limits_{z\to 0}\frac1{\ST{}z} &=& \int x\,{\rm d}\Prob{}x
\end{eqnarray*}
\end{table}
The S-transform is decreasing on the real line and strictly decreasing unless the distribution is a single point mass.

It was shown in \cite{mueller:01b} that Silverstein's formula \eq{PRM} for products of some asymptotic random matrices is equivalent to applying the S-transform.
Thus, the S-transform is not restricted to free random matrices, but also applies to any asymptotic random matrices which obey the conditions of Section~\ref{ProRanMat}.

\subsection{Non-Crossing Partitions}
\label{NonCroPar}

As explained in Section \ref{FreCenLim}, the classical central limit theorem can easily be proven along the same lines as \eqs{eq118}{eq121}.
The limit distribution is called the {\em normal} distribution, in general, and {\em Gaussian} and {\em semicircle} distribution for commutative and non-commutative random variables, respectively.

And just like there is an isomorphism between the $m^{\rm th}$ moment of a Gaussian random variable and the number of ways a set of cardinality $m$ can be partitioned into subsets, there is also an isomorphism between the $m^{\rm th}$ moment of a semicircular random variable and the number of ways a set of cardinality $m$ can be partitioned into subsets in a certain, so-called {\em non-crossing} way. This isomorphism can be used to convert moments series into cumulant series and vice versa.
The present section provides an introduction to this theory.

A {\em partition}
$\pi \define \{ V_1, \ldots, V_k \}$ of the set $\{1, \ldots, m\}$ is a
decomposition of  $\{1, \ldots, m\}$ into disjoint and non-empty sets $V_i$
such that $\bigcup_{i=1}^k V_i =\{1, \ldots, m\}$. The elements $V_i$ are called
the blocks of the partition $\pi$. We will denote the set of all partitions of
$\{1, \ldots, m\}$ by ${\cal P}(m)$.
If for two distinct partitions of the same set, each block of the first partition is a union of the blocks of the second partition, we call the first partition greater than the second partition.
We denote the smallest and largest partition of $\{1,2,\dots,m\}$ by
\begin{eqnarray}
0_m &=& \{(1),(2), \ldots, (m) \} \\
1_m &=& \{(1,2, \ldots, m)\},
\end{eqnarray}
respectively.

A partition $\pi \in {\cal P}(m)$
is called {\em crossing}, or, more specifically  {\em $1$-crossing}\footnote{Partitions with more complicated crossings between blocks  are recent subject of investigation (see e.g. \cite[and refs. therein]{bona:99}) for different modeling purposes.}, if there exist four numbers $1 \leq i_1 < i_2 < i_3 < i_4 \le m$ such that $i_1$
and $i_3$  are in some block $V\in \pi$ and $i_2$ and $i_4$ are in a different block $V^\prime\in\pi$.
The set of all non-crossing partitions of ${\cal P}(m)$ is
denoted by $\NC(m)$.


%


A toy example where non-crossing partitions arise as solution to a combinatorial problem is shown  in the following.

\textbf{Problem:} {\em Consider $m$ balls arranged on a circle,
numbered from $1$ to $m$ clockwise.
Among all the partitions of the $m$ balls into $k$ disjoint sets, such that
the cardinalities of these sets are $m_1, \ldots, m_k$, how many among them are non-crossing? The solution is given by
the coefficient \begin{equation}
c(m_1, \ldots, m_k) = {m! \over (m-k+1)! \cdot f(m_1, \ldots, m_k)}\,,
\end{equation}
with
\[
f(m_1, \ldots, m_k)  = f_1! \cdots f_n!\,,
\]
where a vector of $k$ integers $m_1, \ldots, m_k$
is partitioned into $n$ equivalence classes under the equivalence
relation $a=b$, and the cardinalities of
the equivalence classes are given by $f_1, \ldots, f_n$.
}

The proof of this statement, originally given in \cite{li:01b} for a combinatorial problem posed in \cite{honig:01}, is shown below.

The problem is stated in such a way that sets that contain the same number of elements are considered
indistinguishable. For sake of simplicity, we start from the case in which these sets are distinguishable and
prove that the number of non-crossing partitions is given by
\begin{equation}
{m! \over (m-k+1)!}.
\end{equation}

Afterwards, to make the sets distinguishable, we label them by numbering  from $1$ to $k$, where the cardinality
associated to set  $S_i$ is $m_i$ ($1 \leq i \leq k$).
Selecting a ball, denoted by $b_1$, from the $m$ balls, then another, say $b_2$, from the remaining $m-1$ balls, and so on
until we select $b_{k-1}$, it is easy to see that there are $m! / (m-k+1)!$
ways to make this selection.

Once we have carried on the selections, we describe a procedure that generates a non-crossing
partition from any such selection. Then we prove that non-crossing partitions
generated from different selections must also be different, and for each
non-crossing partition, there must exist a selection that generates it.
Assume, without
loss
of generality, that $b_2$ is located after $b_1$ on the circle, and so
on.

(i) Suppose $b_1, \ldots, b_{k-1}$ are selected.
Being $m_1 + \cdots + m_k = m$, there must exist $i$ such that there are at
least
$m_i$ balls between $b_i$ and $b_{i+1}$, with $b_i$ included and $b_{i+1}$
excluded.
We assign the $m_i$ balls starting from $b_i$ ($b_i$ included) to set $S_i$, and
 we take away the already assigned $m_i$ balls from the circle. Now we have
a smaller
circle of $m - m_i$ balls.
Following the same steps, there exists $j$ such that there are at least $m_j$
balls between $b_j$ and the next selected ball after $b_j$, with $b_j$ included
and the
next selected ball after $b_j$ excluded. We then assign $m_j$ balls starting
from $b_j$ to set $S_j$. Then the already assigned $m_j$ balls are taken
away from the circle. By iterating the procedure  until the $k-1$ selected
balls are used up, we get $m - m_1 - \cdots - m_{k-1} = m_k$
balls left,
and we naturally assign them to set $S_k$.

Since at each stage of the procedure described above, we always assign a
continuous block
of balls to a set, it is readily shown by induction that a partition
generated by this procedure is non-crossing.

%

(ii) The fact that two different selections of balls, say  $b_1, \ldots,
b_{k-1}$
and  \\ $b_1', \ldots, b_{k-1}'$, generate different partitions can be easily proven by contradiction.

(iii) It remains to  be proven that for every non-crossing partition, there exists a set of
$k-1$
selected balls that generates it. This can be done through the application of the following


{\bf Lemma:} {\em For a given non-crossing partition, there exists
a set $S_i$ such that all the balls in $S_i$ form a continuous block on the original
circle.}


Suppose that $S_i$ is such a set, we select the first ball (in clockwise
direction) of $S_i$ as $b_i$. Then we take away $S_i$ from the circle,
getting a partition of $m-m_i$ balls into $k-1$ disjoint sets which is non-
crossing.
The lemma above can be iterated until
$k-1$ balls are selected. It is easy to verify that this set of selected
balls indeed
generates the same partition that we start with.

From  (i), (ii), and (iii), a bijection  from
the selections
of $k-1$ balls to the non-crossing partitions has been established. Therefore,
the number
of non-crossing partitions is given by

\begin{equation}
{m! \over (m-k+1)!}
\end{equation}
when sets are distinguishable.

Moving to our case, where sets are considered indistinguishable and all that matters is their cardinality, the number of non-crossing
partitions
will be reduced, with combinatorial arguments,  by a factor $f(m_1, \ldots, m_k)$ as previously described.
The factor $f(m_1, \ldots, m_k)$ makes the sets of the same size
indistinguishable
by mixing up these sets.

Two particularly meaningful operations can be defined over the $\NC(m)$ lattice. In particular, observe first that
 given any two partitions $\pi, \, \sigma \in \NC(m)$  with $\pi \leq \sigma$, one can define an interval of $\NC(m)$ as follows:
 $$
 [\pi, \sigma]=\{\rho \in \NC(m): \pi \leq \rho \leq \sigma\}\,.
 $$
 Such an interval can be always factorized, i.e.  there exist
canonical natural numbers $k_1,k_2, \ldots,  $ such that
\begin{equation}
[\pi, \sigma]  \cong \NC(1)^{k_1} \times \NC(2)^{k_2} \times \cdots
\end{equation}

From this factorization property  one can define a {\em
multiplicative function } $f$  (for non-crossing partitions)
corresponding to a sequence $(a_1,a_2, \ldots)$ of complex numbers by requirement
that:
\begin{equation}
f (\pi, \sigma) \define
\left \{
\begin{array}{clll}
a_1^{k_1} a_2^{k_2} \cdots  &     \\
0 &  \mbox{whenever} \, \, \,  \pi  \not \leq \sigma
\end{array}
\right.
\label{eq:moltifunct}
\end{equation}
if $[\pi,\sigma]$ has a factorization as above. Via this operation over the above-considered partition lattice, the spectrum of sub-matrices has been characterized in \cite{TCSVit2010}.

The second interesting operation one can define over $\NC(m)$ is called { \em (combinatorial)
convolution } $(\cdot \ast \cdot)$  of
 $f$ and $g$ defined
as in (\ref{eq:moltifunct}), and  is given by
 \begin{equation}
(f \ast g ) \define
\sum_{ \begin{array}{ccc}
\tau \in \NC(m)   \\
\pi \leq \tau \leq \sigma
\end{array} }
f (\pi , \tau ) g(\tau, \sigma)  \label{eq:conv}
\end{equation}
for $ \pi \leq \sigma \in \NC(m)$.

We may further notice that the lattice $\NC(m)$ is self-dual, and the isomorphism
$K:\NC(m)\rightarrow\NC(m)$ implementing such a self-duality (also called \emph{complementation map} \cite{Kreweras:65}) is defined as follows: Let $\pi \in \NC(m)$, and consider also numbers $\overline{1},\ldots,\overline{m}$, ordered in alternating way such that
\[
1,\overline{1},2,\overline{2},\ldots,m,\overline{m}.
\]
The \emph{Kreweras complement} $K(\pi)$ of $\pi \in \NC(m)$ is by definition the biggest $\sigma \in \NC(\overline{1},\ldots,\overline{m})$ such that $\pi\cup\sigma \in \NC(1,\overline{1},2,\overline{2},\ldots,m,\overline{m})$, and can be usefully exploited to simplify computation of  the free convolution, as already mentioned in  (see Sec.\ref{AddFreCon}).

Examples of the exploitation of the abovementioned combinatorial operations in the field of wireless communications are mainly \cite{TCSVit2010} and \cite{DRyan}. In \cite{TCSVit2010} the authors use (\ref{eq:conv}) to provide a characterization of the capacity of 
channels affected by severe impairments in time and frequency. In \cite{DRyan} the focus is on the study of the 
spectral properties of matrices arising from the collection of sample measurments in sensor networks with randomly placed nodes.

\subsection{1st Example: Channel Modeling}
\label{Exa4}

Asymptotic freeness of certain large random matrices can be used to characterize the properties of communication channels with antenna arrays at both ends of the communication link in terms of the singular values of the channel's transfer matrix.
This approach was studied by M\"uller \cite{mueller:00c} for the first time decomposing the channel matrix into a product of three asymptotically free matrices: the steering matrix at transmitter side, a scattering matrix, and a steering matrix at receiver side.
The approach was generalized to matrix products of more factors by M\"uller in \cite{mueller:01b} and dispersive channels with decaying power-delay profile in \cite{mueller:01f} and confirmed by measurements in \cite{mueller:01d}.
As an example the asymptotic eigenvalue distribution associated with a channel composed of an arbitrary number of free matrix factors is calculated by means of the S-transform.

Consider a communication channel with $K_0$ transmitting and $K_N$ receiving antennas grouped into a transmitter and a receiver array, respectively.
Let there be $N-1$ clusters of scatterers each with $K_n, 1\le n\le N-1$, scattering objects.
Assume that the vector-valued transmitted signal propagates from the transmitter array to the first cluster of scatterers, from the first to the second cluster, and so on, until it is received from the $(N-1)^{\rm st}$ cluster by the receiver array.
Such a channel model is discussed and physical motivation is given in \cite[Sec.\ 3]{andersen:02}.
Indoor propagation between different floors, for instance, may serve as an environment where multifold scattering can be typical, cf.\ \cite[Sec.\ 13.4.1]{saunders:99}.

The communication link outlined above is a linear vector channel that is canonically described by a channel matrix 
\begin{equation}
\label{eq1}
\matr H_N = \matr M_N \matr M_{N-1} \cdots \matr M_2 \matr M_1 \define \prod\limits_{n=1}^N \matr M_n
\end{equation}
where the matrices $\matr M_1$, $\matr M_{1<n<N}$, and $\matr M_N$ denote the subchannels from the transmitter array to the first cluster of scatterers, from the $(n-1)^{\rm st}$ cluster of scatterers to the $n^{\rm th}$ cluster, and from the $(N-1)^{\rm st}$ cluster to the receiving array, respectively. 
This means that $\matr M_n$ is of size $K_n\times K_{n-1}$.

The performance of communication via linear vector channels described as in \eq{eq1} is determined by the $K_N$ eigenvalues of the covariance matrix $\matr C_N\define\matr H_N\matr H_N^{\dagger}$ for many practically relevant cases. 
In the following, the asymptotic distributions of these eigenvalues are calculated.

Assume that the family $\big(\{\matr M_1^{\dagger}\matr M_1\}$, $\{\matr M_2^{\dagger}\matr M_2\}$, $\dots$, $\{\matr M_N^{\dagger}\matr M_N\}\big)$ is asymptotically free as all sizes $K_n$ tend to infinity with the ratios
\begin{eqnarray}
\chi_n \define \frac {K_{n-1}}{K_n}, \qquad 1\le n\le N,
\end{eqnarray}
remaining constant.
Consider the random covariance matrices 
\begin{equation}
\matr C_i \define \matr H_i\matr H_i^{\dagger} = \left(\prod\limits_{n=1}^i\matr M_n\right)\left(\prod\limits_{n=1}^i\matr M_n\right)^{\dagger}.
\end{equation}
In the following, the asymptotic eigenvalue distribution of $\matr C_N$ is recursively calculated applying the S-transform.

For that purpose, note that the  non-zero eigenvalues of the matrices
\begin{equation}
\matr{\tilde{C}}_i \define \left(\prod\limits_{n=1}^{i-1}\matr M_n\right)\left(\prod\limits_{n=1}^{i-1}\matr M_n\right)^{\dagger} \matr M_i^{\dagger} \matr M_i.
\end{equation}
and $\matr C_i$ are identical and that
\begin{equation}
\matr{\tilde{C}}_{i+1} = \matr C_{i} \matr M_{i+1}^{\dagger}\matr M_{i+1}\qquad \forall i.
\label{iplus1}
\end{equation}

Since the non-zero eigenvalues of the matrices $\matr C_i$ and $\matr{\tilde C}_i$ are identical, their empirical distributions differ only by a scaling factor and a point mass at zero.
In the Stieltjes domain, this translates with Table~\ref{PropStT} into 
\begin{equation}
\label{eq13}
G_{\matr C_i}\left( s \right)+\frac1s = \chi_{i} G_{\matr{\tilde{C}}_i}\left( s \right)+\frac{\chi_{i}}s.
\end{equation}
It is straightforward from \eq{defYTS} and \eq{defST} that \eq{eq13} reads in terms of $\Upsilon(s)$ and $S(z)$ as
\begin{eqnarray}
\Upsilon_{\matr C_i}(s)&=&\chi_{i}\Upsilon_{\matr{\tilde{C}}_i}(s)\\
S_{\matr C_i}(z)&=&\frac{z+1}{z+\chi_{i}}S_{\matr{\tilde{C}}_i}\left(\frac z{\chi_{i}}\right),
\label{rotfor}
\end{eqnarray}
respectively.
Let the entries of $M_{1\le n\le N}$ be independent and identically distributed with zero-mean and respective variances $1/K_n$.
Then, Table~\ref{tabST} yields
\begin{eqnarray}
\label{eq17}
S_{\matr M_i\matr M_i^{\dagger}}(z) &=& \frac1{z+\chi_i}\\
S_{\matr M_i^{\dagger}\matr M_i}(z) &=& \frac1{1+z\chi_i}.
\label{eq16}
\end{eqnarray} 
Now, we proof by induction:
\begin{lemma}
Define the ratios
\begin{equation}
 \label{eq18}
\rho_n\define \frac{K_n}{K_N}.
\end{equation}
Let $K_n\to\infty$, but the ratios $\rho_n$ remain fixed for all $0\le n\le N$.
Then,
\begin{equation}
\label{lem41}
S_{\matr C_N}(z) = \prod\limits_{n=1}^N \frac{\rho_n}{z+ \rho_{n-1}}.
\end{equation}
\end{lemma}
{\em Proof:}
Since the ratios $\rho_n$ depend on $N$ via \eq{eq18}, \eq{lem41} is re-written into
\begin{equation}
\label{eq44}
S_{\matr C_N}(z) = \prod\limits_{n=1}^N \frac{K_n}{zK_N+ K_{n-1}}.
\end{equation}

First, \eq{eq44} is verified for $N=1$.
\eq{eq17} directly gives
\begin{equation}
S_{\matr C_1}(z) = \frac1{z+\chi_{_1}} = \frac{K_1}{zK_1+K_0}
\end{equation}
which proofs \eq{eq44} for $N=1$.

Second, assuming \eq{eq44} holding for the $i$-fold product, \eq{eq44} is shown to also hold for the $(i+1)$-fold product.
Transforming \eq{iplus1} into the S-domain gives
\begin{equation}
\label{eq47}
S_{\matr{\tilde{C}}_{N+1}}(z) = S_{\matr C_N}(z) \frac1{1+zK_N/K_{N+1}}.
\end{equation}
The rotation formula \eq{rotfor} yields
\begin{equation}
S_{\matr C_{N+1}}(z) = \frac{z+1}{z+K_N/K_{N+1}}\, S_{\matr{\tilde{C}}_{N+1}}\left(z\,\frac{K_{N+1}}{K_N}\right)
\end{equation}
which gives with \eq{eq47}
\begin{equation}
S_{\matr C_{N+1}}(z) = \frac{K_{N+1}}{zK_{N+1}+K_N} \, S_{\matr{C}_N}\left(z\,\frac{K_{N+1}}{K_N}\right).
\end{equation}
We make use of the assumption that \eq{eq44} is valid for the $N$-fold product and find
\begin{eqnarray}
S_{\matr C_{N+1}}(z) &=& \frac{K_{N+1}}{zK_{N+1}+K_N} \, \prod\limits_{n=1}^N\frac{K_n}{zK_{N+1}+K_{n-1}}\hphantom{fff}\\
&=& \prod\limits_{n=1}^{N+1}\frac{K_n}{zK_{N+1}+K_{n-1}}.
\end{eqnarray}
Hereby, the induction is complete.
\hfill $\Box$

The lemma yields with \eq{defST} and \eq{defYTS}
\begin{equation}
 s\big(\Upsilon_{\matr C_N}(s)+1\big)\prod\limits_{n=1}^N\frac{\Upsilon_{\matr C_N}(s)+\rho_{n-1}}{\rho_n} = \Upsilon_{\matr C_N}(s)
\end{equation}
\begin{equation}
\label{eq21}
G_{\matr C_N}(s)\prod\limits_{n=1}^N\frac{\rho_{n-1}-1-sG_{\matr C_N}(s)}{\rho_n} - sG_{\matr C_N}(s) = 1.
\end{equation}
The Stieltjes transform of the asymptotic eigenvalue distribution of $\matr C_N$ is determined in \eq{eq21} by a polynomial equation of $(N+1)^{\rm st}$ order.
For $N>3$, it cannot be resolved with respect to the Stieltjes transform, in general. 
However, it is shown in \cite{mueller:01b} how to obtain an infinite power series for $\Upsilon_{\matr C_N}(s)$ and calculate the moments of the asymptotic eigenvalue distribution.

\subsection{2nd Example: Correlated MIMO Channels}
\label{Exa5}


As shown above, random matrices with independent entries
are nowadays quite well understood and there exist many analytic results
on their asymptotic eigenvalue distribution.
Some special cases of the correlated situation have also been treated in the literature.
One such case with relevance in communications engineering is separable correlation, when the channel matrix of a MIMO system can be expressed as
\begin{eqnarray}
\label{eq:separable} \Hm=\ThetaR^{1/2} \Wm \ThetaT^{1/2},
\end{eqnarray}
where the ($N$$\times$$K$) matrix $\Wm$ has i.i.d. zero-mean
unit-variance complex Gaussian random entries, and the entries of
$\ThetaT$ and $\ThetaR$ indicate, respectively, the correlation
between transmit and between receive antennas. This is equivalent to say that the correlation among the entries of $\Hm$ factorizes so that
$\expect[h_{p,j}h_{q,k}^*]=(\ThetaT)_{j,k}(\ThetaR)_{p,q}$.

In this section, we describe a realistic MIMO setting with ISI (inter-symbol-interference), where the separability assumption (\ref{eq:separable}) does not hold. In this case, the channel impulse response between the transmitter antenna $j$ and the receiver antenna $i$ is a (length-$L$)  vector, say $\mathbf{h}^{i,j}=[h_1^{i,j},h_2^{i,j},\ldots,h^{i,j}_{L-1},h^{i,j}_L]$
 where $L$ is the number of the channel taps. The channel matrix, for a length-$M$ frame, may then be written as
 \begin{eqnarray}
\label{eq:unseparable} \Hm=\left[\begin{array}{ccccccccc} \matr A_1 & \matr A_2 & \ldots & \matr A_L &  \mathbf{0} & \mathbf{0} & \ldots & & \mathbf{0}\\
 \mathbf{0}&  \matr A_1 & \matr A_2 & \ldots & \matr A_L & \mathbf{0} & \ldots & &  \mathbf{0}\\
 \mathbf{0} &\mathbf{0} & \matr A_1 & \matr A_2 & \ldots & \matr A_L &  \mathbf{0}& \ldots & \mathbf{0}\\
 \vdots & \vdots & \ddots & \ddots & \ldots & & \ddots & \ldots \\
\mathbf{0}& \ldots & \ldots & \mathbf{0} &\mathbf{0} & \matr A_1 & \matr A_2 & \ldots & \matr A_L
 \end{array}\right]
\end{eqnarray} where there are $M-1$ null matrices in each row and $\matr A_\ell=(h^{i,j}_\ell)^{i=1,\ldots,N}_{j=1,\ldots,K}$. Notice that, due to the ISI modeling, the role of single-channel usage is herein played by $\matr A_\ell$, which plays then the role of $\Hm$ in (\ref{AWGN}).

This kind of channel matrix does not fall into the class of separable correlation, thus one needs to develop new tools to deal with the problem of evaluating the eigenvalue distribution of $\Hm\Hm^\dagger$. The needed characterization is given by the following Theorem \cite{far:08}.

\begin{theorem} Let a real-valued covariance
function  $\tau(i,k;j,l)$  be such that $\tau(i,k;j,l)=\tau(j,l;i,k)$, for $i,j,=1,\ldots,a$ and $k,l=1,\ldots,b$, where $a$ and $b$ are fixed natural numbers.
 Assume, for $n \in \NN$, that $\{h_{rp}^{(i,j)}|i=1,\ldots,a; k=1,\ldots,b; r,p=1,\ldots,n\}$
are jointly Gaussian
complex random variables, with the prescription of mean zero
and covariance
\begin{equation}\label{cov:unsep}
\expect[h_{rp}^{(i,k)}\overline{h}_{sq}^{(j,l)}]=\frac{1}{(b+a)n}\delta_{rs}\delta_{pq}\tau(i,k;j,l).
\end{equation}
Assume further circular complex Gaussian law, i.e., $\expect[(h_{rp}^{(i,k)})^2]=0$, and consider block matrices $\Hm_n=(\Hm^{(i,k)})_{i=1,\ldots,a}^{j=1,\ldots,b}$, whose blocks are given by $\Hm^{(i,k)}=(h_{rp}^{(i,k)})^n_{r,p=1}$ for $i=1,\ldots,a$ and $k=1,\ldots,b$.

It follows that, as $n\rightarrow\infty$, the square $an$-dimensional matrix $\Hm_n\Hm_n^\dagger$ has almost surely a limiting eigenvalue distribution whose Stieltjes transform is given by $\StT{\!}z={\rm Tr}_a(\mathcal{G}_1(z))$. Here,  $\mathcal{G}_1(z)$ is an analytic function (in the upper half of the complex plane) with values in the set of square $a$-dimensional matrices with complex entries. Such function is   uniquely determined by both $\lim_{|z|\rightarrow\infty,\Im(z)>0}z\mathcal{G}_1(z)=\I_a$ together with
\begin{equation}\label{sixspeicher}
z\mathcal{G}_1(z)=\I_a+\eta_1\left(\left(\I_b-\eta_2(\mathcal{G}_1(z))\right)^{-1}\right)\,. 
\end{equation}
In (\ref{sixspeicher}),  $\eta_1$ and $\eta_2$ are complex, matrix-valued, covariance mappings defined by
\begin{equation}
[\eta_1(D)]_{i,j}=\frac{1}{b+a}\sum_{k,l=1}^b\tau(i,k;j,l)[D]_{k,l}\,,\nonumber
\end{equation}
\begin{equation}
[\eta_2(D)]_{k,l}=\frac{1}{b+a}\sum_{i,j=1}^a\tau(i,k;j,l)[D]_{j,i}\,,
\end{equation}
and, defining $M_d(\CC)$  as the set of complex $d$-dimensional square matrices, $\eta_1:M_b(\CC)\rightarrow M_a(\CC)$ and  $\eta_2:M_a(\CC)\rightarrow M_b(\CC)$. 
\end{theorem}

Notice that the  sought-for Stieltjes transform $\StT{\!}z$ is given as a function (trace) of a matrix-valued quantity, i.e. $\mathcal{G}_1(z)$. The corresponding eigenvalue distribution for the channel matrix of interest is then recovered through  (\ref{Stinv}). The proof of this statement involves very advanced arguments from operator-valued free probability theory and is out of the scope of this tutorial paper\footnote{*We refer the interested reader to \cite{far:08} for the complete proof with some toy examples.}*, where we rather aim at remarking the possibility of characterizing the asymptotic eigenvalue distribution needed in (\ref{chcap}) through the inversion formula even in the non-separable case.
\subsection{3rd Example: Cellular Systems}
\label{Exa6}


Random band matrices, already exploited in the previous Section for  the analysis of correlated MIMO channels with ISI, are well suited to model a setting where multiple users and base stations (BS) coexist in a geographic area and the BS are arranged forming a regular cellular pattern. In this context, the band structure reproduced in (\ref{eq:unseparable}) 
adequately models a one-dimensional@ \footnote{@ We refer to this cellular scenario as one-dimensional in that all cells are assumed to be located on a line and no grid, hexagonal coverage or more complicated structures are analyzed hereinafter.} cellular communication setting governed by the linear relationship (\ref{AWGN})
\begin{equation}
\label{MUcell}
\matr y=\matr H \matr x + \matr n,
\end{equation}
where in this context
\begin{itemize}
\item $\matr x$ represents a $K(M+1)$-dimensional vector of symbols transmitted by the $K$ active users present at each of the $M+1$ cells constituting the system (irrespective of scheduling policy),
\item $\matr y$ represents a $MN$-dimensional vector of symbols received by the $M$ BS in the system, each equipped with the same number $N$ of receive antennas,
\item $\matr H$ represents the $MN\times K(M+1)$ channel matrix (\ref{eq:unseparable}),
\item $\matr n$ represents a $MN$-dimensional vector of additive noise,
 \item the number of meaningful interfering cells for a reference communication is $L-1$.
 \end{itemize}
 The actual number of interfering cells maps directly onto the number of secondary non-zero diagonals in the overall system matrix.

Tools from operator-valued free-probability theory can be exploited to investigate the spectral behavior of non-separable correlation matrices and provide an implicit characterization of the spectrum of the channel matrix Gramian  in terms of its Stieltjes transform. We would like to obtain a somewhat more explicit solution, in order to ease the channel capacity evaluation for the cellular setting. However,  no general solution is available in closed-form for arbitrary values of $N$ (number of receive antenna at each BS)  and $L-1$ (number of interfering cells). Meanwhile, high-$SNR$ affine expansions of the  capacity can be provided in the simplest case of a single interfering cell  and single-antenna equipped BS ($N=1$ and $L=2$) \cite{ZLSTIT}. In this case, let us  assume that the block matrices $\matr A_i$ (which reduce to row vectors of length $K$, say $\matr a_i$) have i.i.d. entries, and that the marginal law of the elements of $\matr A_1$ is $\pi_a$, while we denote by $\pi_b$ the law of the entries  of $\matr A_2$ (by assumption, only matrix blocks over the main diagonal and the second main diagonal are non-zero, that is $\matr A_l=0$ for $l$ \textgreater 2)*. {\footnote{* We remark that this is strictly related with the celebrated model of Wyner \cite{Wyner:94}.}}
 The main performance index we are interested in is the per-cell sum-rate, which can be expressed in slightly different form from (\ref{chcap}) due to the normalizations over the number of users/cells, as follows
 \begin{equation}
 C_M(P)=\frac{1}{M}\expect_{\matr H}\log\det\left(\mbox{\bf I} + \rho \matr H\matr H^\dagger\right)\,,
 \end{equation}
 where $P=K\rho$ is the average transmit power per-cell and  $\rho$ the transmit power per single user.
 Borrowing results from \cite{ZLSTIT}, we can state the following:
 \begin{theorem}
 Assume \begin{itemize}
 \item[i] $\expect_{\pi_a}\log( |x|)^2< \infty\,,\qquad \expect_{\pi_b}\log( |x|)$;
 \item[ii] $\pi_a$ and $\pi_b$ are absolutely continuous with respect to Lebesgue measure on $\CC$;
 \item[iii]There exists a real $\mathcal{M}$ such that if $x\sim\pi_a$ (resp. $x\sim\pi_b$), then the density of $|x|^2$ is strictly positive on the interval $\left[\mathcal{M},+\infty\right)$\footnote{\textit{**}This assumption should in some cases be strengthened, see \cite[Sec. II.A]{ZLSTIT}.};**
 \item[iv]There exists a ball in $\CC$ such that the Lebesgue measure outside that ball is absolutely continuous with respect to $\pi_a$ and $\pi_b$.
 \end{itemize}
 Then, as $M$ grows large, $C_M(P)$ converges, $\forall\,P>0$, to a nonrandom limit, say $C(P)$, whose behavior is bounded as follows
 \begin{eqnarray}\label{bounds:mcellsr}
 \!\!\max(\expect_{\pi_a}\log(1 + P |x|^2), \expect_{\pi_b}\log(1 + P|y|^2)) \leq C(P) \leq \expect_{\pi_a,\pi_b}
\log(1 + P(|x|^2+ |y|^2)).\nonumber
 \end{eqnarray}
 Moreover, as $P$ goes to infinity,
 \[
 C(P)=\log P+2\max(\expect_{\pi_a}\log( |x|),\expect_{\pi_b}\log(|x|))+o(1).
 \]
 \end{theorem}
The proof hinges on arguments from Markov chains theory, and its length keeps us from including it in this work. Moreover, it relies on the fact that, for $L=2$, $\matr H\matr H^{\dagger}$ turns out to be a tri-diagonal (Jacobi) matrix (\cite[p.28]{horn:85}), and a particular expression for its determinant (\cite[and references therein]{ZLSTIT}) is available. This, unfortunately, cannot be extended to a matrix with more than three non-zero diagonals.
The result given above holds under ideal intra-cell Time Division Multiple Access scheduling
enforcing the additional constraint of $K=1$ (only one user at time is transmitting). Simple steps allow the analysis to be extended to the general case of simultaneous transmissions with arbitrary $K>1$, and the sum-rate behavior is analyzed in \cite[Thm. 2]{ZLSTIT}.

 Slightly more complicated is the case of $L>2$, for which (still assuming the number of BS receive antenna to be limited to $N=1$) one can rely on the theory of product of random matrices \cite[and references therein]{prodfursten} to prove the following behavior for the sum-rate capacity in the high-$SNR$ regime:
 \begin{theorem}
Assume \begin{itemize}
 \item[i] All rows of $\matr H$,  say $\matr a_i$'s, $i=1,\ldots,L$ and all their  sub-vectors of fixed length form a stationary and ergodic sequence;
 \item[ii] There exists  $\varepsilon>0$ such that for $0\leq i \leq L-1$, $\expect_{\pi_i}|\log|x||^{1+\varepsilon}< \infty$;
 \item[iv] If $(x_1,\ldots,x_L)$ is distributed according  to $\pi$, then almost surely $x_1x_L^\dagger\neq 0$.
 \end{itemize}
 Then, as $M$ grows large, $C_M(\rho)$ converges, $\forall\,\rho>0$, to a nonrandom limit, say $C(\rho)$, whose behavior is
 \[
  C(\rho)=\log\rho+\expect_\pi\log|\matr a_1\matr a_L|+\frac{1}{L}\gamma(\rho)
 \]
 Moreover, as $\rho$ goes to infinity,
 \[
 C(\rho)=\log\rho+\expect_\pi\log|\matr a_1\matr a_L|+\frac{1}{L}\gamma(\rho)+o(1)\,,
 \]
 where $\gamma(\rho)$ is a function of the entries of the channel matrix and both its definition as well as its convergence to a bounded value as $\rho\rightarrow\infty$ are proven in \cite[Sec. III and Appendices]{Levy:10}.
 \end{theorem}

We have to stress that, unlike the case of tri-diagonal matrices, here the entries in the rows of $\matr H$ are allowed to be arbitrarily correlated, once they fulfill the ergodicity assumption. This reverts into a condition of arbitrary correlation of the fading process, while the previous theorem only dealt with independent fading.

An even easier way to cope with the difficulties arising from the multiple-cell setting is that of  asymptotically approximating the eigenvalues of the matrix $\matr H \matr H^{\dagger}$,  by the root of a properly built matrix polynomial as per \cite[Theorem V]{blockeigen}, and numerically evaluate the ergodic capacity via (\ref{chcap}). While conceptually this method would be easier than resorting to the operator-valued free probability counterpart, a computational efficiency comparison is yet to be performed. It is worth noting that this approach would, however, only work for complex Gaussian fading, for which explicit results are somewhat expected. 

The results above depend on the actual distribution of the matrix entries, joint or marginal, depending on the desired degree of generality.
This is a feature of band matrices with relatively few non-zero diagonals. Above a certain number of non-zero diagonals, the asymptotic spectral behavior of the matrix resembles that of a Wigner matrix, and indeed converges towards it as recently proven \cite{Phtrans12}.


\section{Complex-Valued Distributions}
\label{QuaFrePro}

The treatment of eigenvalue distributions by means of moments and cumulants in previous sections is restricted to random matrices with real eigenvalues. 
It cannot be applied to the full circle law and other random matrices with complex-valued eigenvalues straightforwardly.
A method to deal with circularly symmetric complex eigenvalue distributions was proposed in \cite{haagerup:00}.
In the sequel, we present the ideas proposed in \cite{jarosz:06,jarosz:11}, which allow for the treatment of general complex eigenvalue distributions.

The method of choice when dealing with real-valued eigenvalue distributions in previous sections is the use of complex analysis, i.e.\ presenting real-valued eigenvalue distributions as a limit of a complex-valued holomorphic function, cf.\ \eq{Stinv}. 
Complex-valued eigenvalue distributions are often circularly symmetric and thus not holomorphic.
They can be represented by a pair of holomorphic functions representing the real and imaginary parts.
Or, instead of the real and imaginary part of a complex variable $z$, one can also consider $z$ and its complex conjugate $z^\ast$, and then apply the Wirtinger rule \cite{remmert:91} for differentiation, i.e.\
\begin{equation}
\label{wirtingerrule}
\frac{\partial z^\ast}{\partial z} = \frac{\partial z}{\partial z^\ast} = 0.
\end{equation}

\subsection{Stieltjes Transform}
 \label{quaternions}
 
In order to generalize the Stieltjes transform introduced in Section \ref{StiTra} to two complex variables $z$ and $z^\ast$, 
we first rewrite \eq{defSt} as
\begin{equation}
\StT{}s = -\frac{\rm d}{{\rm d}s} \int \log(x-s){\rm d}\Prob{}x.
\end{equation}
Note that the Dirac function of a complex argument can be represented as the limit
\begin{equation}
\delta (z) = \frac 1\pi \lim\limits_{\epsilon \to 0} \frac {\epsilon^2} {(|z|^2+\epsilon^2)^2}.
\end{equation}
Thus, we have 
\begin{equation}
\prob{}z = \frac1\pi \lim\limits_{\epsilon\to0} \frac{\partial^2}{\partial z\partial z^\ast} \int \log\left[|z^\prime-z|^2+\epsilon^2\right]{\rm d}\Prob{}{z^\prime}.
\end{equation}
We define the bivariate Stieltjes transform by
\begin{equation}
\label{defbiSt}
\StT{}{s,\epsilon} = \frac{\partial}{\partial s}  \int \log\left[|z-s|^2+\epsilon^2\right]{\rm d}\Prob{}z
\end{equation}
and get the bivariate Stieltjes inversion formula to read
\begin{equation}
\label{diffpz}
\prob{}z = \frac1\pi \lim\limits_{\epsilon\to0} \frac\partial{\partial z^\ast}\StT{}{z,\epsilon}.
\end{equation}

At first sight, the bivariate Stieltjes transform looks quite different from \eq{defSt}.
However, we can rewrite \eq{defbiSt} as
\begin{equation}
\label{defmSt}
\StT{}{s,\epsilon} = \int \left[ \left(
\begin{array}{cc}
z-s & {\rm j}\epsilon\\
{\rm j}\epsilon & z^\ast - s^\ast
\end{array}
\right)^{-1}\right]_{1,1} {\rm d}\Prob{}z ,
\end{equation}
which clearly resembles the form of \eq{defSt}. 
To get an even more striking analogy with \eq{defSt}, we can introduce the Stieltjes transform with {\em quaterionic} argument $q=v+{\rm i}w, (v,w)\in\CC^2, {\rm i}^2=-1, {\rm ij}=-{\rm ji}$
\begin{equation}
\label{defqSt}
\StT{}q = \int \frac {{\rm d}\Prob{}z}{z-q}  
\end{equation}
with the respective inversion formula
\begin{equation}
\prob{}z = \frac 1\pi \lim\limits_{\epsilon\to 0}\Im \StT{}{z+{\rm i}\epsilon} 
\end{equation}
and the definition $\Im(v+{\rm i}w) = w$.
Quaternions are inconvenient to deal with, since multiplication of quaternions does not generally commute.
However, any quaternion $q=v+{\rm i} w$ can be conveniently represented by the complex-valued $2\times 2$ matrix
\begin{equation}
\left(
\begin{array}{cc}
v & w\\
-w^\ast & v^\ast
\end{array}
\right).
\end{equation}
This matrix representation directly connects \eq{defmSt} with \eq{defqSt}.

\subsection{R-Diagonal Random Matrices}
R-diagonal random matrices are an important subset of random matrices with complex eigenvalues.
\begin{definition}
A random matrix $\matr X$ is called R-diagonal, if it can be decomposed as $\matr X=\matr{UY}$ where $\matr Y=\sqrt{\matr {XX}^{\rm H}}$ and $\matr U$ is Haar distributed and free of $\matr Y$.
\end{definition}
All asymptotically large bi-unitarily invariant random matrices are R-diagonal. Furthermore, their eigenvalue distribution is circularly symmetric in the complex plane.  For R-diagonal random matrix, the quaternion-valued calculus introduced in Section~\ref{quaternions} can be avoided, as outlined below.

Let the random matrix $\matr H$ be R-diagonal and denote  its eigenvalue distribution by $\prob{\matr H}z$.
Let $\ST{\matr {HH}^{\dagger}}s$ denote the S-transform of the asymptotic eigenvalue distribution of $\matr {HH}^{\dagger}$ and define the function 
\begin{equation}
f(s) = \frac1{\sqrt{\ST{\matr {HH}^{\dagger}}{s-1}}}.
\end{equation} 
Then, we have
\begin{equation}
 \prob{\matr H}z = \frac1{2\pi s f^\prime\left[f^{-1}(z)\right]}
\end{equation}
with $f^\prime(s) = {\rm d}f(s)/{\rm d}s$ wherever the density is positive and continuous \cite{haagerup:00}.

Let the random matrices $\matr A$ and $\matr B$ be asymptotically free and R-diagonal and denote their respective asymptotic singular value distributions by $\Prob{\matr A}x$ and $\Prob{\matr B}x$.
Define the symmetrization of a density by
\begin{equation}
\tilde{\rm p}(x) = \frac{\prob{}x+\prob{}{-x}}2.
\end{equation}
Then, we have for the respective R-transforms of the symmetrized singular value distributions
\begin{equation}
\tilde{\rm R}_{\matr A+\matr B}(w) = \tilde{\rm R}_{\matr A}(w) +\tilde {\rm R}_{\matr B}(w).
\end{equation}

Let the matrices $(\{\matr H_1\},\dots,\{\matr H_L\})$ form a free family and follow the full circle law.
Define
\begin{equation}
\matr H = \prod\limits_{\ell=1}^L\matr H_\ell.
\end{equation}
Then, we have
\begin{equation}
\prob{\matr H}z = \cases{\frac1{L\pi |z|^{L-1}}& $|z|\le 1$\cr 0 & \mbox{elsewhere}
}.
\end{equation}
With the help of \eq{diffpz} and the Wirtinger rule \eq{wirtingerrule}, we find
\begin{equation}
\StT{}{s,s^\ast} = \int\limits_{\frac{-1}{s^\ast}}^{\frac1{s^\ast}} \frac{{\rm d}s^\ast}{L|s|^{L-1}} = 2 
\end{equation}
Using this result, concatenated vector channels as in Section \ref{Exa4} can be analyzed much more easily. In fact, the analysis becomes straightforward for square matrices. The above calculus was used to characterize multiuser MIMO channels with line-of-sight in \cite{mueller:12}.

\subsection{Free Convolution}

We define the R-transform of quaternion argument $p$ in complete analogy to the complex case \eq{defRT} as
\begin{equation}
\RT{}p \define \StTi{}{-p}-\frac 1p
\label{defqRT}
\end{equation}
and obtain for free random matrices $\matr A$ and $\matr B$, with $\RT{\matr A}p$ and $\RT{\matr B}p$ denoting the R-transforms of the respective asymptotic eigenvalue distributions,
\begin{equation}
\RT{\matr A+\matr B}p = \RT{\matr A}p +\RT{\matr B}p.
\end{equation}

The scaling law of the R-transform generalizes as follows
\begin{equation}
\RT{z \matr H}p = z \RT{\matr H}{pz}
\end{equation}
for $z\in\CC$. Note that the order of factors does matter here, since $pz\ne zp$, in general.

Multiplicative free convolution does not generalize as straightforwardly as additive free convolution does. The interested reader is referred to \cite{burda:11}.

\section{The Replica Method}
\label{RepMet}

In physics it is accepted that there are microscopic variables (possibly random) and macroscopic variables (whose evolution is deterministic).
Microscopic variables represent physical properties of small systems, e.g.\ the speed of a gas molecule or the (z-projection) spin of an electron.
Macroscopic variables represent the collective properties of ensembles made up of many such small systems, e.g.\ the temperature or pressure of a gas, the radiation of a hot object, or the magnetic field of a piece of ferromagnetic material.
From a physics point of view, it is clear which variables are macroscopic and which ones are microscopic depending on the properties they represent.

The study of the evolution and relationships between properties described by macroscopic variables is known as thermodynamics. Thermodynamics does not take into consideration the microscopic structure of the system, it is an experimental science that concerns itself only with macroscopic properties.
Statistical physics, however, is a theoretical construct that allows (under a series of assumptions) for the determination of thermodynamic variables when the value (or the statistics) of microscopic variables can be determined.

In the following subsections we shall introduce how the machinery of statistical physics, in particular the so-called {\em replica method}, may be used as an alternative to and as an extension of RMT and FPT.

\subsection{The Thermodynamic Limit}
\label{SelAve}

As described above, statistical mechanics is the bridge between the microscopic and the thermodynamic domains; it explains why thermodynamics works. It turns out thermodynamics can afford to ignore microscopic variables because it only considers systems that are so large as not to be affected by any microscopic event. These systems which allow themselves to be accurately described by thermodynamics are known as thermodynamic systems, and the size that a system must approach to be thermodynamic is known as the \textit{thermodynamic limit}.

Random matrix theory and free probability theory prove the (almost sure) convergence of some random variables to deterministic values in the large matrix limit \cite{hiai:00,thorbjoernsen:99}.
However, statistical physics does not always do so; it is just assumed that all microscopic variables converge to their own average in the thermodynamic limit. An explicit proof that a particular variable is {\em self-averaging} is a nice result, but not of great important to the physics community.

The replica method is a tool of statistical physics to compute macroscopic properties by averaging over microscopic variables. Therefore, when applying the replica method, random variables are often assumed (without proof) to be self-averaging.


\subsection{Free Energy}
\label{FreEne}

The second law of thermodynamics demands the entropy of any physical system with conserved energy to converge to its maximum as time evolves.
If the system is described by a probability distribution $\prob{\matr x}{\matr x}$ of states $\matr x\in\RR^{K\times1}$, this means that in the thermodynamic equilibrium the (normalized) entropy
\begin{equation}
{\rm H}(\matr x)=-\frac1K\sum\limits_{\matr x} \Pr(\matr x)\log\Pr(\matr x)
\end{equation}
is maximized while keeping the (normalized) energy
\begin{equation}
\label{energy}
E(\matr x) = \frac1K\sum\limits_{\matr x} \Pr(\matr x) ||\matr x||
\end{equation}
constant.
Hereby, the energy function $||\matr x||$ can be any measure which is uniformly bounded from below.

The distribution at thermodynamic equilibrium is easily shown by the method of Lagrange multipliers to be
\begin{equation}
\label{Lagmul}
\Pr(\matr x) = \frac{{\rm e}^{-\frac1T||\matr x||}}{\sum\limits_{\matr x}{\rm e}^{-\frac1T||\matr x||}}
\end{equation}
and called the Boltzmann distribution.
The parameter $T$ is called the temperature of the system and determined by \eq{energy}.
For a Euclidean energy measure, the Boltzmann distribution takes on the form of a Gaussian distribution, which is well-known in information theory to maximize entropy for given average energy.

A helpful quantity in statistical mechanics is the (normalized) {\em free energy}\footnote{*The {\em free} energy is not related to {\em free}ness in free probability theory.}* defined as
\begin{eqnarray}
\label{defFE}
{\rm F}(\matr x) &\define& E(\matr x) - T {\rm H}(\matr x)\\
&=& -\frac TK  \log \left(\sum\limits_{\matr x} {\rm e}^{-\frac1T||\matr x||}\right).
\label{defFE2}
\end{eqnarray}
In thermodynamic equilibrium the entropy is maximized, and the free energy minimized since the energy is constant.
The free energy normalized to the dimension of the system is a self averaging quantity.

\subsection{The Replica Trick}
\label{RepCon}

The explicit evaluation of the free energy turns out to be very complicated in many cases of interest.
One major obstacle is the occurrence of the expectation of the logarithm of a function of a random variable
\begin{equation}
\expect\limits_z \log f(z,T).
\end{equation}
In order to circumvent this expectation, which also appears frequently in information theory, we can exploit the Riesz equality \cite{riesz:30}
\begin{equation}
\label{LHP}
\expect\limits_z\log f(z,T)= \lim\limits_{n\to0}\frac{\expect\limits_z\left[f(z,T)\right]^n-1}{n},
\end{equation}
which for convenience we write as \footnote{**The mild assumption is made that limit in $n$ and expectation in $z$ commute.}**
\begin{equation}
\label{eq512}
\expect\limits_z\log f(z,T) = \lim\limits_{n\to0} \frac\partial{\partial n}\, \log \expect\limits_z \left[f(z,T)\right]^n.
\end{equation}

\label{RepCon_2}

Equation \eq{eq512} reduces the problem to the calculation of the $n^{\rm th}$ moment in the neighborhood of $n=0$.
However, in order to perform the limit operation, the expectation must be calculated for real-valued variables $n$,
which is far from trivial. At this point the assumption is made that, although $n$ is treated as a positive integer, we may nevertheless continually extend the result of the expectation to $n \rightarrow 0$.
This may sound odd at first, but it can indeed be proven to hold rigorously provided that the function $\log f(z,T)$ allows for a series expansion in $T$ \cite{dotsenko:01}.
This surprising result is connected to the well-known fact that knowing an analytic function in any arbitrarily tiny neighborhood about some point is equivalent to knowing it everywhere.

The invocation of analytic continuity from integer to real $n$ will be referred to as {\em replica continuity}.
The combination of expression \eq{eq512} and replica continuity is known as the \textit {replica trick}, and it is the cornerstone of the so called {\em replica method}. The name \textit{replica} stems from the expectation in the right hand side of equation \eq{eq512}, which may be seen as that of a product of $n$ identical replicas $(1,2,\dots,n)$ of $f(z,T)$:
\begin{equation}
\label{repcont}
\expect\limits_z \left[f(z,T)\right]^n=\expect\limits_z \prod_{a=1}^n {f_{a}(z,T)}.
\end{equation}

To help see how replica continuity can prove helpful, let
\begin{equation}
y=\int\int \e ^{f(x,z)} \,{\rm d}x{\rm d} z
\end{equation}
for some function $f(x,z)$.
Since the variable of integration is arbitrary, analytic continuity on $n$ implies
\begin{eqnarray}
\label{eq177}
y^n &=& \left(\int\int \e^{f(x,z)}{\rm d}x{\rm d}z\right)^n\\
& =& \int\prod\limits_{a=1}^n \int \e^{f(x_a,z)}{\rm d}x_a{\rm d}z\\
&=& \int \int \cdots \int \e^{\, \sum\limits_{a=1}^nf(x_a,z)}\prod\limits_{a=1}^n {\rm d}x_a{\rm d}z.
\label{eq179}
\label{rm}
\end{eqnarray}

Assume now that in \eq{eq177} the integration over $x$ can be performed in closed form, while the integration over $z$ cannot.
Without replica continuity, one would be stuck with the $n^{\rm th}$ power of an integral over $z$.
However, with the help of replica continuity, we can perform the integrations over all replicated variables of $x$ and finally arrive to \eq{eq179} only with an integral over $z$ left, but without an $n^{\rm th}$ power on top of it.

The replica method consists of generating replicas of $x$ instead of getting stuck with calculating the $n^{\rm th}$ power of $y$.
These replicated variables $x_a$ are arbitrary and, as we shall discuss in the next section, they may be assigned helpful properties such as symmetries.

This ingenious trick was first proposed by Kac in 1968 while analyzing lattice vibrations at the Trondheim Theoretical Physics Seminar \cite{kac:68}. However, its proving ground was the spin glass (a disordered magnetic system in which the local couplings vary randomly in sign and magnitude) where Edwards and Anderson boiled the problem down to its physical essentials in 1975 \cite{edwards:75}.

\subsection{Replica Symmetry}
\label{RepSym}

Integrals arising from the replica method are typically solved by saddle point integration.

The general idea of saddle point integration is as follows:
Consider an integral of the form
\begin{equation}
\frac1K\log \int\limits_{{\cal I}^{n\times n}} {\rm e}^{Kf(\matr Q)} {\rm d}\matr Q.
\end{equation}
In the limit $K\to\infty$ the integral is dominated by that matrix $\matr Q$ which maximizes the function $f(\matr Q)$.
Thus, we have
\begin{equation}
\label{eq180}
\lim\limits_{K\to\infty}\frac1K\log \int\limits_{{\cal I}^{n\times n}} {\rm e}^{Kf(\matr Q)} {\rm d}\matr Q = \max\limits_{\matr Q\in{{\cal I}^{n\times n}}} f(\matr Q).
\end{equation}
That means, the integral can be solved taking the derivative of the argument of the exponential function.

Since the function in the exponent is multivariate, one would need to find the extremum of an $n^2$-variate function for an arbitrary number of arguments.
This can easily become a hopeless task, unless one can exploit some properties of the exponential argument.

When it was first developed, the replica method considered all copies of the system $(1,2,\dots,n)$ as identical, in what is known as the {\em replica symmetry (RS)} ansatz.
Assuming RS means that one imposes a certain structure
onto the matrix $\matr Q$, i.e.\

\begin{equation}
\label{eq181}
\matr Q = q_0 \1_{n\times n} + \chi_0 \I_{n\times n}T
\end{equation}
where $\1_{n\times n}$ denotes the $n\times n$ all-one matrix and $T$ denotes temperature.
Then, the multivariate optimization problem reduces to a two-variate one, i.e.\ $\max_{q_0,\chi_0} f(q_0 \1 + \chi_0 \I\, T)$.
Replica symmetry  imposes that all diagonal entries and all off-diagonal entries respectively equal each other. It is a critical assumption.

\subsection{Replica Symmetry Breaking}
\label{RepSymBre}

The assumption of replica symmetry was, from its dawn, successful in finding known correct results for some simple spin glass models, see e.g.\ \cite{kosterlitz:76}.
However, the infinite ranged model by Sherrington and Kirkpatrick \cite{sherrington:75,kirkpatrick:78} proved extremely difficult to treat and unphysical solutions were being produced.

In 1978, Almeida and Thouless suggested the possibility of assuming that, somehow, perhaps different copies of the system would not be regarded as identical \cite{almeida:78}.
In 1980, Parisi formally proposed the {\em replica symmetry breaking (RSB)} approach \cite{parisi:80} which took many years to interpret,  see e.g. \cite{mezard:87,nishimori:01,dotsenko:01} and has only recently been proven to be the correct solution \cite{talagrand:06}.

Replica symmetry breaking comes in steps. The philosophy behind it is somehow similar to that of Taylor or Laurent series expansions of analytic functions with replica symmetry taking the role of the expansion of first order.

For 1-step replica symmetry breaking (1RSB) assume that the matrix $\matr Q$ has the, compared to \eq{eq181},  slightly more general structure
\begin{equation}
\label{eq182}
\matr Q = q_1 \1_{n\times n} +  p_1 \I_{\frac n{\mu_1 T}\times \frac n{\mu_1 T}} \otimes \1_{\mu_1 T\times \mu_1 T}+ \chi_1 \I_{n\times n}T
\end{equation}
with $\otimes$ denoting the Kronecker product. Now, there are four parameters $\{q_1,p_1,\mu_1,\chi_1\}$ to optimize.
The 1RSB ansatz leads to correct results in some cases where the RS ansatz fails.
Nevertheless, the 1RSB ansatz may fail as well in certain cases. Such failures typically occur for NP-complete problems at low temperature.
To improve accuracy of the solution, one can move onward to $r$-step RSB defined by the ansatz
\begin{equation}
\label{eq183}
\matr Q = q_r \1_{n\times n} +  \sum\limits_{i=1}^r p_r^{(i)} \I_{\frac n{\mu_r^{(i)} T}\times \frac n{\mu_r^{(i)} T}} \otimes \1_{\mu_r^{(i)} T\times \mu_r^{(i)} T}+ \chi_r \I_{n\times n}T.
\end{equation}
The number of parameters to optimize increases linearly with $r$.
The saddle point equations determining these parameters are typically not linear and one often reaches the limits of analytical tractability for low values of $r$.
The limit $r\to\infty$ is called {\em full} replica symmetry breaking.
It is believed---and actually proven for certain cases---that full RSB always gives correct results.


\subsection{An Engineering Perspective of the Hamiltonian}
\label{MisDet}

The free energy is clearly related in statistical mechanics to the entropy of the system at a given energy due to \eq{defFE}.
This establishes the usefulness of the free energy for information theoretic tasks like calculations of channel capacities.
Moreover, the free energy is a tool to analyze various types of multiuser detectors.
In fact, the free energy is such a powerful concept that it needs not any coding to be involved in the communication system to yield striking results.
All it requires is the existence of macroscopic variables, microscopic random variables and the existence of an energy function.
For communication systems, this requires, in practice, nothing more than their size growing above all bounds.

The broad applicability of the statistical mechanics approach to communication systems stems form the validity of \eq{defFE} for any definition of the energy function.
The performance measures of a detector need not only be based on the eigenvalues of the channel matrix in the large system limit.
Instead, the energy function can be interpreted as the metric of a detector.
Thus, any detector parameterized by a certain metric can be analyzed with the tools of statistical mechanics in the large system limit.
However, there is a practical limit to the applicability of the statistical mechanics framework to the analysis of large communication systems: The analytical calculations required to solve the equations arising from \eq{defFE} are not always feasible.
The replica method was introduced to circumvent such difficulties in certain cases.
Many other cases, however, have remained intractable until present time.

Consider a communication channel uniquely characterized by a conditional probability density $\prob{y|x}{y,x}$ and a source uniquely characterized by a prior density $\prob{x}x$.
Consider a detector for the output of this channel characterized by an assumed channel transition probability $\proba{y|x}{y,x}$ and an assumed prior distribution $\proba{x}{x}$.
Let the detector minimize some kind of cost function, e.g.\ bit error probability, subject to its hypotheses on the channel transition probability $\proba{y|x}{y,x}$ and the prior distribution $\proba{x}{x}$.
If the assumed distributions equal the true distributions, the detector is optimum with respect to its cost function.
If the assumed distributions differ from the true ones, the detector is mismatched in some sense.
Many popular detectors can be described with this framework, see Sections~\ref{UncErrPro} and \ref{LinDet} for a few examples.

The minimization of a cost function subject to some hypothesis on the channel transition probability and some hypothesis on the prior distribution defines a metric which is to be optimized.
This metric corresponds to the energy function in thermodynamics and determines the distribution of the microscopic variables at thermodynamic equilibrium.
In analogy to \eq{Lagmul}, we find
\begin{equation}
\proba{\matr x|\matr y,\matr H}{\matr x,\matr y,\matr H} = \frac{{\rm e}^{\,-\frac1T||\matr x||}}{\int{\rm e}^{\,-\frac1T||\matr x||}\,{\rm d}{\matr x}}
\label{lagmul}
\end{equation}
where the dependency on $\matr y$, $\matr H$, and the assumed distributions is implicit via the definition of the energy function $||\cdot||$.
The energy function reflects the properties of the detector.
Using Bayes' law, the appropriate energy function corresponding to particular hypotheses on the channel transition function and the prior distribution can be calculated via \eq{lagmul}.

In order to study macroscopic properties of the system, we must calculate the free energy of the system.
For that purpose, we make use of the self-averaging property of thermodynamic equilibrium and \eq{defFE2}:
\begin{eqnarray}
{\rm F}\left(\matr x\right)&=& {\rm F}\left(\matr x|\matr y,\matr H\right)\\
& =& \expect\limits_{\matr y,\matr H} {\rm F}\left(\matr x|\matr y,\matr H\right)\\
&=& \frac TK\expect\limits_{\matr H}\int\log\frac1{\int{\rm e}^{\,-\frac1T||\matr x||}\,{\rm d}{\matr x} }\, {\rm d}\Prob{\matr y}{\matr y}.
\label{FrEn}
\end{eqnarray}
Note that, inside the logarithm, expectations are taken with respect to the assumed distribution via the definition of the energy function, while, outside the logarithm, expectations are taken with respect to the true distribution.

In the case of matched detection, i.e.\ the assumed distributions equal the true distributions, the argument of the logarithm in \eq{FrEn} becomes $\prob{\matr y}{\matr y}$ up to a normalizing factor.
Thus, the free energy becomes the entropy of $\matr y$ up to a scaling factor and an additive constant.

Statistical mechanics provides an excellent framework to study not only matched, but also mismatched detection.
The analysis of mismatched detection in large communication systems without exploiting the tools provided by statistical mechanics, and purely based on asymptotic properties of large random matrices, has been very limited so far.
One exception is the asymptotic SINR of linear MMSE multiuser detectors with erroneous assumptions on the powers of interfering users by M\"uller and Caire in \cite{mueller:02a}.



\subsection{1st Example: CDMA Systems}
\label{Exa51}

The replica method was introduced into multiuser communications by the landmark paper of Tanaka \cite{tanaka:02} for the purpose of studying the performance of the maximum a-posteriori detector.
Subsequently his work was generalized and extended to other areas of communications by Guo and Verd\'u \cite{guo:02,guo:03,guo:03a,guo:04} and M\"uller et al.\ \cite{mueller:02,mueller:02b,mueller:02c,mueller:03,mueller:03a,mueller:03b,mueller:03c}.
An analysis of an asymptotically large CDMA system with arbitrary joint distribution of the variances of the random chips is given in \cite{mueller:09}. It includes the practically important case of multi-carrier CDMA transmission with users of arbitrary powers in frequency-selective fading as a special case. In the sequel, we give an overview of the initial treatment of the subject by Tanaka \cite{tanaka:02}.

Consider a vector-valued real additive white Gaussian noise channel characterized by the conditional probability distribution.
\begin{equation}
\prob{\matr y|\matr x,\matr H}{\matr y,\matr x,\matr H}=\frac{{\rm e}^{\,-\frac1{2\sigma_0^2} \left(\matr y-\matr{Hx}\right)^{\rm T}\left(\matr y-\matr{Hx}\right)}}{\left(2\pic\sigma_0^2\right)^{\frac N2}}
\label{disawgn}
\end{equation}
with $\matr x,\matr y,N,\sigma_0^2,\matr H$ denoting the channel input, channel output, the latter's number of components, the noise variance, and the channel matrix, respectively.
Moreover, let the detector be characterized by the assumed conditional probability distribution
\begin{equation}
\proba{\matr y|\matr x,\matr H}{\matr y,\matr x,\matr H}=\frac{{\rm e}^{\,-\frac1{2\sigma^2} \left(\matr y-\matr{Hx}\right)^{\rm T}\left(\matr y-\matr{Hx}\right)}}{\left(2\pic\sigma^2\right)^{\frac N2}}
\end{equation}
and the assumed prior distribution $\proba{\matr x}{\matr x}$.
Let the entries of $\matr H$ be independent zero-mean with vanishing odd order moments and variance $1/N$.
 Applying Bayes' law, we find
\begin{equation}
\proba{\matr x|\matr y,\matr H}{\matr x,\matr y,\matr H}=\frac{ {\rm e}^{\,-\frac1{2\sigma^2} \left(\matr y-\matr{Hx}\right)^{\rm T}\left(\matr y-\matr{Hx}\right)+\log \proba{\matr x}{\matr x}}}{\int {\rm e}^{\,-\frac1{2\sigma^2} \left(\matr y-\matr{Hx}\right)^{\rm T}\left(\matr y-\matr{Hx}\right)}\, {\rm d} \Proba{\matr x}{\matr x}}.
\end{equation}
Since \eq{Lagmul} holds for any temperature $T$, we set without loss of generality $T=1$ in \eq{Lagmul} and find the appropriate energy function to be
\begin{equation}
\label{aeef}
||\matr x|| = \frac1{2\sigma^2} \left(\matr y-\matr{Hx}\right)^{\rm T}\left(\matr y-\matr{Hx}\right)-\log \proba{\matr x}{\matr x}.
\end{equation}
This choice of the energy function ensures that the thermodynamic equilibrium models the detector defined by the assumed conditional and prior distributions.

Let $K$ denote the number of users, that is the dimensionality of the input vector $\matr x$.
Applying successively \eq{FrEn} with \eq{disawgn}, \eq{eq512}, replica continuity \eq{repcont}, \eq{rm}, and \eq{aeef} we find for the free energy per user
\begin{eqnarray}
\frac{{\rm F}(\matr x)}K &=& -\frac1K\expect\limits_{\matr H} \int\int\limits_{\RR^N}\frac{{\rm e}^{\,-\frac1{2\sigma_0^2} \left(\matr y-\matr{Hx}\right)^{\rm T}\left(\matr y-\matr{Hx}\right)}}{\left(2\pic\sigma_0^2\right)^{\frac N2}}\times \nonumber\\
&&\log\int\limits_{\RR^K}{\rm e}^{\,-||\matr x||}{\rm d}\matr x {\rm d}\matr y {\rm d}\Prob{\matr x}{\matr x}\\
&=&-\frac1K\lim\limits_{n\to0}\frac\partial{\partial n}\log \expect\limits_{\matr H} \int\int\limits_{\RR^N}\left( \int\limits_{\RR^K}{\rm e}^{\,-||\matr x||}{\rm d}{\matr x}\right)^n \nonumber \times \\
&&\qquad \frac{{\rm e}^{\,-\frac1{2\sigma_0^2} \left(\matr y-\matr{Hx}\right)^{\rm T}\left(\matr y-\matr{Hx}\right)}}{\left(2\pic\sigma_0^2\right)^{\frac N2}}\, {\rm d}\matr y {\rm d}\Prob{\matr x}{\matr x}\label{eq518}\\
&=& -\frac1K\lim\limits_{n\to0}\frac\partial{\partial n}\log \underbrace{\frac{\int\limits_{\RR^N}\expect\limits_{\matr H}\prod\limits_{a=0}^n\int{\rm e}^{\,-\frac1{2\sigma_a^2} \left(\matr y-\matr{Hx_a}\right)^{\rm T}\left(\matr y-\matr{Hx_a}\right)}{\rm d}\Prob{a}{\matr x_a}{\rm d}\matr y}{\left(2\pic\sigma_0^2\right)^{\frac N2}}}_{\define \Xi_n}\nonumber
\end{eqnarray}
with $\sigma_a=\sigma,\forall a\ge1$, $\Prob{0}{\matr x}=\Prob{\matr x}{\matr x}$, and $\Prob{a}{\matr x}=\Proba{\matr x}{\matr x}, \forall a\ge1$.

The integral in \eq{eq518} is given by
\begin{equation}
\label{intX}
\Xi_n=\int
\left(
\frac{\displaystyle\int\limits_{\RR}\expect\limits_{\matr H}
\prod\limits_{a=0}^n\exp\left[-\frac1{2\sigma^2_a}\left(y-\sum\limits_{k=1}^K h_{1k}x_{ak}\right)^2\right]
{\rm d}y}{\sqrt{2\pic}\sigma_0}\right)^N{\rm d}\Prob a{\matr x_a},
\end{equation}
with $x_{ak}$, and $h_{1k}$ denoting the $k^{\rm th}$ component of $\matr x_a$, and the $(1,k)^{\rm th}$ entry of $\matr H$, respectively.
The integrand depends on $\matr x_a$ only through
\begin{equation}
v_{a} \define \frac1{\sqrt \beta}\sum\limits_{k=1}^Kh_{1k}x_{ak}, \qquad a=0,\dots,n
\end{equation}
with the load $\beta$ being defined as $\beta\define K/N$.
Following \cite{tanaka:02}, the quantities $v_{a}$ can be regarded, in the limit $K\to\infty$ as jointly Gaussian random variables with zero mean and covariances
\newcommand{\skp}{\bullet}
\begin{equation}
\label{defQR}
Q_{ab} = \expect\limits_{\matr H}{v_{a}v_{b}} = \frac1K\, \matr x_a \skp \matr x_b
\end{equation}
where we defined the following inner products
\begin{equation}
\matr x_a \skp \matr x_b \define \sum\limits_{k=1}^K x_{ak}x_{bk}.
\end{equation}

In order to perform the integration in \eq{intX}, the $K(n+1)$-dimensional space spanned by the replicas and the vector $\matr x_0$ is split into subshells
\begin{equation}
S\{Q\} \define \left\{\matr x_0,\dots,\matr x_n\left|\matr x_a\skp\matr x_b=KQ_{ab}\right.\right\}
\end{equation}
where the inner product of two different vectors $\matr x_a$ and $\matr x_b$ is constant*.\footnote{*The notation $f\{Q\}$ expresses the dependency of the function $f(\cdot)$ on all $Q_{ab}, 0\le a\le b\le n$.}
With this splitting of the space, we find**\footnote{**The notation $\prod_{a\le b}$ is used as shortcut for $\prod_{a=0}^n\prod_{b=a}^n$.}
\begin{equation}
\label{domsh}
\Xi_n= \int\limits_{\RR^{N(n+1)(n+2)/2}}{\rm e}^{K{\cal I}\{Q\}}{\rm e}^{N{\cal G}\left\{Q\right\}}\prod\limits_{a\le b}{\rm d}Q_{ab},
\end{equation}
where
\begin{equation}
{\rm e}^{\beta {\cal I}\{Q\}} = \int  \left[ \prod\limits_{a\le b} \deltaf\left( \frac{\matr x_a\skp \matr x_b}N - \beta Q_{ab}\right)\right]\prod\limits_{a=0}^n {\rm dP}_a(\matr x_a)
\end{equation}
denotes the probability weight of the subshell and
\begin{equation}
{\rm e}^{{\cal G}\{Q\}}=\frac1{\sqrt{2\pic}\sigma_0}\int\limits_{\RR}\expect\limits_{\matr H}
\prod\limits_{a=0}^n\exp\left[-\frac\beta{2\sigma^2_a}\left(\frac{y}{\sqrt{\beta}}-v_{a}\{Q\}\right)^2\right]{\rm d}y.
\end{equation}
This procedure is a change of integration variables in multiple dimensions where the integration of an exponential function over the replicas has been replaced by integration over the variables $\{Q\}$.
In the following the two exponential terms in \eq{domsh} are evaluated separately.

First, we turn to the evaluation of the measure ${\rm e}^{\beta{\cal I}\left\{Q\right\}}$.
Since for some $t\in\RR$, we have the Fourier expansions of the Dirac measure
\begin{eqnarray}
\nonumber
\deltaf\left(\frac{\matr x_a\skp\matr x_b}N - \beta Q_{ab}\right) =
\frac1{2\pic {\rm j}}\int\limits_{\cal J}\exp\left[\tilde{Q}_{ab}\left( \frac{\matr x_a\skp\matr x_b}N-\beta Q_{ab}\right)\right] {\rm d}\tilde{Q}_{ab}
\end{eqnarray}
with ${\cal J}=(t-{\rm j}\infty;t+{\rm j}\infty)$, the measure ${\rm e}^{\beta{\cal I}\left\{Q\right\}}$ can be expressed as
\begin{eqnarray}
{\rm e}^{\beta{\cal I}\left\{Q\right\}} &=&
\int \left[\prod\limits_{a\le b} \int\limits_{\cal J}{\rm e}^{\tilde{Q}_{ab}\left(\frac{\matr x_a\skp\matr x_b}N-\beta Q_{ab}\right)}\frac{{\rm d}\tilde{Q}_{ab}}{2\pic {\rm j}}\right] \prod\limits_{a=0}^n{\rm dP}_a(\matr x_a) \\
&=&  \int\limits_{{\cal J}^{N(n+2)(n+1)/2}} \!\!\!\!\hspace*{-5mm}{\rm e}^{-\beta\sum\limits_{c=1}^N\sum\limits_{a\le b}\tilde{Q}_{ab}Q_{ab}} \left(M\left\{\tilde{Q}\right\}\right)^K \prod\limits_{a\le b} \frac{{\rm d}\tilde{Q}_{ab}}{2\pic {\rm j}}
\label{saddlemu}
\end{eqnarray}
with
\begin{equation}
M\left\{\tilde{Q}\right\} = \int  \exp \left( \sum\limits_{a\le b}\tilde{Q}_{ab} x_{a}x_{b}\right)\prod\limits_{a=0}^n {\rm dP}_a(x_{a}).
\end{equation}
In the limit of $K\to\infty$ one of the exponential terms in \eq{domsh} will dominate over all others.
Thus, only the maximum value of the correlation $Q_{ab}$ is relevant for calculation of the integral, as shown in Section~\ref{RepSym}.

At this point, we assume that the replicas have a certain symmetry, as outlined in Section~\ref{RepSym}.
This means, that in order to find the maximum of the objective function, we consider only a subset of the potential possibilities that the variables $Q_{ab}$ could take.
Here, we restrict them to the following four different possibilities
$Q_{00}=p_{0}$, $Q_{0a}=m,\forall a\ne0$, $Q_{aa} = p,\forall a\ne0$, $Q_{ab}=q,\forall 0\ne a\ne b\ne0$.
One case distinction has been made, as zero and non-zero replica indices correspond to the true and the assumed distributions, respectively, and thus will differ, in general.
Another case distinction has been made to distinguish correlations $Q_{ab}$ which correspond to correlations between different and identical replica indices. This gives four cases to consider in total.
We apply the same idea to the correlation variables in the Fourier domain and set $\tilde{Q}_{00}=G_{0}/2$, $\tilde{Q}_{aa}=G/2,\forall a\ne0$, $\tilde{Q}_{0a}=E,\forall a\ne0$, and $\tilde{Q}_{ab}=F,\forall 0\ne a\ne b\ne0$.

At this point the crucial benefit of the replica method becomes obvious.
Assuming replica continuity, we have managed to reduce the evaluation of a continuous function to sampling it at integer points.
Assuming replica symmetry, we have reduced the task of evaluating infinitely many integer points to calculating eight different correlations (four in the original and four in the Fourier domain).

The assumption of replica symmetry leads to
\begin{equation}
\sum\limits_{a\le b} \tilde{Q}_{ab}Q_{ab} = nEm + \frac{n(n-1)}2\, Fq + \frac{G_{0}p_{0}}2 + \frac n2\, Gp
\label{lineq}
\end{equation}
and
\begin{equation}
M\{E,F,G,G_0\} = \int  {\rm e}^{\frac{G_{0}}2x_{0}^2+\sum\limits_{a=1}^nEx_{0}x_{a}+\frac{G}2x_{a}^2+\sum\limits_{b=a+1}^n Fx_{a}x_{b}}\prod\limits_{a=0}^n {\rm dP}_a(x_{a})
\label{lastMk}
\end{equation}
Note that the prior distribution enters the free energy only via \eq{lastMk}.
We will focus on this later on after having finished with the other terms.

%
%
%
%

For the evaluation of ${\rm e}^{{\cal G}\{Q\}}$ in \eq{domsh}, we can use the replica symmetry to construct the correlated Gaussian random variables $v_{a}$ out of independent zero-mean, unit-variance Gaussian random variables $u,t,z_{a}$ by
\begin{eqnarray}
v_{0} &=& u\sqrt{p_{0}-\frac{m^2}{q}} - t\frac{m}{\sqrt{q}}\\
v_{a} &=& z_{a} \sqrt{p-q} - t\sqrt{q}, \quad a>0.
\end{eqnarray}
With that substitution, we get
\begin{eqnarray}
{\rm e}^{{\cal G}(m,q,p,p_{0})}&=&\\
&\hspace*{-3.4cm}=&\hspace*{-1.7cm}\nonumber\frac1{\sqrt{2\pic}\sigma_0}\int\limits_{\RR^2}\int\limits_{\RR}\exp\left[-\frac\beta{2\sigma_0^2}\left(u\sqrt{p_{0}-\frac{m^2}{q}}-\frac{tm}{\sqrt{q}}-\frac{y}{\sqrt\beta}\right)^2\right]{\rm D}u\\
&\hspace*{-3.4cm}&\hspace*{-1.7cm}\quad\times\left[\int\limits_{\RR}\exp\left[-\frac \beta{2\sigma^2}\left(z\sqrt{p-q}-t\sqrt{q}-\frac{y}{\sqrt\beta}\right)^2\right]{\rm D}z\right]^n{\rm D}t\,{\rm d}y \nonumber\\
&\hspace*{-3.4cm}=&\hspace*{-1.7cm} \sqrt{\frac{(1+\frac\beta{\sigma^2}(p-q))^{1-n}}{1+\frac\beta{\sigma^2}(p-q)+n\frac\beta{\sigma^2}\left(\frac{\sigma_0^2}\beta+p_{0}-2m+q\right)}}
\label{ext1}
\end{eqnarray}
with the Gaussian measure ${\rm D}z=\exp(-z^2/2)\,{\rm d}z/\sqrt{2\pic}$.
Since the integral in \eq{domsh} is dominated by the maximum argument of the exponential function, the derivatives of
\begin{equation}
{\cal G}\{Q\} - \beta \sum\limits_{a\le b} \tilde{Q}_{ab}Q_{ab}
\label{tedr}
\end{equation}
with respect to $m,q,p$ and $p_{0}$ must vanish as $N\to\infty$.
Taking derivatives after plugging \eq{lineq} and \eq{ext1} into \eq{tedr}, solving for $E,F,G$, and $G_{0}$ and letting $n\to 0$ yields
\begin{eqnarray}
\label{resE}
E &=&\frac 1{\sigma^2+\beta(p-q)}\\
\label{resF}
F &=&\frac {\sigma_0^2 +\beta\left(p_{0}-2m+q\right)}{[\sigma^2+\beta(p-q)]^2}\\
\label{resG}
G &=&F-E\\
\label{resG0}
G_{0} &=& 0.
\end{eqnarray}
In the following, the calculations are shown explicitly for Gaussian and binary priors. Additionally,  a general formula for arbitrary priors is given.

\subsection{Gaussian Prior Distribution}

Assume a Gaussian prior distribution
\begin{equation}
\prob a{x_{a}} = \frac1{\sqrt{2\pic}}\, {\rm e}^{-x_{a}^2/2} \qquad \forall a.
\end{equation}
Thus, the integration in \eq{lastMk} can be performed explicitly and we find with \cite[(87)]{tanaka:02}
\begin{equation}
\label{surprise}
M\left\{E,F,G,G_0\right\} = \sqrt{\frac{\left(1+ F - G \right)^{1-n}}{\left(1- G_{0}\right)\left(1+ F- G-n F\right)-{n E^2}}}.
\end{equation}

In the large system limit, the integral in \eq{saddlemu} is also dominated by that value of the integration variable which maximizes the argument of the exponential function.
Thus, partial derivatives of
\begin{equation}
\label{derMk}
\log M\left\{E,F,G,G_{0}\right\} -  nEm - \frac{n(n-1)}2\, Fq - \frac{G_{0}p_{0}}2 - \frac n2\, Gp
\end{equation}
with respect to $E,F,G,G_{0}$ must vanish as $N\to\infty$.
An explicit calculation of these derivatives yields
\begin{eqnarray}
\label{resm}
m &=&  \frac{ E}{1+ E}\\
q &=&  \frac{ E^2 + F}{\left(1+ E\right)^2} \\
\label{resp}
p &=&  \frac{E^2 + E +  F +1}{\left(1+ E\right)^2}\\
p_{0} &=& 1
\label{resp0}
\end{eqnarray}
in the limit $n\to0$ with \eq{resG} and \eq{resG0}.
Surprisingly, if we let the true prior to be binary and only the replicas to be Gaussian we also find \eq{resm} to \eq{resp0}.
This setting corresponds to linear MMSE detection \cite{verdu:98}.

Returning to our initial goal, the evaluation of the free energy, and collecting our previous results, we find$^{@}$\footnote{$^{@}$Eq. \eq{fineqr2} is inconsistent with \cite[(8.62)]{mueller:09} due to a typo in \cite{mueller:09}.}
\begin{eqnarray}
\frac{{\rm F}(\matr x)}K &=&
-\frac 1K \lim\limits_{n\to0}\frac\partial{\partial n} \log \Xi_n\!\!\!\! \\
&=&\lim\limits_{n\to0}\frac{\partial}{\partial n}  \frac 1\beta \left[-{\cal G}(m,q,p,p_{0}) +\frac{\beta n(n-1)}2\, Fq+\right.  \nonumber\\
&& \left. + \beta n Em + \frac {\beta n}2\,Gp\right] -  \log M\left\{E,F,G, 0\right\}\\
&=&\frac1{2\beta} \lim\limits_{n\to0}\left[  \log\left(1+\frac\beta{\sigma^2}(p-q)\right) + 2\beta Em +\beta Gp\right.\nonumber\\
&& \left. + \beta (2n-1)Fq+\frac{\sigma_0^2+\beta(p_{0}-2m+q)}{\sigma^2+\beta(p-q)+n\sigma_0^2+n\beta(p_{0}-2m+q)} \vphantom{\sum\limits_{c=1}^N}\right] \nonumber\\
&&+\frac1{2} \lim\limits_{n\to0} \log\left(1+ E\right) -\frac{ E^2 +  F}{1+ E-n E^2-n F}\\
&=& \frac1{2\beta}\left[ \log\left(1+\frac\beta{\sigma^2}(p-q)\right)+\frac{F}{E}+ \beta E(2m-p) +\beta F(p-q) \right] \nonumber\\ && +\frac1{2}\log\left(1+ E\right) -\frac{ E^2 +  F}{2(1+ E)}.
\label{fineqr2}
\end{eqnarray}
This is the final result for the free energy of the mismatched detector assuming noise variance $\sigma^2$ instead of the true noise variance $\sigma^2_0$.
The five macroscopic parameters $E, F, m,q,p$ are implicitly given by the simultaneous solution of the system of equations \eq{resE} to \eq{resF} and \eq{resm} to \eq{resp}.
This system of equations can only be solved numerically.

Specializing our result to the matched detector assuming the true noise variance by letting $\sigma\to\sigma_0$, we have $F\to E$, $q\to m$, $p\to p_{0}$.
This makes the free energy simplify to
\begin{eqnarray}
\frac{{\rm F}(\matr x)}K & =&\frac1{2\beta}\left[ \sigma_0^2E-\log\left(\sigma_0^2E\right)\right] +\frac1{2} \log\left(1+ E\right).
\end{eqnarray}
with
\begin{equation}
\frac 1E= \sigma_0^2 + \frac\beta{1+E}
\label{fpeec}
\end{equation}
This result is more compact and it requires only to solve \eq{fpeec} numerically which is conveniently done by fixed-point iteration.

It can be shown that the parameter ${E}$ is actually the signal-to-interference and noise ratio.
It has been derived independently by Tse and Hanly \cite{tse:99b} in context of CDMA using results from random matrix theory, cf.\ \eq{sinrfix} and \eq{tsehanlyequation}.

The similarity of free energy with the entropy of the channel output mentioned at the end of Section~\ref{MisDet} is expressed by the simple relationship
\begin{equation}
\label{mi}
\frac{{\rm I}(\matr x,\matr y)}K = \frac{{\rm F}(\matr x)}K -\frac1{2\beta}
\end{equation}
between the (normalized) free energy and the (normalized) mutual information between channel input signal $\matr x$ and channel output signal $\matr y$ given the channel matrix $\matr H$.
Assuming that the channel is perfectly known to the receiver, but totally unknown to the transmitter, \eq{mi} gives the channel capacity per user.
\subsection{Binary Prior Distribution}
\label{BinPri}

Now, we assume a binary prior distribution
\begin{equation}
\label{nubin}
\prob a{x_{a}} = \frac{1}2 \,\deltaf(x_{a}-1) + \frac{1}2 \,\deltaf(x_{a}+1).
\end{equation}
For uniform binary prior distributions, the calculations can be simplified assuming without loss of generality that all users have transmitted the symbol $x_{0k}=+1$. In statistical mechanics this simplification is called the {\em gauge transformation}.

Plugging the prior distribution into \eq{lastMk}, we find
\begin{eqnarray}
M\left\{E,F,G,G_0\right\} &=& \nonumber\\
&\hspace*{-4cm}=&\hspace*{-4cm}\int\limits_{\RR^{n}}  {\rm e}^{\frac{ G_{0}+n G}2+\sum\limits_{a=1}^n Ex_{a}+\sum\limits_{b=a+1}^n  Fx_{a}x_{b}}\prod\limits_{a=1}^n {\rm dP}_a(x_{a})\\
&\hspace*{-4cm}=&\hspace*{-4cm}\textstyle {\rm e}^{\frac12\left( G_{0}+n G-n F\right)}\hspace*{-7mm}\sum\limits_{\{x_{a},a=1,\dots,n\}}  \left\{\exp \left[\frac { F}2\left(\sum\limits_{a=1}^nx_{a}\right)^2+ E\sum\limits_{a=1}^nx_{a}\right]\right\} \prod\limits_{a=1}^n\Pr(x_{a})
\label{noteqc}
\end{eqnarray}
where we can use the following property of the Gaussian measure
\begin{equation}
\label{HST}
\exp\left( F\frac {S^2}2\right) = \int\exp\left(\pm\sqrt { F} z S\right){\rm D}z\qquad \forall S\in\RR
\end{equation}
which is also called the Hubbard-Stratonovich transform to linearize the exponents
\begin{eqnarray}
M\left\{E,F,G,G_{0}\right\}&=&\nonumber\\
&\hspace*{-4cm}=&\hspace*{-4cm}e^{\frac12\left( G_{0}+n G-n F\right)}\textstyle\sum\limits_{\{x_{a},a=1,\dots,n\}}  \int \exp \left[\left(z\sqrt{ F}+ E\right)\sum\limits_{a=1}^nx_{a}\right]
{\rm D}z \prod\limits_{a=1}^n\Pr(x_{a}) .
\end{eqnarray}
Since
\begin{eqnarray}
f_n&\define& \sum\limits_{\{x_{a},a=1,\dots,n\}}\exp\left[\left(z\sqrt{ F}+ E\right)\sum\limits_{a=1}^nx_{a}\right]\prod\limits_{a=1}^n\Pr(x_{a})\\
&=& \sum\limits_{x_{n}} \Pr(x_{n}) f_{n-1} \exp\left[\left(z\sqrt{ F}+ E\right) x_{n}\right]\\
&=&  f_{n-1} \, {\cosh\left(z\sqrt{ F}+E\right)}\\
& =& {\cosh^n\left(z\sqrt{F}+ E\right)},
\end{eqnarray}
we find
\begin{eqnarray}
M\left\{E,F,G,G_0\right\}=
\frac{\int \cosh^n\left(z\sqrt{ F}+ E\right)   {\rm D}z}{\exp\left(\frac{n F- G_{0}-nG}2\right)}.
\end{eqnarray}

In the large system limit, the integral in \eq{saddlemu} is dominated by that value of the integration variable which maximizes the argument of the exponential function.
Thus, partial derivations of \eq{derMk} with respect to $E,F,G,G_{0}$ must vanish as $N\to\infty$.
An explicit calculation of these derivatives gives
\begin{eqnarray}
\label{eqmc}
 m &=&  \int \tanh\left(z\sqrt{ F}+ E\right) {\rm D}z\\
\label{eqqc}
q &=&  \int \tanh^2\left(z\sqrt{F}+ E\right) {\rm D}z\\
\label{eqpc}
p &=& p_{0} = 1
\end{eqnarray}
in the limit $n\to0$. In order to obtain \eq{eqqc}, note from \eq{noteqc} that the first order derivative of $M\{\cdot\} \exp(n F/2)$ with respect to $F$ is identical to half of the second order derivative of $M\{\cdot\} \exp(n F/2)$ with respect to $E$.

Returning to our initial goal, the evaluation of the free energy, and collecting our previous results, we find$^{@@}$\footnote{$^{@@}$\eq{fineqr} is inconsistent with \cite[(8.62)]{mueller:09} due to a typo in \cite{mueller:09}.}
\begin{eqnarray}
\nonumber
\frac{{\rm F}(\matr x)}K &=& - \frac 1K \lim\limits_{n\to0} \frac\partial{\partial n} \log \Xi_n \nonumber \\
&=&\lim\limits_{n\to0} \frac{\partial}{\partial n}\frac1\beta \left[-{\cal G}(m,q,1,1)
+ \beta n Em + \frac{\beta n(n-1)}2\, Fq+
 \frac {\beta n}2\,G\right] \nonumber\\ && - \log M\left\{E,F,G, 0\right\}\\
&=&\frac1{2\beta} \left[ \log\left(1+\frac\beta{\sigma^2}(1-q)\right)
+ \frac{F}{E}  +\beta E(2m -1) +\beta F(1-q)
\right] \nonumber\\  && + \frac{ E}2 -F - \int \log\cosh\left(z\sqrt{ F}+ E\right) {\rm D}z
\label{fineqr}
\end{eqnarray}
This is the final result for the free energy of the mismatched detector assuming noise variance $\sigma^2$ instead of the true noise variance $\sigma^2_0$.
The four macroscopic parameters $E, F, m,q$ are implicitly given by the simultaneous solution of the system of equations \eq{resE}, \eq{resF} and \eq{eqmc} to \eq{eqqc}.
This system of equations can only be solved numerically.
Moreover, it can have multiple solutions.
In case of multiple solutions, the correct solution is that one which minimizes the free energy, since in the thermodynamic equilibrium the free energy is always minimized, cf.\ Section~\ref{FreEne}.

Specializing our result to the matched detector assuming the true noise variance by letting $\sigma\to\sigma_0$, we have $F\to E$, $G\to G_{0}$, $q\to m$ which makes the free energy simplify to
\begin{equation}
\frac{{\rm F}(\matr x)}K = \frac1{2\beta} \left[\sigma_0^2 E-\log\left(\sigma_0^2E\right) \right]  -  \label{febin} \int \log\cosh\left(z\sqrt{ E}+ E\right) {\rm D}z
\end{equation}
where the macroscopic parameter $E$ is given by
\begin{eqnarray}
\hspace*{-10mm} \frac1{E}
&=&\sigma_0^2 + \beta  \int {1-\tanh\left(z\sqrt{ E}+ E\right)} {\rm D}z.
\label{eqec}
\end{eqnarray}

Similar to the case of Gaussian priors, $E$ can be shown to be a kind of signal-to-interference and noise ratio, in the sense that the bit error probability is given by
\begin{equation}
\Pr(\hat x_k\ne x_k) = \int\limits_{\sqrt{E}}^{\infty} {\rm D}z.
\label{eqber}
\end{equation}
In fact, it can even be shown that in the large system limit, an equivalent additive white Gaussian noise channel can be defined to model the multiuser interference \cite{mueller:03,guo:04}.
Constraining the input alphabet of the channel to follow the non-uniform binary distribution \eq{nubin} and assuming channel state information being available only at the transmitter, channel capacity is given by \eq{mi} with the free energy given in \eq{febin}.

Large system results for binary prior (even for uniform binary prior) have not yet been able to be derived by means of rigorous mathematics despite intense effort to do so.
Only for the case of vanishing noise variance a fully mathematically rigorous result was found by Tse and Verd\'u \cite{tse:00} which does not rely on the replica method.
\subsection{Arbitrary Prior Distribution}
\label{ArbPri}
Consider now an arbitrary prior distribution.
As shown by Guo and Verd\'u \cite{guo:04}, this still allows to reduce the multi-dimensional integration over all replicated random variables to a scalar integration over the prior distribution.
Consider \eq{lastMk} giving the only term that involves the prior distribution and apply the Hubbard-Stratonovich transform \eq{HST}
\begin{eqnarray}
M\{E,F,G,G_0\} &=& \nonumber\\
&\hspace*{-4.4cm}=& \hspace*{-2.2cm}\int {\rm e}^{\frac{ G_{0}}2x_{0}^2+\sum\limits_{a=1}^n Ex_{0}x_{a}+\frac{ G}2x_{a}^2+\sum\limits_{b=a+1}^n  Fx_{a}x_{b}}\,\prod\limits_{a=0}^n {\rm dP}_a(x_{a})\\
\label{anoteqc}
&\hspace*{-4.4cm}=& \hspace*{-2.2cm}\int {\rm e}^{\frac{ G_{0}}2x_{0}^2+\frac{ F}2\left(\sum\limits_{a=1}^n x_{a}\right)^2+\sum\limits_{a=1}^n Ex_{0}x_{a}+\frac{ G- F}2x_{a}^2}\prod\limits_{a=0}^n {\rm dP}_a(x_{a})\\
&\hspace*{-4.4cm}=& \hspace*{-2.2cm}\int\!\!\! \int {\rm e}^{\frac{ G_{0}}2x_{0}^2+\sum\limits_{a=1}^n\tilde Ex_{0}x_{a}+\sqrt{F}zx_{a}+\frac{G- F}2\,x_{a}^2}{\rm D}z\prod\limits_{a=0}^n {\rm dP}_{a}(x_{a})\\
\nonumber
&\hspace*{-4.4cm}=& \hspace*{-2.2cm} \int {\rm e}^{\frac{ G_{0}}2x^2}\int\!\!\left(\int{\rm e}^{ Ex\breve x+\sqrt{F}z\breve x+\frac{ G- F}2\,\breve x^2}{\rm d}\Proba{\breve x\,}{\breve x}\right)^n{\rm D}z{\rm dP}_{x}(x)\\
&&
\end{eqnarray}

In the large system limit, the integral in \eq{saddlemu} is dominated by that value of the integration variable which maximizes the argument of the exponential function.
Thus, partial derivations of \eq{derMk} with respect to $E,F,G,G_{0}$ must vanish as $N\to\infty$.
While taking derivatives with respect to $E,G$ and $G_{0}$ straightforwardly lead to suitable results, the derivative with respect to $F$ requires a little trick:
Note for the integrand $I$ in \eq{anoteqc}, we have
\begin{equation}
\label{trickdiff}
\frac{\partial I}{\partial F} = \frac1{2x_{0}^2}\frac{\partial^2 I}{\partial E^2} -\frac{\partial I}{\partial G}.
\end{equation}

With the help of \eq{trickdiff}, an explicit calculation of the four derivatives gives the following expressions for the macroscopic parameters $m,q,p$ and $p_{0}$
\begin{eqnarray}
 m &=& {\int\!\!\!\!\int x\frac{\int \breve x{\rm e}^{E\left(x\breve x-\frac{\breve x^2}2\right)+\sqrt{F}z\breve x}{\rm d}\Proba{\breve x}{\breve x}}{\int{\rm e}^{ E\left(x\breve x-\frac{\breve x^2}2\right)+\sqrt{ F}z\breve x}{\rm d}\Proba{\breve x}{\breve x}}\,{\rm D}z{\rm dP}_{x}(x)}
\label{aeqmc}\\
q &=& {\int\!\!\!\!\int\!\left[\frac{\int\breve x{\rm e}^{E\left(x\breve x-\frac{\breve x^2}2\right)+\sqrt{ F}z\breve x}{\rm d}\Proba{\breve x}{\breve x}}{\int{\rm e}^{ E\left(x\breve x-\frac{\breve x^2}2\right)+\sqrt{ F}z\breve x}{\rm d}\Proba{\breve x}{\breve x}}\right]^2\!{\rm D}z{\rm d}\Prob{x}{x}}
\label{aeqqc}\\
p &=& {\int\!\!\!\! \int\frac{\int\breve x^2{\rm e}^{E\left(x\breve x-\frac{\breve x^2}2\right)+\sqrt{F}z\breve x}{\rm d}\Proba{\breve x}{\breve x}}{\int {\rm e}^{E\left(x\breve x-\frac{\breve x^2}2\right)+\sqrt{F}z\breve x}{\rm d}\Proba{\breve x}{\breve x}}\,{\rm D}z{\rm dP}_{x}(x)
}\\
\label{aeqpc}\\
p_{0} &=& {\int x^2{\rm dP}_{x}(x)}
\end{eqnarray}
with \eqs{resG}{resG0} in the limit $n\to0$.

Returning to our initial goal, the evaluation of the free energy, and collecting our previous results, we find$^{*}$\footnote{$^{*}$This equation is inconsistent with \cite[(8.74)]{mueller:09} due to a typo in \cite{mueller:09}.}
\begin{eqnarray}
\nonumber
\frac{{\rm F}(\matr x)}K
&=&\frac1{2\beta} \left[ \log\left(1+\frac\beta{\sigma^2}(p-q)\right)  +\frac{F}{E}
+\beta E(2m -p) +\beta F(p-q) \right]\\ &&-
\int \!\!\!\int\log\int{\rm e}^{ E\left(x\breve x-\frac{\breve x^2}2\right)+\sqrt{F}z\breve x}{\rm d}\Proba{\breve x}{\breve x}{\rm D}z{\rm dP}_{x}(x)
\end{eqnarray}
This is the final result for the free energy of the mismatched detector assuming noise variance $\sigma^2$ instead of the true noise variance $\sigma^2_0$.
The five macroscopic parameters $E, F, m,q,p$ are implicitly given by the simultaneous solution of the system of equations \eq{resE}, \eq{resF} and \eq{eqmc} to \eq{eqpc}.
This system of equations can only be solved numerically.
Moreover, it can have multiple solutions.
In case of multiple solutions, the correct solution is that one which minimizes the free energy, since in the thermodynamic equilibrium the free energy is always minimized, cf.\ Section~\ref{FreEne}.


\subsection{2nd Example: Vector Precoding}
\label{Exa52}


In the following we shortly outline the application of the replica analysis to the setting described in Section \ref{VecPre}, and summarize some of the main results in \cite{zaidel:11}. 
We start by focusing on the energy penalty, and recall that the task of the nonlinear stage at the transmitter (namely, the nonlinear precoding stage) can be described as follows. It is equivalent to the minimization of an objective function, which we interpret here as the \emph{Hamiltonian}, having the quadratic form 
\begin{equation}\label{eq: General definition of the minimization objective function}
\Hc(\wv) =   \wv^\dag \Jm \wv \quad,
\end{equation}
where $\Jm$ is a random matrix of dimensions $N \times N$.
Thus, the minimum energy penalty per symbol can be expressed as 
\begin{equation}\label{eq: min_energy_penalty}
\frac{1}{N} \min_{\wv\in \B_{\sv}} \Hc(\wv)  \quad ,
\end{equation}
where we use the shortened notation $\B_{\sv} \triangleq \B_{s_1}\times \cdots \times \B_{s_N}$ as in equation (\ref{eq: Definition of x as argmin}) above.
Note that in order to comply with (\ref{eq: Definition of x as argmin}) one should take $\Jm= \Tm^\dag\Tm$, however since the results in the sequel hold for a more general class of matrices, we proceed with the formulation suggested in (\ref{eq: General definition of the minimization objective function}).

To calculate the minimum of the objective function (\ref{eq: General definition of the minimization objective function}), we use notions from statistical physics as defined in Section \ref{FreEne} above, albeit with a slight adaptation to the problem at hand. In particular, we define the Boltzmann distribution on the set of state vectors $\set{\wv}$  as
\begin{equation}\label{eq: Boltzmann distribution for the problem in hand}
P_{\Bc}(\wv) = \frac{1}{\Zc} \e^{-\beta \Hc(\wv)}  \quad ,
\end{equation}
where the parameter $\beta>0$ is referred to as the {\em inverse temperature} $\beta=1/T$ (cf.\ (\ref{Lagmul})), while the normalization factor $\Zc$ is the {\em partition function}
\begin{equation}\label{eq: definition of Z}
\Zc = \sum_{\wv\in \B_{\sv}} \e^{-\beta \Hc(\wv)} \quad .
\end{equation}
The \emph{energy} of the system is given by
\begin{equation}
\label{eq: Thermodynamics definition of energy1}
\Ec(\beta) = \sum_{\wv\in \B_{\sv}} P_{\Bc}(\wv) \Hc(\wv) \quad .
\end{equation}
The definitions above hold for both discrete and continuous alphabets $\B_{\sv}$. The only difference is that for continuous alphabets the sums over $\wv \in  \B_{\sv}$ are replaced by integrals.
At thermal equilibrium, 
the \emph{free energy} can be expressed as
\begin{equation}
\label{eq: Thermodynamics free energy at equilibrium}
\Fc(\beta) = - \frac{1}{\beta} \log \Zc \quad.
\end{equation}
With that in mind, and while relying on the self-averaging property (see Section \ref{FreEne}), the limiting energy penalty (per symbol) can be represented as
\begin{equation}\label{eq: Representation of the energy penalty through F}
\begin{array}{cll}
\bE &= \lim\limits_{N\to\infty}\frac{1}{N} \min\limits_{\wv\in \B_{\sv}} \wv^\dag \Jm \wv \\ &=
  - \lim\limits_{N\to\infty}  \lim\limits_{\beta \rightarrow \infty} \frac{1}{\beta N} \E\set{\log \sum\limits_{\wv \in \B_{\sv}} \e^{-\beta \wv^\dag\Jm\wv }} \\ 
  &
  =  \lim\limits_{N\to\infty} \lim\limits_{\beta \rightarrow \infty} \E\set{\frac{\Fc(\beta)}{N}}\quad .
\end{array}
\end{equation}
The rationale behind the above formulation is that as $\beta\rightarrow \infty$, the partition function becomes dominated by the terms corresponding to the \emph{minimum} energy. Hence, taking the logarithm and further normalizing with respect to $\beta$ produces the desired limiting quantity 
at the minimum energy subspace of $\B_{\sv}$. Note that the expectation in (\ref{eq: Representation of the energy penalty through F}) is over all realizations of $\Jm$ and $\sv$, and it is the self-averaging property that ensures convergence of the normalized free energy (as $N\to\infty$) to a \emph{deterministic} quantity. We further note here that even if the energy minimizing vector is not unique, or in fact even if the number of such vectors is exponential in $N$, one still gets the desired quantity when taking the limit $\beta \rightarrow \infty$.

Another useful characteristic of the precoding scheme is the joint input-output distribution of the nonlinear precoding stage. We first define the \emph{empirical} joint distribution (for general $\beta$) of the precoder input $s$ and output $w$ as
\begin{eqnarray}\label{eq: joint empirical definition}
 P^{(N)}_{W,S}(\xi,\upsilon)=\frac{1}{N} \sum_{\wv\in \B_{\sv}}P_{\Bc}(\wv) \sum_{k=1}^N  1\left\{(w_k,s_k)=(\xi,\upsilon)\right\} \quad ,
\end{eqnarray}
where $1\set{\cdot}$ denotes the indicator function.
To obtain the limiting empirical joint distribution $P_{W,S}$, we introduce a dummy variable $h\in \mathbb R$ \cite{Mezard-Montanari-Book-Final-2009}, as well as the function 
\begin{equation}
\label{defsumV}
V(h) = -h \sum_{k=1}^N  1\left\{(w_k,s_k)=(\xi,\upsilon)\right\} \quad.
\end{equation}
Adding this term to the Hamiltonian $\Hc(\wv)$, the partition function gets modified to
\begin{equation}\label{eq: 1RSB: Definition of script-Z1}
\Zc({h})= \sum_{\wv \in \B_{\sv}} \e^{-\beta \left(\Hc(\wv)+V(h)\right)} \ ,
\end{equation}
which using (\ref{eq: Thermodynamics free energy at equilibrium}) yields a modified expression for the free energy. Now, differentiating with respect to $h$, setting $h= 0$, and letting $\beta\rightarrow\infty$ we get
\begin{equation}\label{eq: joint empirical as a function of partial derivative of log Z-h}
\begin{array}{lll}
P_{W,S}(\xi,\upsilon) &= \lim\limits_{N\to\infty} P^{(N)}_{W,S}(\xi,\upsilon)\\ 
&=  \lim\limits_{N\to\infty}\limits\frac{1}{N} \lim\limits_{\beta\rightarrow\infty} \E\set{\frac{\partial \Fc(\beta,{h})}{\partial h}\bigg|_{h= 0}} \quad
\label{eq317}
\end{array}
\end{equation}
where $\mathcal F(\beta, h)$ denotes the free energy that corresponds to the modified partition function $\mathcal Z(h)$. The self-averaging property of the free energy implies that the empirical joint distribution is self-averaging as well.


We next invoke the following underlying assumption on the matrix $\Jm$.
\begin{assumption}
\label{Assm: Assumption on J}
Let the random matrix $\Jm$ be decomposable into
\begin{equation}\label{eq: Definition of the decomposability property}
\Jm = \Um\Dm\Um^\dag \quad,
\end{equation}
where $\Um$ is a unitary \emph{Haar distributed} matrix, and $\Dm$ is a diagonal matrix with its diagonal elements being the eigenvalues of $\Jm$ whose empirical distribution converges to a nonrandom distribution 
uniquely characterized by its $\Rrm$-transform $\Rrm(\cdot)$, which is assumed to exist. 
\end{assumption}
Denoting the limiting normalized energy for some inverse temperature $\beta$ as
\begin{equation}
\Esc(\beta) \triangleq \lim_{N \rightarrow \infty} \frac{1}{N} \E\set{\Ec(\beta)} \quad , \label{eq: Definition of normalized internal energy}
\end{equation}
then applying the replica method we get that the limiting energy penalty at \emph{zero temperature} ($\beta\to\infty$) can be expressed as (cf.\ (\ref{eq: Representation of the energy penalty through F}) and (\ref{LHP})--(\ref{eq179}))
\begin{eqnarray}
\bE &= \lim_{\beta \rightarrow \infty} \Esc(\beta) & \nonumber\\
&= -  \lim_{\beta \rightarrow \infty} \lim_{n\rightarrow 0^+}  \lim_{N \rightarrow \infty} \frac{1}{n \beta N} \log \E\set{\left(\sum_{\wv \in \B_{\sv}} \e^{-\beta \wv^\dag\Jm\wv } \right)^n}  \nonumber\\
&= -  \lim_{\beta \rightarrow \infty}  \lim_{n\rightarrow 0}  \lim_{N \rightarrow \infty} \frac{1}{n \beta N}     \log \E \set{  \sum_{\set{\wv_a}}
 \e^{\, \sum_{a=1}^n -\beta \wv_a^\dag\Jm\wv_a } } \label{eq: Representation of the energy penalty in terms of replicas} \\
&= -  \lim_{\beta \rightarrow \infty} \lim_{n\rightarrow 0} \lim_{N \rightarrow \infty} \frac{1}{n \beta N} \log \E \set{     \sum_{\set{\wv_a}}   \e^{- \Tr (\beta \Jm \sum_{a=1}^n \wv_a \wv_a^\dag )}  } \ , \nonumber
\end{eqnarray}
where we use the notation $\sum_{\set{\wv_a}} = \sum_{\wv_1 \in \B_{\sv}} \cdots \sum_{\wv_n \in \B_{\sv}}$,  $\Tr(\cdot)$ denotes the trace operator, and it is assumed that the limits with respect to $N$ and $n$ can be interchanged (which is the common practice in replica analyses).

The summation over the replicated precoder output vectors $\set{\wv_a}_{a=1}^n$ in (\ref{eq: Representation of the energy penalty in terms of replicas}) is performed by splitting the replicas into \emph{subshells},  defined through a matrix $\Qm_{[n\times n]}$ as follows
\begin{equation}\label{eq: Definition of the subshell S-Q}
S(\Qm) \triangleq 
\set{\wv_1,\dots,\wv_n \big| \wv_a^\dag \wv_b = NQ_{ab}} .
\end{equation}
The limit  $N\to \infty$ allows the application of saddle point integration, which yields the following general result \cite{zaidel:11}.
\begin{prop}[{\cite[Proposition III.2]{zaidel:11}}]
\label{prop:general_saddle_point_free_energy}
For any inverse temperature $\beta$, any structure of $\Qm$ consistent with (\ref{eq: Definition of the subshell S-Q}), and any random matrix fulfilling Assumption~\ref{Assm: Assumption on J}, 
the energy  is given by
\begin{equation}
\label{eq: General expression for energy}
\Esc(\beta) =  \lim\limits_{n\to 0}\frac{1}{n} \Tr \left[\Qm \, \Rrm(-\beta \Qm)\right] \quad ,
\end{equation}
where $\Qm$ is the solution to the saddle point equation
\begin{equation}
\Qm = \int \frac{\sum\limits_{{\bf w}\in \B_{s}^n} {\bf ww^\dagger}{\rm e}^{\,-\beta{\bf w}^\dagger \Rrm(-\beta\Qm)\bf w}}
{\sum\limits_{{\bf w}\in \B_{s}^n} {\rm e}^{\,-\beta {\bf w}^\dagger \Rrm(-\beta\Qm)\bf w}}
\,{\rm d} F_S(s) \quad 
\label{prop31b}
\end{equation}
with $\B_{s}^n$ denoting the $n$-fold Cartesian product of $\B_s$.
\end{prop}
Note that in (\ref{prop31b}) $\bf w$ is an $n$-dimensional vector and its components represent replicas of the \emph{same} user, while the elements of $\wv$ in (\ref{eq: Representation of the energy penalty in terms of replicas}) represent precoder outputs of \emph{different} users.

To produce explicit results, certain structures were imposed in \cite{zaidel:11} onto the matrix $\Qm$. More specifically, the limiting energy penalty and the limiting conditional distribution of the precoder's outputs were derived while employing the 1RSB ansatz, as well as the simpler RS ansatz (see, respectively, (\ref{eq182}) and (\ref{eq181})). In the following we only review the 1RSB results, and the reader is referred to \cite{zaidel:11} for the full scope of the analysis. As demonstrated therein, when compared to simulation results at finite numbers of antennas, 1RSB gives quite accurate approximations for the quantities of interest, while the RS ansatz does so only in special cases.

As discussed in Section  \ref{RepSymBre}, when applying the 1RSB ansatz the limiting properties of the precoder output are characterized by means of four macroscopic parameters $q_1, p_1, \chi_1, \mu_1 \in (0,\infty) $, which are determined for the vector precoding setting as specified below. Let $\Jm$ be an $N \times N$ random matrix satisfying the decomposability property (\ref{eq: Definition of the decomposability property}), and let $\Rrm(\cdot)$ denote the $\Rrm$-transform of its limiting eigenvalue distribution. Consider now the following function of complex arguments
\begin{equation}
\sFyz \triangleq   \e^{- \mu_1 \min\limits_{w \in \B_{s}} \varepsilon_1  \abs{w}^2 -2  \Re\set{w (f_1 z^* + g_1y^*)}   } \ , \ (y, z) \in \mathbb{C}^2 \ ,
\label{eq: 1RSB: Definition of the function sF}
\end{equation}
where $\Re\set{\cdot}$  takes the real part of the argument, and the parameters $\varepsilon_1$, $g_1$ and $f_1$ are defined as
\begin{eqnarray}
\varepsilon_1 & = &   \Rrm(-\chi_1) \label{eq: 1RSB: Expression for epsilion after taking derivative wrt q p and chi} \ ,\\
g_1 & = &
\sqrt{\frac{\Rrm(-\chi_1) - \Rrm(-\chi_1 - \mu_1 p_1)}{\mu_1}}
\label{eq: 1RSB: Expression for g after taking derivative wrt q p and chi} \ ,\\
f_1 & = & \sqrt{q_1 \Rrm'(-\chi_1-\mu_1 p_1)} \ ,
\label{eq: 1RSB: Expression for f after taking derivative wrt q p and chi}
\end{eqnarray}
where $\Rrm^\prime(\cdot)$ denotes the derivative of the $\Rrm$-transform.
Furthermore, we denote the normalized version of $\sFyz$ by
\begin{equation}
\tsFyz = \frac{\sFyz}{\int_\mathbb{C} \sFypz {\rm d}\tilde y} \quad .
\end{equation}
Then, using the two following integral notations
\begin{equation}\label{eq: Shortened notation for the Dz integral}
\int_\mathbb{C} (\cdot) \, \rD z \triangleq \int\limits_{-\infty}^\infty \int\limits_{-\infty}^\infty (\cdot) \, \frac{\e^{-\abs{z}^2}}{\pi} \, \rd z_\re \, \rd z_\im \  
\end{equation}
and
\begin{equation}\label{eq: Shortened notation for the DyDz integral}
\int_{\mathbb{C}^2} (\cdot) \, \rD y \, \rD z \triangleq \int_\mathbb{C}\int_\mathbb{C} (\cdot) \, \rD y \, \rD z \ 
\end{equation}
(with $z_\re,z_\im$ $\in$ $\mathbb{R}$ and $z$ $\triangleq$ $z_\re$$+j z_\im$), the parameters $\set{q_1, p_1, \chi_1, \mu_1}$ are solutions to the four following coupled equations:
\begin{eqnarray}
\chi_1  +  p_1\mu_1 &=
 \frac{1}{f_1} \iint_{\mathbb{C}^2}  \Re\set{z^*\argmin_{w \in \B_{s}} \abs{f_1 z + g_1 y - \varepsilon_1 w}} \nonumber \\ 
  & \phantom{\chi_1  +  p_1\mu_1 = \frac{1}{f_1} \iint_{\mathbb{C}^2}  \Re} \hfill \cdot \tsFyz  \rD y
 \, \rD z \, \rd F_S(s) \ , \label{eq: 1RSB: Equation after partial derivative wrt f - beta inf h zero}
\end{eqnarray}
\begin{eqnarray} 
 \chi_1 + (q_1 + p_1)\mu_1 &=
\frac{1}{g_1}  \iint_{\mathbb{C}^2}  \Re\set{y^*\argmin_{w \in \B_{s}} \abs{f_1 z + g_1 y - \varepsilon_1 w}}  \nonumber \\
& \phantom{\chi_1 + (q_1 + p_1)\mu_1 =  } \hfill \cdot \tsFyz   \rD y
 \, \rD z \, \rd F_S(s) \ , \label{eq: 1RSB: Equation after partial derivative wrt g  - beta inf h zero}
\end{eqnarray}
\begin{eqnarray}
q_1+p_1 &=
\iint_{\mathbb{C}^2} \abs{\argmin_{w \in \B_{s}} \abs{f_1 z + g_1 y - \varepsilon_1 w}}^2 \nonumber \\
& \phantom{q_1+p_1 = \iint_{\mathbb{C}^2} \argmin_{w \in \B_{s}} f_1 z +} \hfill \cdot \tsFyz  \rD y
 \, \rD z \, \rd F_S(s) \ , \label{eq: 1RSB: Equation after partial derivative wrt epsilon - beta inf h zero}
\end{eqnarray}
and
\begin{eqnarray}
 \int\limits_{\chi_1}^{\chi_1+\mu_1 p_1}
 \Rrm(-\omega)\, \rd \omega & =
  \iint_{\mathbb{C}} \log\left( \int_{\mathbb{C}}  \sFyz  \, \rD y \right)\, \rD z \, \rd F_S(s) \nonumber \\ 
  & \ -2\chi_1\Rrm(-\chi_1) \nonumber \\ 
  & \  +(\mu_1 q_1  +2 \chi_1 +2\mu_1 p_1) \Rrm(-\chi_1 -\mu_1p_1) \nonumber \\  
  & \ - 2\mu_1 q_1(\chi_1 + \mu_1 p_1)\Rrm^\prime(-\chi_1 - \mu_1p_1) \ .\label{eq: 1RSB: Equation after partial derivative wrt mu - beta inf and h zero}
\end{eqnarray}
These coupled equations have in general multiple solutions, 
which were found by minimizing the free energy with respect to the crosscorrelation matrix $\Qm$ at the saddle point. Since the energy penalty coincides with the free energy at zero temperature ($\beta\to\infty$), in case of multiple sets of solutions one has to choose that set which minimizes the energy penalty.

The limiting properties of the precoder outputs can now be summarized by means of the following two propositions. 
\begin{prop}[{\cite[Proposition IV.1]{zaidel:11}}]
\label{prop: 1RSB: Limiting Energy Penalty}
Suppose the random matrix $\Jm$ satisfies the decomposability property (\ref{eq: Definition of the decomposability property}). Then under some technical assumptions, including in particular \emph{one-step replica symmetry breaking}, 
the effective energy penalty per symbol $ \Esc^\tot / N$ converges in probability as $K,N \rightarrow\infty$, $N/K\rightarrow\alpha<\infty$, to
\begin{eqnarray}
\bE_\rsb &\triangleq
  \left(q_1+p_1+\frac{\chi_1}{\mu_1}\right)\Rrm(-\chi_1-\mu_1p_1) \nonumber \\
  & \quad - \frac{\chi_1}{\mu_1}\Rrm(-\chi_1)  - q_1(\chi_1 + \mu_1 p_1) \Rrm^\prime(-\chi_1 - \mu_1 p_1) \ .
 \label{eq: 1RSB: Limit of the effective energy penalty}
\end{eqnarray}
\end{prop}
\begin{prop}[{\cite[Proposition IV.2]{zaidel:11}}]\label{prop: 1RSB: Marginal limiting conditional distribution of x given u}
With the same underlying assumptions as in Proposition \ref{prop: 1RSB: Limiting Energy Penalty}, the limiting conditional empirical distribution of the nonlinear precoder's outputs given an input symbol $s$ satisfies
\begin{eqnarray}
P_{W|S}(\xi|\upsilon) & = \int_{\mathbb{C}^2} 1{\set{\xi = \argmin_{{w} \in \B_{\upsilon}} \abs{f_1 z + g_1 y - \varepsilon_1{w}}}} \nonumber \\
& \phantom{P_{W|S}(\xi|\upsilon) = \int_{\mathbb{C}^2} 1 \xi = \argmin_{{w} \in \B_{\upsilon}}  } \hfill \cdot \tsFyzup  \rD y \, \rD z  \ . \label{eq: 1RSB: Equation for the limiting conditional distribution - Infinite beta}
\end{eqnarray}
\end{prop}

The above properties were used in \cite{zaidel:11} to conduct an information-theoretic analysis of vector precoding. Particularizing to a zero-forcing (ZF) linear front-end and to Gaussian channel transfer matrices,  the spectral efficiency of vector precoding was expressed via the input-output mutual information of
the \emph{equivalent single-user channel} observed by each of the receivers.
The analysis was then applied to particular families of
extended alphabet sets (following \cite{mueller:08}), focusing on
a quaternary phase shift-keying (QPSK) input (while demonstrating
the RSB phenomenon). Numerical results therein indicate significant
performance enhancement over linear ZF
for medium to high SNRs. Furthermore, performance enhancement
is also revealed compared to a generalized THP approach
(which is a popular practical nonlinear precoding alternative
for MIMO GBCs). The reader is referred to \cite{zaidel:11} for the full details.

\subsection{3rd Example: Replica-based Estimation for Compressed Sensing}
\label{Exa53bis}

The recovery of a sparse vector through linear observation as in (\ref{AWGN}) can be cast  (see Section \ref{ComSen}) as a  noisy  compressive sensing problem, i.e. 
\[
\min_{\matr x \in \RR^K}||\matr x||_0\qquad \rm{subject\, to}\qquad \matr H \matr x+\matr n =\matr y\,. 
\]
Assume the observation matrix to have a product form, i.e. $\matr H =\matr A\matr S^{1/2}$. Here the matrix $\matr A$ is an $N\times K$ measurement matrix with i.i.d. entries with finite moments of any order, $A_{i,j}=1/\sqrt{N}A$, for some zero-mean unit variance random variable $A$. And the matrix $\matr S$ is a diagonal matrix containing positive scale factors, $s_j\,,j=1,\ldots,K$, which scale up the variance of the $j$-th component of $\matr x$ by $s_j$, and which are represented by random variable with finite moments. This model is able to capture variations in power of the sensed signal which are \emph{apriori} known to the estimator, while possible unknown  variations should be embodied in the distribution of $\matr x$.

Throughout this Section, we will assume that the components $x_j$ of $\matr x$ are i.i.d. with probability distribution $p_0(x_j)$, whose moments are all finite. The noise is such that $\matr w\sim \mathcal{N}(0,\sigma_0^2\matr I_N)$, and $\matr w$, $\matr x$, $\matr A$ and $\matr S$ are independent of each other.

The $\ell_0$-norm regularized estimator for $\matr x$ can be written as 
\begin{equation}\label{est0}
\widehat{\matr x}^0(\matr y)=\argmin_{\matr x \in \RR^K}\frac{1}{2\gamma}||\matr y -\matr A\matr S^{1/2}\matr x||^2_2+||\matr x||_0\,,
\end{equation} 
or, equivalently, as 
\begin{equation}\label{est0map}
\widehat{\matr x}^0(\matr y)=\argmin_{\matr x \in \RR^K}\frac{1}{2\gamma}||\matr y -\matr A\matr S^{1/2}\matr x||^2_2+\sum_{j=1}^Kf(x_j)\,,
\end{equation} 
with $\gamma >0$ an algorithm parameter and $f(x)$ the support function of the vector $\matr x$ (i.e. $f(x_j)=0$ if $x_j=0$ and $f(x_j)=1$ otherwise).  In the form (\ref{est0map}), $\widehat{\matr x}^0$ can be seen as a \emph{postulated MAP} (PMAP) estimator \cite[Sec. IV]{rangan:09}, where a prior distribution on $\matr x$, say $p_u(\matr x)$ driven by a positive parameter $u$, is postulated, and the estimator performance and complexity depend on the choice of the scalar valued, non-negative cost function $f:\Xi\rightarrow \RR$, where $\Xi$ is a measurable set such that $\matr x \in \Xi^K$.

The performance analysis of the estimator in (\ref{est0}) can be carried over via Replica method, assuming the symmetry is not broken. 

The possibility of decoupling the estimation of a vector of considerable size into the scalar estimation of each of its components separately is a crucial feature in the choice of the estimator to be implemented, since it clearly leads to enormous gain in terms of complexity. In \cite{guo:04}, such a decoupling is stated for the \emph{postulated MMSE} (PMMSE) estimators in the large-system limit and under several assumptions regarding the replica symmetry and the cost function behavior. 

In \cite{rangan:09}, instead, a framework for Replica analysis of PMAP estimators is provided. However, while in \cite{guo:04} an analytic test for the replica symmetry to hold is formulated, no such test is yet available for checking validity of symmetry assumptions in the MAP case.

The Replica analysis of  PMAP estimators is obtained  in \cite{rangan:09} as a limiting expression of the corresponding analysis for the  PMMSE estimators, asymptotically in the driving parameter $u$.

Following \cite{rangan:09}, we herein list the assumptions to be fulfilled in order for the PMAP estimator of $\matr x$ to be decoupled into $K$ scalar estimation problems, separately for each of its components.
\begin{itemize}
\item
Given a prior 
$$
p_u(\matr x)=\left[\int_{\matr x \in \Xi^K}e^{-uf(\matr x)}d\matr x\right]^{-1}e^{-uf(\matr x)}
$$
and a postulated noise level $\sigma_u^2=\gamma/u$, assume that for sufficiently large $u$ the \emph{postulated MMSE} estimator 
\begin{equation}\label{estmmse0}
\widehat{\matr x}^{pmmse}(\matr y)=\int_{\matr x \in \Xi^K}\matr x p_{\matr x|\matr y}(\matr x|\matr y; p_{post}, \sigma^2_{post}) d\matr x
\end{equation} 
satisfies the Replica Symmetric  decoupling property, i.e. there exists scalar MMSE estimators separately for each $x_j$ component \cite{guo:04};
\item
Let $\sigma^2_{eff}(u)$ and $\sigma^2_{p,eff}(u)$ be the effective noise levels$^{@}$\footnote{$^{@}$The effective noise level can be thought of as the noise seen by the linear scalar system when, while estimating a single component $x_j$ of the vector $\matr x$, all the contributions related to the remaining components $x_i\, i\neq j$ are aggregate to the Gaussian noise and hence perceived as noise. It is reminiscent of the concept of effective interference in multiuser detection, see \cite{hanly:99a}.}, i.e. the noise levels associated with the resulting decoupled scalar estimation problems,  when using the postulated prior $p_u$ and noise level $\sigma^2_u$, assuming that there exist both $\sigma^2_{eff,map}(u)=\lim_{u\rightarrow\infty}\sigma^2_{eff}(u)$ and $\gamma_p=\lim_{u\rightarrow\infty}u\sigma^2_{p,eff}(u)$;
\item
Let $\widehat{x_j^u}(K)$ be the MMSE estimate of $x_j$ with postulated $p_u$ and $\sigma^2_u$, and assume that the following holds in distribution:
$$
\lim_{u\rightarrow\infty}\lim_{K\rightarrow\infty}\widehat{x_j^u}(K)=\lim_{K\rightarrow\infty}\lim_{u\rightarrow\infty}\widehat{x_j^u}(K);
$$
\item
For every $\matr A$, $\matr S$, and $K$, and for almost every $\matr y$, (\ref{est0map}) achieves a unique essential minimum; 
\item
Assume $f(x)$ is non-negative and satisfies $\lim_{|x|\rightarrow\infty}\frac{f(x)}{\log|x|}=\infty$ for all sequences in $\Xi$ with $|x|\rightarrow\infty$;
\item
Defined the \emph{scalar postulated MAP} estimator 
\begin{equation}\label{smap}
\widehat{x}^{pmap}(z;\lambda)=\argmin_{x \in \Xi} F(x,z,\lambda)\,,
\end{equation} 
with $F(x,z,\lambda)=\frac{1}{2\lambda}|z-x|^2+f(x)$ and $\lambda$ and algorithm parameter, assume that (\ref{smap}) has a unique essential minimum and that, for almost all $\lambda$ and $z$, there exists $\sigma^2(z,\lambda)$ such that
$$
\lim_{x\rightarrow\widehat{x}}\frac{|x-\widehat{x}|^2}{2( F(x,z,\lambda)- F(\widehat{x},z,\lambda))}=\sigma^2(z,\lambda)\,,
$$
with $\widehat{x}=\widehat{x}^{pmap}(z;\lambda)$. 
\end{itemize}

Under the abovelisted assumptions and as $K\rightarrow\infty$, the random triplet $(x_j,s_j,\widehat{x_j}^{pmap})$ converges in distribution to the triplet $(x,s,\widehat{x})$ with independent components, distributed according $x\sim p_0(x)$, $s\sim p_S(s)$, and $v\sim\mathcal{N}(0,1)$, with
$\widehat{x}=\widehat{x}^{pmap}(z;\lambda_p)$, $z=x+\sqrt{\mu}v$, $\mu=\sigma^2_{eff,map}/s$ and $\lambda_p=\gamma_p/s$. 

The limiting effective noise levels can be determined from the set of coupled equations
\begin{equation}\label{sigmaeff}
\sigma^2_{eff,map}=\sigma^2_0+\beta\expect[s|x-\widehat{x}|^2]\,, 
\end{equation}
\begin{equation}\label{gammaeff}
\gamma_p=\gamma+\beta\expect[s\sigma^2(z;\lambda_p)]\,,
\end{equation}
with expectations taken over the laws of $s$, $x$ and $v$, in turn. 

In our specific case, i.e. as the cost function coincides with the $\ell_0$-norm, the scalar MAP estimator is given by
$$
\widehat{x}^{pmap}(z;\lambda)=\cases{z & $|z|>\sqrt{2\lambda}$ \cr 0 & elsewhere}.
$$
It can be shown \cite{rangan:09} that the solution to the set of fixed point equations
\begin{equation}\label{sigmaeff0}
\sigma^2_{eff,map}=\sigma^2_0+\beta\expect[s|x-\widehat{x}|^2]\,,
\end{equation}
\begin{equation}\label{gammaeff0}
\gamma_p=\gamma+\beta\gamma_p\Pr(|z|>\sqrt{2\lambda})\,,
\end{equation}
with expectations over $x\sim p_0(x)$, $s\sim p_S(s)$,  $z=x+\sqrt{\mu}v$, are the effective noise levels such that the random triplet $(x_j,s_j,\widehat{x_j})$ converges in distribution to $(x,s,\widehat{x}(z))$, with $z=x+\sqrt{\mu}v$. These fixed-point equations can be numerically solved.  
\subsection{Phase Transitions}
\label{PhaTra}

In thermodynamics, the occurrence of phase transitions, i.e.\ melting ice becomes water, is a well-known phenomenon.
In digital communications, however, such phenomena are less known, though they do occur.
The similarity between thermodynamics and multiuser detection pointed out in Section~\ref{Exa51}, should be sufficient to convince the reader that phase transitions in digital communications do occur.
Phase transitions in turbo decoding and detection of CDMA were found in \cite{agrawal:01} and \cite{tanaka:02}, respectively.

The phase transitions in digital communications are similar to the hysteresis in ferro-magnetic materials.
They occur if the equations determining the macroscopic parameters, e.g.\ $E$ determined by \eq{eqec}, have multiple solutions.
Then, it is the free energy who decides which of the solutions correspond to thermodynamic equilibrium.
If a system parameter, e.g.\ the load or the noise variance, changes, the free energy may shift its favor from the present to another solution.
Since each solution corresponds to a different macroscopic property of the system, changing the valid solution means that a phase transition takes place.

In digital communications, a popular macroscopic property is the bit error probability.
It is related to the macroscopic property $E$ in \eq{eqec} by \eq{eqber} for the case considered in Section~\ref{Exa51}.
Numerical results are depicted in Fig.~\ref{PhaTraL}.
\begin{figure}
\centerline{\epsfig{file=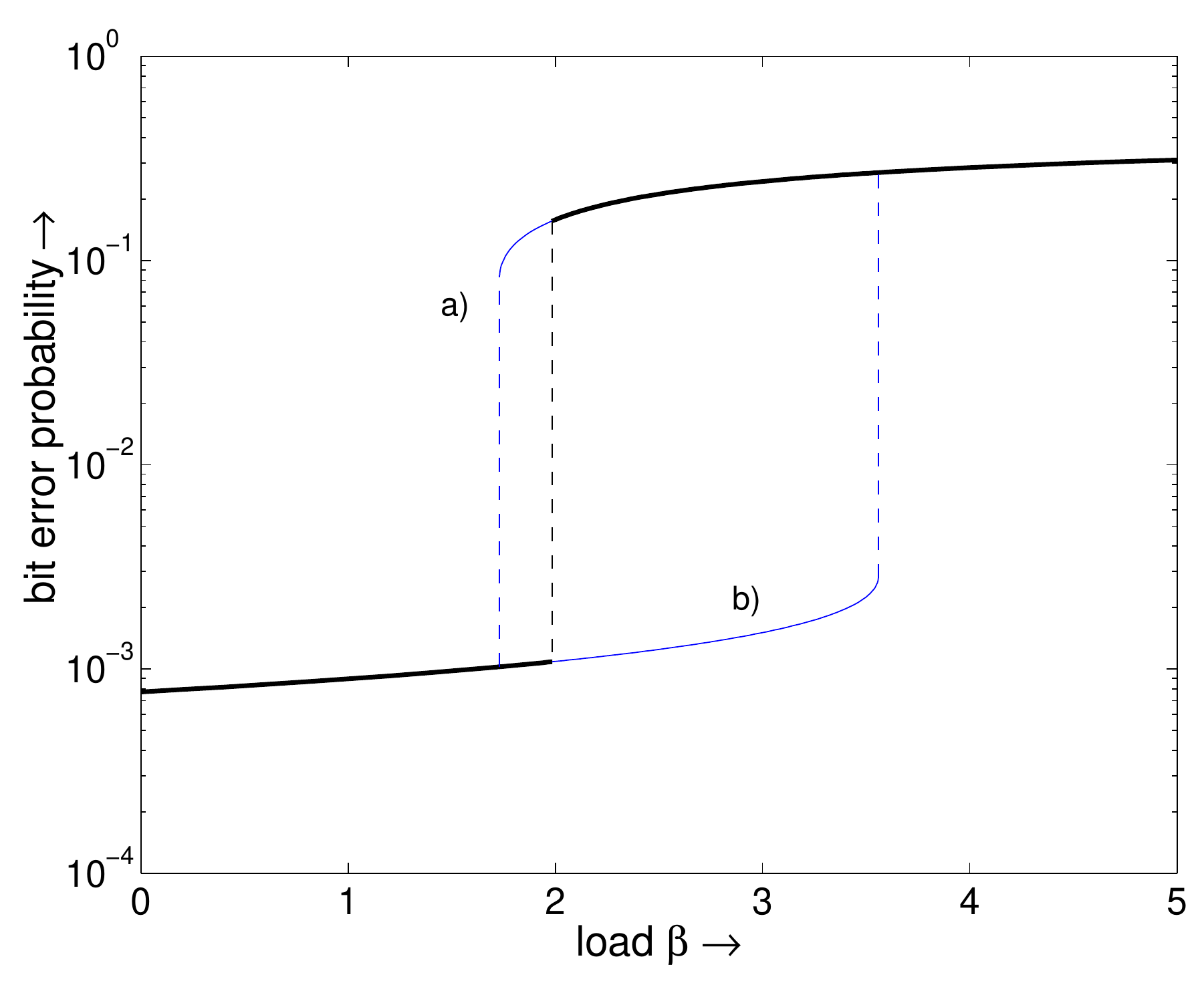,width=\columnwidth}}
\caption{\label{PhaTraL} Bit error probability for the individually optimum detector with uniform binary prior distribution versus system load for $10\log_{10}(E_{\rm s}/N_0) =$ {\rm 6~dB}.}
\end{figure}
The thick curve shows the bit error probability of the individually optimum detector as a function of the load.
The thin curves show alternative solutions for the bit error probability corresponding to alternative solutions to the equations for the macroscopic variable $\tilde E_k$.
Only for a certain interval of the load, approximately $1.73\le\beta\le3.56$ in Fig.~\ref{PhaTraL}, multiple solutions occur.
As expected, the bit error probability increases with the load.
At a load of approximately $\beta=1.986$ a phase transition occurs and lets the bit error probability jump.
Unlike ferromagnetic materials, there is no hysteresis effect for the bit error probability of the individually optimum detector, but only a phase transition.

In order to observe a hysteresis behavior, we can expand our scope to neural networks.
Consider a Hopfield neural network \cite{hopfield:82} implementation of the individually optimum multiuser detector which is an algorithm based on non-linear gradient search maximizing the energy function associated with the detector.
Its application to the problem of multiuser detection is discussed in \cite{kechriotis:96a}.
With an appropriate definition of the energy function, such a detector will achieve the performance of the upper curve in Fig.~\ref{PhaTraL} in the large system limit.
Thus, in the interval $1.73\le\beta\le1.986$ where the free energy favors the curve with lower bit error probability, the Hopfield neural network is suboptimum (labeled with {\em a}).
The curve labeled with {\em b} can also be achieved by the Hopfield neural network, but only with the help of a genie.
In order to achieve a point in that area, cancel with the help of a genie as many interferers as needed to push the load below the area where multiple solutions occur, i.e.\ $\beta<1.73$.
Then, initialize the Hopfield neural network with the received vector where the interference has been canceled and let it converge to the thermodynamic equilibrium.
Then, slowly add one by one the interferers you had canceled with the help of the genie while the Hopfield neural network remains in the thermodynamic equilibrium by performing iterations.
If all the interference suppressed by the genie has been added again, the targeted point on the lower curve in area {\em b} is reached.
The Hopfield neural network follows the lower curve, if interference is added, and it follows the upper line, if it is removed.

It should be remarked that a hysteresis behavior of the Hopfield neural network detector does not occur for any definition of the energy function and any prior distribution of the data to be detected, but additional conditions on the microscopic configuration of the system need to be fulfilled.
\subsection{3rd Example: Compressed Sensing}
\label{Exa53}

 An interesting instance of a phase transition is observed in compressed sensing (Section \ref{ComSen}). Such a transition occurs in the dependence of the sparsity of the signal. As mentioned in Section \ref{ComSen}, most results are available for IID measurement matrices, but  most physical phenomena indeed produce correlated measurements. An index of statistical independence among entries or columns of a random matrix, particularly meaningful in the Compressed Sensing setting, is  the so-called \emph{coherence}.  It is defined as  the largest magnitude of the off-diagonal entries of the sample correlation (resp. covariance) matrix generated from the random matrix itself.
Let us denote the correlation-based coherence of the $N$$\times$$K$ random matrix $\matr H$ as
\[
L_N=\max_{1\leq i\leq j\leq K}|\rho_{i,j}|\,,
\]
and  the covariance-based, respectively,  as
\[
\widetilde{L}_N=\max_{1\leq i\leq j\leq K}|\widetilde{\rho}_{i,j}|\,.
\]
The condition that ensures both exact recovery  in the noiseless case as well as  and the stable recovery in the noisy case, for  an $s$-sparse signal is  \cite{Donoho:01,Fuchs:04}
\begin{equation}\label{eq:coherence}
(2s-1)\widetilde{L}_N< 1\,. 
\end{equation}

Clearly, $\widetilde{L}_N$ is a random variable, driving the reconstruction performance. Interestingly, its  law undergoes phase transition depending on the scaling of the dimension of the sensed signal $K$ versus the sample size $N$. This basically means that an increasing value of the coherence yields to ever worsening reconstruction performance, up to a certain point where phase transition occurs, and traditional sampling theory should be used therebeyond. 

In particular, it can be shown that \cite[Theorems IV and V]{Phtrans12}
\begin{theorem}
Assume the $K$ columns of $\matr H$ are independent $N$-dimensional random vectors with a common spherical distribution (which may depend on $N$) having no probability masses in the origin, and let $\widetilde{T}_N=\log(1-\widetilde{L}^2_N)$.
Then,
\begin{itemize}
\item If we assume that $(\log K)/N\rightarrow 0$ as $n\rightarrow\infty$, then $\widetilde{L}_N\rightarrow 0$ in probability and
$$
N\widetilde{T}_N+4\log K-\log\log K
$$
converges weakly in distribution  to the law
$$
F(y)=1-\exp\left\{-\frac{e^{y/2}}{\sqrt{8\pi}}\right\}\,,
$$
for any real $y$.
\item
If we are in the transitional case for which $(\log K)/\sqrt{N}\rightarrow \alpha \in [0,\infty)$ as $N\rightarrow\infty$, then
$$
N\widetilde{L}^2_N-4\log K+\log\log K
$$
converges weakly to the distribution function
$$
F(y)=\exp\left\{-\frac{e^{-(y+8\alpha^2)/2}}{\sqrt{8\pi}}\right\}\,,
$$
for any real $y$.
\item
If $(\log K)/N\rightarrow c \in (0,\infty)$, then $\widetilde{L}_N\rightarrow \sqrt{1-e^{-4c}}$ in probability and
$$
N\widetilde{T}_N+4\log K-\log\log K
$$
converges weakly in distribution  to the law
$$
F(y)=1-\exp\left\{-\sqrt{\frac{c}{2\pi(1-e^{-4c})}}e^{(y+8c)/2}\right\}\,,
$$
for any real $y$.
\item
If $(\log K)/N\rightarrow \infty$ as $N\rightarrow\infty$, then $\widetilde{L}_N\rightarrow 1$ in probability and $\frac{N\widetilde{T}_N}{\log K}\rightarrow -4$ always in probability as $N\rightarrow\infty$, while
$$
N\widetilde{T}_N+\frac{4N\log K}{N-1}-\log N
$$
converges weakly in distribution  to the law
$$
F(y)=1-\exp\left\{\frac{e^{y/2}}{\sqrt{2\pi}}\right\}\,,
$$
for any real $y$.
\end{itemize}
\end{theorem}
By the above-listed results, explicit bounds to signal sparsity can be stated. In particular, taking as an instance the sub-exponential case, e.g. $(\log K)/N\rightarrow 0$ as $N\rightarrow\infty$, we recover
$$
\widetilde{L}^2\sim 2\sqrt{\frac{\log K}{N}}\,,
$$
and in order to meet the condition on the coherence behavior (\ref{eq:coherence}), the signal sparsity $s$ should satisfy
$$
s < \frac{1}{4}\sqrt{\frac{N}{\log K}}.
$$


\subsection{4th Example: Low-Density Parity Check Codes}
\label{Exa54}

The analysis of error-correcting codes by means of the replica method was first performed by Sourlas \cite{sourlas:89}. A helpful tutorial treatment can be found in \cite{nishimori:01}. Both works describe error-correcting codes from a purely physics perspective. In the following, we try to present the analysis keeping a balance between physics and engineering perspectives.

Define the operator $\tilde\cdot$ as follows
\begin{equation}
\tilde x = 1-2x.
\end{equation}
For vectors the operator works component by component.
It is helpful to express modulo-2 arithmetic with standard algebra since
\begin{equation}
\widetilde{x\oplus y} = \tilde x \tilde y
\end{equation}
where $\oplus$ denotes modulo-2 addition.
Consider a binary symmetric channel.
Define the inverse temperature
\begin{equation}
\beta = \frac 12 \ln \frac{1-\sigma_0^2}{\sigma_0^2}.
\end{equation}
With the help of the inverse temperature, the binary symmetric channel can be described as
\begin{equation}
\Pr(y_n|{ \matr x} )= \frac{{\e}^{\,\beta \tilde y_n  \prod \limits_{k\in{\cal H}_n} \tilde x_k}}{2\cosh\beta}
\end{equation}
where $ {\cal H}_n = \{ k: (\matr H)_{n,k}=1\}$ is the set that contains the indices of the symbols that enter the $n^{\rm th}$ paritity check equation, $y_n$ the code symbol received at time instant $n$, and $x_k$ the $k^{\rm th}$ source symbol.
For a memoryless channel, we have
\begin{eqnarray}
\Pr( {\matr y}|{\matr x}) & = & \prod \limits_{n=1}^N \Pr( y_n|{\matr x}) = \frac{{\e}^{\,\beta \sum\limits_{n=1}^N \tilde y_n \prod\limits_{k\in{\cal H}_n} \tilde x_k}}{(2\cosh\beta)^N}.
\end{eqnarray}
Using Bayes law, the posterior probability is given by
\begin{eqnarray}
\label{bl}
\Pr( {\matr x}|{\matr y}) &=& \Pr(\matr y|\matr x) \cdot \frac{\Pr(\matr x)}{\Pr(\matr y)}
\end{eqnarray}
On symmetric channels, the highest data rate is achieved for uniform input distribution (also referred to as uniform prior), i.e.\ $\Pr(\matr x) = 2^{-K}$.
For uniform prior, we get
\begin{eqnarray}
\label{pp}
\Pr( {\matr x}|{\matr y})&=& \frac{{\e}^{\,\beta \sum\limits_{n=1}^N \tilde y_n \prod\limits_{k\in{\cal H}_n} \tilde x_k}}{\sum\limits_{\matr \x}{\e}^{\,\beta \sum\limits_{n=1}^N \tilde y_n \prod\limits_{k\in{\cal H}_n} \tilde \x_k}}.
\end{eqnarray}

Consider now a binary input Gaussian channel with received signal $\matr y$ containing noise of variance $\sigma_0^2$ and useful signals components of power $P$. Note that on the Gaussian channel, the received signal $\matr y$ is not binary.
Define the inverse temperature
\begin{equation}
\beta = \frac{\sqrt P}{\sigma_0^2}.
\end{equation}
With the help of the inverse temperature, the binary input Gaussian channel can be described as
\begin{equation}
\Prob{ \matr y |{ \matr x}}{\matr y,\matr x}=  \prod\limits_{n=1}^N \frac{{\e}^{\,-\frac{\beta}{2\sqrt P} \left(y_n  - \sqrt P \prod \limits_{k\in{\cal H}_n} \tilde x_k\right)^2}}{\sqrt{2\pi}\sigma_0}.
\end{equation}
Using Bayes law \eq{bl}, the posterior probability for uniform prior on a memoryless channel is given by
\begin{eqnarray}
\Prob{\matr x|\matr y}{ {\matr x},{\matr y}}&=& \frac{{\e}^{\,\beta \sum\limits_{n=1}^N  y_n \prod\limits_{k\in{\cal H}_n} \tilde x_k}}{\sum\limits_{\matr \x}{\e}^{\,\beta \sum\limits_{n=1}^N  y_n \prod\limits_{k\in{\cal H}_n} \tilde \x_k}}.
\end{eqnarray}
which is a generalization of \eq{pp} for non-binary output.
Apparently, the posterior probability of a binary symmetric channel is a special case of its counterpart for the binary input Gaussian channel under the bijective transformation $y_n \mapsto \tilde y_n$. Thus, there is no need to distinguish these two channels in the sequel.

The $m^{\rm th}$ power of the partition function is given by
\begin{equation}
Z^m = \sum\limits_{\{\matr \x_a\}} {\e}^{\,\beta  \sum\limits_{n=1}^N y_n \sum\limits_{a=1}^m \prod\limits_{k\in{\cal H}_n} \tilde \x_{a,k}}
\end{equation}
and its configurational average for uniform prior, i.e.\ $\Pr(\matr x)=2^{-K}$
\begin{eqnarray}
\expect\limits_{\matr y} Z^m &=&  \int Z^m {\rm d}\Prob{\matr y}{\matr y}\\
&=& \int\sum\limits_{\matr x} Z^m \Pr(\matr x) {\rm d}\Prob{\matr y|\matr x}{\matr y,\matr x}  \\
&=&2^{-K} \int\sum\limits_{\matr x_0} Z^m  {\rm d}\Prob{\matr y|\matr x}{\matr y,\matr x_0}\\
&=& 2^{-K} \int\limits_{\RR^N}\sum\limits_{\matr x_0} \sum\limits_{\{\matr \x_a\}}\prod\limits_{n=1}^N{\e}^{\,\beta  y_n \sum\limits_{a=1}^m \prod\limits_{k\in{\cal H}_n} \tilde \x_{a,k}}\nonumber \\
&& \times
 \frac{{\e}^{\,-\frac{\beta}{2\sqrt P} \left( y_n  - \sqrt P \prod \limits_{k\in{\cal H}_n} \tilde x_{0,k}\right)^2}}{\sqrt{2\pi}\sigma_0}{\rm d}y_n.
\end{eqnarray}
Now, we apply a change of variables.
We substitute
\begin{eqnarray}
\label{gauge1}
\upsilon_n &=&  y_n  \prod\limits_{k\in{\cal H}_n} \tilde x_{0,k}\\
\xi_{a,k} &=&  \tilde \x_{a,k} \tilde x_{0,k}
\label{gauge2}
\end{eqnarray}
and get
\begin{eqnarray}
\expect\limits_{\matr y} Z^m &=&  \sum\limits_{\{\matr \xi_a\}}\int\limits_{\RR^N}\prod\limits_{n=1}^N{\e}^{\,\beta   \upsilon_n \sum\limits_{a=1}^m \prod\limits_{k\in{\cal H}_n}  \xi_{a,k}}
 \cdot\frac{{\e}^{\,-\frac{\beta}{2\sqrt P} \left( \upsilon_n  - \sqrt P \right)^2}}{\sqrt{2\pi}\sigma_0}{\rm d}\upsilon_n\\
 &=& \sum\limits_{\{\matr \xi_a\}}\prod\limits_{n=1}^N{\e}^{\,\frac\beta 2\sqrt P  \left( \sum\limits_{a=1}^m \prod\limits_{k\in{\cal H}_n}  \xi_{a,k}\right)^2+\beta \sqrt P\sum\limits_{a=1}^m \prod\limits_{k\in{\cal H}_n}  \xi_{a,k}}\\
 &=& \sum\limits_{\{\matr \xi_a\}}\prod\limits_{n=1}^N{\e}^{\,\beta \sqrt P\sum\limits_{a=1}^m \prod\limits_{k\in{\cal H}_n}  \xi_{a,k}+ \frac12\sum\limits_{b=1}^m \prod\limits_{k\in{\cal H}_n}  \xi_{a,k} \xi_{b,k}}
\end{eqnarray}
The substitutions \eqs{gauge1}{gauge2} are known as the {\em gauge transformation} in physics literature \cite{nishimori:01}.
In coding theory, it is a well-known property of linear codes that one can assume transmission of the all-zero codeword $\matr x =\matr 0$, i.e.\ $\tilde x_k=+1, \forall k$ without loss of generality \cite{mackay:03}.
Depending on the code properties, i.e.\ the choice of the sets ${\cal H}_n$, a large variety of codes can be analyzed and various results are obtained. For further results, the reader is referred to, e.g.\ \cite{sourlas:01,kabashima:04}.

The replica method is not the only tool from statistical physics that is applied for the analysis of error correction codes. Further approaches can be found in, e.g.\ \cite{montanari:05,kudekar:09}


\section*{Acknowledgment}
The authors would like to thank Antonia Tulino, Toshiyuki Tanaka, Sergio Verd\'u, Giuseppe Caire, Stephen Hanly, Laura Cottatellucci, Dongning Guo, Aris Moustakas, and Nurul Mahmood.

\section*{References}

\bibliography{lit_BZ}
\bibliographystyle{IEEEbib}

\end{document}